\documentclass{article}

\usepackage{arxiv}

\usepackage{tikz}
\usepackage[utf8]{inputenc} 
\usepackage[T1]{fontenc}    
\usepackage{hyperref}       
\hypersetup{colorlinks,allcolors=black}
\usepackage{url}            
\usepackage{booktabs}       
\usepackage{amsfonts}       
\usepackage{nicefrac}       
\usepackage{microtype}      
\usepackage{lipsum}

\usepackage{natbib}
\usepackage{amssymb}
\usepackage{lineno,hyperref}
\usepackage{amsmath}
\usepackage{bm}
\usepackage{subcaption}
\usepackage{graphicx}
\usepackage{svg}
\usepackage{rotating}
\usepackage{lscape}
\usepackage{array,multirow}
\usepackage[title]{appendix}
\usepackage{float}
\usepackage{stmaryrd}
\usepackage{textcomp}
\usepackage{algorithm}
\usepackage{algpseudocode}
\usepackage{listings}
\usepackage{xcolor}
\usepackage{siunitx}
\usepackage{symbols}
\usepackage{pifont}
\usepackage{lineno}
\usepackage{import}
\usepackage{tikz}
\usetikzlibrary{quotes,arrows.meta}
\usetikzlibrary{positioning}

\usepackage{Ball}
\usepackage{Box}
\usepackage{RightBandedBox}

\usetikzlibrary{positioning}
\usetikzlibrary{3d} 

\def\checkmark{\tikz\fill[scale=0.4](0,.35) -- (.25,0) -- (1,.7) -- (.25,.15) -- cycle;}

\newtheorem{remark}{Remark}

\newcommand{\beginapp}{%
 \setcounter{equation}{0}
        \renewcommand{\theequation}{\thesection.\arabic{equation}}%
        \setcounter{table}{0}
        \renewcommand{\thetable}{\thesection.\arabic{table}}%
        \setcounter{figure}{0}
        \renewcommand{\thefigure}{\thesection.\arabic{figure}}%
     }

\title{
Non-intrusive reduced order modeling of natural convection in porous media using convolutional autoencoders: comparison with linear subspace techniques }

\author{
  Teeratorn Kadeethum \\
  Sibley School of Mechanical and Aerospace Engineering \\
  Cornell University, 
   New York, USA \\
  \texttt{tk658@cornell.edu} \\
  
   \And
 Francesco Ballarin\\
  Department of Mathematics and Physics\\
  Catholic University of the Sacred Heart, Brescia, Italy \\
  \texttt{francesco.ballarin@unicatt.it} \\
  
    \And
 Youngsoo Choi \\
  Lawrence Livermore National Laboratory \\
   California, USA \\
  \texttt{choi15@llnl.gov} \\

  \And
 Daniel O'Malley \\
  Los Alamos National Laboratory \\
   New Mexico, USA \\
  \texttt{omalled@lanl.gov} \\

   \And
 Hongkyu Yoon \\
  Sandia National Laboratories \\
   New Mexico, USA \\
  \texttt{hyoon@sandia.gov} \\

   \And
 Nikolaos Bouklas \\
  Sibley School of Mechanical and Aerospace Engineering  \\
  \& Center for Applied Mathematics\\
  Cornell University,
   New York, USA \\
  \texttt{nb589@cornell.edu} \\
  }

\renewcommand{\shorttitle}{Non-intrusive reduced order modeling of natural convection in porous media}

\begin{document}
\maketitle

\begin{abstract}

Natural convection in porous media is a highly nonlinear multiphysical problem relevant to many engineering applications (e.g., the process of $\mathrm{CO_2}$ sequestration). Here, we present a non-intrusive reduced order model of natural convection in porous media employing deep convolutional autoencoders for the compression and reconstruction and either radial basis function (RBF) interpolation or artificial neural networks (ANNs) for mapping parameters of partial differential equations (PDEs) on the corresponding nonlinear manifolds. To benchmark our approach, we also describe linear compression and reconstruction processes relying on proper orthogonal decomposition (POD) and ANNs. We present comprehensive comparisons among different models through three benchmark problems. The reduced order models, linear and nonlinear approaches, are much faster than the finite element model, obtaining a maximum speed-up of $7 \times 10^{6}$ because our framework is not bound by the Courant–Friedrichs–Lewy condition; hence, it could deliver quantities of interest at any given time contrary to the finite element model. Our model's accuracy still lies within a mean squared error of 0.07 (two-order of magnitude lower than the maximum value of the finite element results) in the worst-case scenario. We illustrate that, in specific settings, the nonlinear approach outperforms its linear counterpart and vice versa. We hypothesize that a visual comparison between principal component analysis (PCA) or t-Distributed Stochastic Neighbor Embedding (t-SNE) could indicate which method will perform better prior to employing any specific compression strategy.
\end{abstract}


\section{Introduction}

The problem of natural convection in porous media is involved in various engineering applications ranging from geothermal energy, seawater intrusion, contaminant transport to the storage of nuclear and radioactive waste \citep{taron2009thermal,nick2013reactive,zheng2002applied,rutqvist2005numerical}. Typically, the problem behavior is characterized by the Rayleigh number ($\mathrm{Ra}$), a dimensionless index indicating laminar or turbulent flow regimes. Additionally, an alteration in the fluid's temperature or concentration resulting in variations in fluid density and viscosity (i.e., formation heterogeneity) could, in turn, influence the fluid flow field through fingering instabilities \citep{nield2006convection}. One of many practical examples of interest is the process of carbon dioxide ($\mathrm{CO_2}$) sequestration in aquifers. Depending on system $\mathrm{Ra}$ and the formation heterogeneity, convective mixing can greatly accelerate $\mathrm{CO_2}$ dissolution during geologic carbon storage \citep{hassanzadeh2007scaling,neufeld2010convective}. \par

To approximate the problem of natural convection in porous media, we utilize the mixed finite element method and interior penalty enriched Galerkin approximation for spatial discretization and the 4th-order backward differentiation formula for time-stepping as our full order model (FOM) as previously implemented by the authors \citep{kadeethum2021locally}. This combination is selected to guarantee local mass conservation and accurately capture the gravity-driven flow in the advection-dominated regime \citep{zhang2016mixed,hosseini2020numerical}. However, this FOM requires substantial computational resources because it is highly nonlinear and needs a very small time-step to satisfy Courant–Friedrichs–Lewy (CFL) condition \citep{diersch1988finite,frolkovivc2000numerical,cheng2020upscaling,kolditz1998coupled}. Hence, it is not directly suitable to handle large-scale inverse problems, optimization, or control, in which an extensive set of simulations must be explored \citep{amsallem2015design,mcbane2021component,choi2019accelerating,choi2020gradient,ballarin2019pod,hansen2010discrete,hesthaven2016certified}.

Alternatively, a reduced order model (ROM) has been developed to decrease computational cost while maintaining an acceptable accuracy, i.e., depending on applications, \citep{schilders2008model,schilders2008introduction}. Recently, the ROM methodology has been applied to a range of parameterized problems (i.e., to repeated evaluations of a problem) depending on a set of parameters ($\bm{\mu}$), which could correspond to physical properties, geometric characteristics, or boundary conditions \citep{copeland2021reduced, hoang2020domain,kim2020efficient2,choi2020sns,choi2019space,choi2021space,carlberg2018conservative,ballarin2019pod,venturi2019weighted,hesthaven2016certified}. It is generally composed of two stages, (1) the offline and (2) online stages \citep{hesthaven2016certified}. The offline stage begins with initializing a set of input parameters,  which we call a training set. Then the FOM is solved corresponding to each member in the training set (in the following, we will refer to the corresponding solutions as snapshots). Data compression techniques, which could be linear or nonlinear compression, are then used to compress FOM snapshots to produce basis functions that span a reduced (linear of nonlinear) space of very low dimensionality but still guarantee accurate reproduction of the snapshots \citep{decaria2020artificial,cleary1984data,hijazi2020effort}. The ROM can then be solved during the online stage for any new value of $\bm{\mu}$ by seeking an approximated solution in the reduced space. \par

For complex coupled processes a non-intrusive ROM is attractive because it does not require any cumbersome modifications of FOM source codes \citep[e.g.][]{hoang2021projection,xie2019non,xiao2019domain,hesthaven2018non,mignolet2013review}. This alleviates several code compatibility complications since many of the source codes used to build FOMs may not be available or easily accessible, especially in legacy codes or commercial software \citep{xiao2015non1}. Consequently, the non-intrusive variants of ROM can provide more flexibility in coupling to any existing FOM platforms. We focus on two ways of building a ROM; the first path in, the linear approach, follows the recent development of \cite{hesthaven2018non}, which has been adapted and applied to a wide range of problems \citep[e.g.][]{girfoglio2020nonA,ortali2020gaussian,demo2020efficient,gadalla2020comparison,kadeethum2021non}. This approach utilizes a set of reduced basis functions, which are linearly extracted from high-fidelity FOM solutions through proper orthogonal decomposition (POD). Then, the ROM solution, which belongs in a linear subspace of the solution manifold, is reconstructed from these reduced basis functions. In the second path, the nonlinear approach, we propose the compression of the high-fidelity FOM solutions through an autoencoder (AE in short) \citep[e.g.][]{kim2020fast, kim2020efficient, phillips2020autoencoder, nikolopoulos2021non, lee2020model, fresca2021comprehensive} learning a nonlinear manifold for the ROM solution. During the prediction phase, we reconstruct our ROM using nonlinear reconstruction (i.e., using a trained decoder to reconstruct our quantities of interest with a given nonlinear manifold). We want to emphasize that in contrast to the work presented in \cite{kim2020fast, phillips2020autoencoder, lee2020model}, our framework relies on the non-intrusive approach similar to \cite{nikolopoulos2021non,fresca2021comprehensive}, which avoids any intrusive modifications of FOM source code. This type of non-intrusive approach for ROM using AE is termed Deep-Learning ROM (DL-ROM) by \cite{fresca2021comprehensive}. Hence, the framework could be easily extended to other engineering problems. This work has three main novelties

\begin{enumerate}
    \item We propose a data-driven framework  to approximate natural convection in porous media. The framework provides an online speed-up of $7 \times 10^{3}$ - $7 \times 10^{6}$, depending on how many evaluations of the ROM are required. As the framework stacks time-domains into the training set, it does not require a large number of training parameters $\bm{\mu}$ to achieve the desired accuracy and avoid overfitting (see Remark \ref{remark:2} for more details). We also provide a comprehensive performance comparison between linear and nonlinear compression and reconstruction techniques. 
    
    \item The framework can handle data provided from FOM with adaptive time-stepping, which is essential for an advection-dominated problem (i.e., to satisfy CFL condition). The framework can also deliver the solution at any time, including times that do not exist in the training snapshots. This is an asset of our ROM because it is not bound by CFL conditions and can evaluate our quantities of interest at any timestamps that we require. For instance, we may be interested in a temperature field at one, two, and three hours with given $\mathrm{Ra}$ value. To achieve that using FOM, you may need to go through many steps in between. However, through ROM, you only need to evaluate it at those three times. 
    
    \item We provide a qualitative way to select between linear and nonlinear approaches using latent space visualization. We note that even though governing equations remain the same throughout this study, the linear approach might be a better choice than its nonlinear counterpart and vice-versa depending on geometry and boundary conditions.
\end{enumerate}


The rest of the manuscript is summarized as follows. We begin with the model description and corresponding governing equations in Section \ref{sec:governing_equations}. The presentation of the discretization of the natural convection problem and its solver could be found in appendix \ref{sec:fem}. The two variants of the ROM framework following the linear and nonlinear approaches are discussed in Section \ref{sec:data_driven}. We then illustrate our ROM framework using three numerical examples in Section \ref{sec:numer_examples}. We provide discussions on the ROM's performance, comparison between linear and nonlinear compression and reconstruction, and qualitative guidelines for selecting the subspace (between linear subspace and nonlinear manifold) in Section \ref{sec:discussion}.



\section{Problem description} \label{sec:governing_equations}

This section briefly describes all the equations used in this study, namely mass balance and heat advection-diffusion equations. Let $\Omega \subset \mathbb{R}^d$ ($d \in \{1,2,3\}$) denote the computational domain and $\partial \Omega$ denote the boundary. $\bm{X}^{*}$ are spatial coordinates in $\Omega$, e.g. $\bm{X}^{*}=[x^{*}, y^{*}]$ when $d=2$. The time domain is denoted by $\mathbb{T} = \left(0,\tau\right]$ with $\tau>0$. Primary variables used in this paper are $\bm{u}^* (\cdot , t^*) : \Omega \times  \mathbb{T} \to \mathbb{R}^d$, which is a vector-valued Darcy velocity (\si{m/s}), $p^* (\cdot ,  t^*) : \Omega \times  \mathbb{T} \to \mathbb{R}^d$, which is a scalar-valued fluid pressure (\si{Pa}), and $T^* (\cdot , t^*) : \Omega \times \mathbb{T} \to \mathbb{R}^d$, which is a scalar-valued fluid temperature (\si{C}). Time is denoted as $t^*$ (\si{s}). 

Applying the Boussinesq approximation \citep{joseph2013stability} to the mass balance equations results in the density difference only appearing in the buoyancy term. The mass balance equations are

\begin{equation} \label{eq:mass_dim}
\bm{u}^{*}+\bm{\kappa}\left(\nabla p^{*}+{\mathbf{y}}\left(\rho-\rho_{0}\right) g\right)=0,
\end{equation}

\noindent
and

\begin{equation} \label{eq:div_0_dim}
\nabla \cdot \bm{u}^{*}=0,
\end{equation}

\noindent
where $\bm{\kappa}=\bm{k}/\mu_f$ is the porous medium conductivity, $\mu_f$ is the fluid viscosity, $\mathbf{y}$ is a unit vector in the direction of gravitational force, $g$ is the constant acceleration due to gravity, $\rho$ and $\rho_0$ are the fluid density at current and initial states, respectively. In addition, $\bm{k}$ is the matrix permeability tensor defined as

\begin{equation} \label{eq:permeability_matrix}
\bm{k} :=\left[ \begin{array}{ll}{{k}_{xx}} & {{k}_{xy}}  \\ {{k}_{yx}} & {{k}_{yy}} \\ \end{array}\right], \quad  \text{if} \ d = 2,  
\end{equation}


\noindent
Here, we only forcus on 2D domain, $d=2$, and simplify our problem by assuming $\bm{k}$ is homogeneous and isotropic, which means $\bm{k} \to k$ and $\bm{\kappa} \to \kappa$.

We assume that $\rho$ is a linear function of $T^{*}$ \citep{chen2006computational, zhang2016mixed} following

\begin{equation} \label{eq:rho}
\rho=\rho_{0}\left(1-\alpha\left(T^{*}-T_{0}^{*}\right)\right),
\end{equation}

\noindent
where $\alpha$ is the thermal expansion coefficient, and $T_{0}^{*}$ is the reference fluid temperature. We note that Equation \eqref{eq:rho} is the simplest approximation, and one may easily adapt the proposed method when employing a more complex relationship provided in \cite{lake2014fundamentals}. The heat advection-diffusion equation reads

\begin{equation} \label{eq:temp_dim}
\gamma \frac{\partial T^{*}}{\partial t^{*}}+\bm{u}^{*} \cdot \nabla T^{*}-K \nabla^{2} T^{*} - {f_c}^* = 0.
\end{equation}

\noindent
Here, $\gamma$ is the ratio between the porous medium heat capacity and the fluid heat capacity, $K$ is the effective thermal conductivity, and ${f_c}^*$ is a sink/source. We follow \cite{nield2006convection} and define dimensionless variables as follows

\begin{equation} \label{eq:dim_less}
\bm{X}:=\frac{1}{H} \bm{X}^{*}, \quad t:=\frac{\kappa}{\mu \gamma H^{2}} t^{*}, \quad p:=\frac{\kappa}{K} p^{*}, \quad \bm{u}:=\frac{H}{K} \bm{u}^{*}, \quad T:=\frac{T^{*}-T_{0}^{*}}{\Delta T^{*}}, \quad f_c:=\frac{t^*}{\Delta T^{*}}{f_c}^*,
\end{equation}

where $H$ is the dimensional layer depth, and $\Delta T^{*}$ is the temperature difference between two boundary layers. Consequently, we could rewrite our Equation \eqref{eq:mass_dim} and Equation \eqref{eq:div_0_dim} using the dimensionless variables as


\begin{equation} \label{eq:mass_dimless}
\begin{split}
\bm{u}+\nabla p-\mathbf{y} \operatorname{Ra} T=0, &\text { \: in \: } \Omega \times \mathbb{T}, \\ 
\nabla \cdot \bm{u}=0,  &\text { \: in \: } \Omega \times \mathbb{T}, \\ 
p=p_{D} &\text { \: on \: } \partial \Omega_{p} \times \mathbb{T}, \\ 
\bm{u} \cdot \mathbf{n}=q_{D} &\text { \: on \:} \partial \Omega_{q} \times \mathbb{T}, \\ 
p=p_{0} &\text { \: in \: } \Omega \text { at } t = 0,
\end{split}
\end{equation}

where $\partial \Omega_{p}$ and $\partial \Omega_{q}$ are the pressure and flux boundaries, respectively, and $\mathrm{Ra}$ is the Rayleigh number defined as

\begin{equation} \label{eq:Ra}
\mathrm{Ra}:=\frac{g \alpha \kappa \Delta T^{*} H}{K}.
\end{equation}

We then write Equation \eqref{eq:temp_dim} in dimensionless form as


\begin{equation} \label{eq:temp_dimless}
\begin{split}
\frac{\partial T}{\partial t}+\bm{u} \cdot \nabla T-\nabla^{2} T - f_c=0,  &\mbox{ \: in \: } \Omega \times (0,\mathbb{T}], \\  
T = T_D  &\mbox{ \: on \: } \partial \Omega_{{T}} \times (0,\mathbb{T}], \\ 
(-\bm{u} T+\nabla T) \cdot \mathbf{n}={T_{\rm in}} \bm{u}\cdot \mathbf{n}  &\mbox{ \: on \: } \partial \Omega_{{\rm in}} \times (0,\mathbb{T}],  \\ 
\nabla T \cdot \mathbf{n} = 0  &\mbox{ \: on \: } \partial \Omega_{{\rm out}} \times (0,\mathbb{T}], \\ 
T={T_{0}}  &\text{ \: in \: } \Omega \text { at } t = 0,
\end{split}
\end{equation}

where $\partial \Omega_{{T}}$ is temperature boundary, $\partial \Omega_{\rm in}$ and $\partial \Omega_{\rm out}$ denote inflow and outflow boundaries, respectively, which are defined as

\begin{equation}
\partial \Omega_{\rm in} := \{ \bm{X} \in \partial \Omega : \bm{u} \cdot \mathbf{n} < 0\} \quad \mbox{ and } \quad \partial \Omega_{\rm out} := \{ \bm{X} \in \partial \Omega : \bm{u} \cdot \mathbf{n} \geq 0\}.
\label{eq:in_and_out}
\end{equation}

The detail of discretization could be found in appendix \ref{sec:fem}.

\section{Non-intrusive reduced order model} \label{sec:data_driven}

\begin{figure}[!ht]
   \centering
    \includegraphics[width=15.0cm,keepaspectratio]{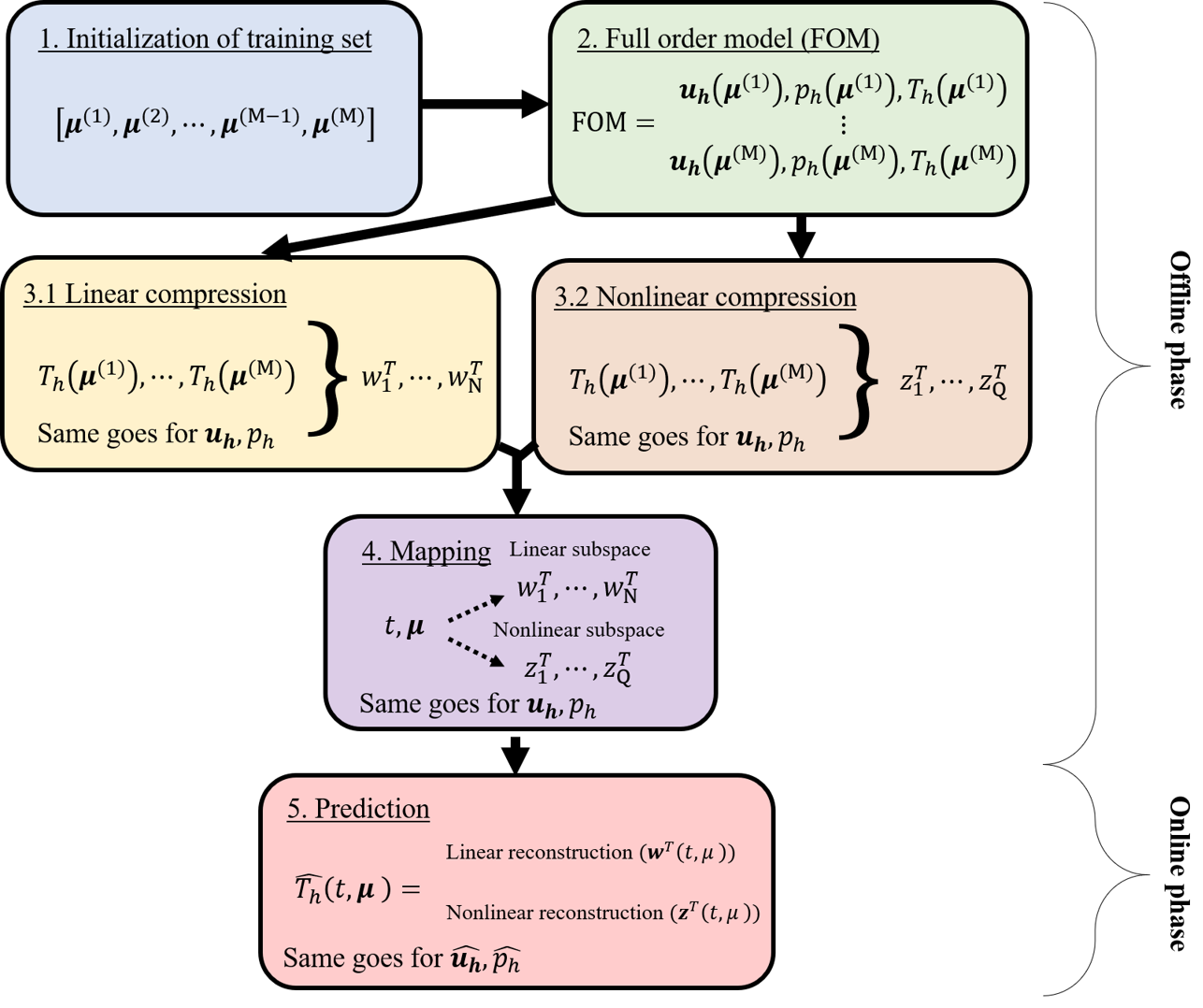} 
   \caption{Summary of data-driven model reduction framework for natural convection in porous media. }
   \label{fig:data_driven}
\end{figure}

The finite element solution scheme described in appendix \ref{sec:fem} is typically a time-consuming operation, making it impractical to query such a solver in a real-time context whenever parametric studies are carried out. Such parametric studies are often of interest to account for uncertain material properties of porous media. Therefore in this work, we propose to employ a reduced order modeling strategy. A graphical summary of the paradigm is presented in Figure \ref{fig:data_driven}: the computations are divided into an offline phase for the ROM construction, which we will show through four consecutive main steps and (single-step) online stage for the ROM evaluation. \par

The first step of the offline stage (blue) represents an initialization of a training set ($\bm{\mu}$) of parameters used to train the framework, of cardinality $\mathrm{M}$. Let $\mathbb{P} \subset \mathbb{R}^P$, $P \in \mathbb{N}$, be a compact set representing the range of variation of the parameters $\bm{\mu} \in \mathbb{P}$. For the sake of notation we denote by $\mu_p$, $p = 1, \hdots, P$, the $p$-th component of $\bm{\mu}$. To explore the parametric dependence of the phenomena, we define a discrete training set of $\mathrm{M}$ parameter instances. Each parameter instance in the training set will be indicated with the notation $\bm{\mu}^{(i)}$, for $i = 1, \hdots, \mathrm{M}$. Thus, the $p$-th component of the $i$-th parameter instance in the training set is denoted by $\mu_p^{(i)}$ in the following. The choice of the value of $\mathrm{M}$, as well as the sampling procedure from the range $\mathbb{P}$, is typically user- and problem-dependent. In this work, we use an equispaced distribution for the training set as done in \cite{kadeethum2021non}. We note that adaptive sampling approaches could be employed and might result in a better model accuracy with a lower number of training instances \citep{paul2015adaptive,vasile2013adaptive,choi2020gradient}. \par


In the second step (green), we query the full order model (FOM), based on the finite element solver discussed in appendix \ref{sec:fem}, for each parameter $\bm{\mu}$ in the training set. After the second step, we have $\mathrm{M}$ snapshots of FOM results associated with the different parametric configurations in $\bm{\mu}$. Since the problem formulation is time-dependent, the output of the FOM solver for each parameter instance $\bm{\mu}^{(i)}$ collects the time series representing the time evolution of the primary variables for each time-step $t$. Thus, each snapshot contains approximations of the primary variables ($\bm{u}_{h}$, $p_h$, and $T_h$) at each time-step of the partition of the time domain $\mathbb{T}$. Therefore, based on the training set cardinality $\mathrm{M}$ and the number $N^t$ of time-steps, we have a total of $N^t \mathrm{M}$ training data to be employed in the subsequent steps. We note that as our finite element solver utilizes an adaptive time-stepping as presented in Equation \eqref{eq:time_mult}, each snapshot may well have a different number of time-steps $N^t$, i.e. $N^t = N^t(\bm{\mu})$. \par

The third step aims to compress the information provided by the snapshots through either linear (yellow) or nonlinear (orange) compression. The linear compression employs the proper orthogonal decomposition (POD) technique \citep{hesthaven2018non,willcox2002balanced,choi2020sns}. POD is used to determine characteristic spatial modes based on relative energy content criteria \citep{chatterjee2000introduction,liang2002proper,wang2018model}. During the linear compression, only the first $\mathrm{N}$ modes are retained \citep{hesthaven2016certified}, and employed as basis functions for the reduced basis spaces $\mathcal{U}_\mathrm{N}$, $\mathcal{P}_\mathrm{N}$, $\mathcal{T}_\mathrm{N}$, used for approximating the velocity, pressure, and temperature fields respectively. The goal is to achieve $\mathrm{N} \ll \mathrm{M} N^t$ as well as $\mathrm{N} \ll N_h^u$, $\mathrm{N} \ll N_h^p$, and $\mathrm{N} \ll N_h^T$. The resulting spatial modes, denoted by $\bm{w}_1^{\bm{u}}, \cdots, \bm{w}_\mathrm{N}^{\bm{u}}$, $\bm{w}_1^p, \cdots, \bm{w}_\mathrm{N}^p$, and $\bm{w}_1^T, \cdots, \bm{w}_\mathrm{N}^T$ span a linear subspace of the velocity, pressure, and temperature fields respectively. \par

For the nonlinear compression, we utilize a deep convolutional autoencoder. The ${z}_1^{\bm{u}}, \cdots, {z}_\mathrm{Q}^{\bm{u}}$, ${z}_1^p, \cdots, {z}_\mathrm{Q}^p$, and ${z}_1^T, \cdots, {z}_\mathrm{Q}^T$ represent the nonlinear manifold (also refered to as nonlinear manifolds) of the velocity, pressure, and temperature fields respectively. More details for $\bm{z}$ will be presented in Section \ref{sec:nonlinear_com}. Again, our goal is to achieve $\mathrm{Q} \ll \mathrm{M} N^t$. Autoencoders could be considered as a nonlinear variation of POD \citep{romero2017quantum,baldi2012autoencoders,hinton1994autoencoders} and have been recently used in physics-based modeling \citep{kim2020fast,phillips2020autoencoder,o2019learning,goh2019solving}. They use the so-called \emph{bottleneck} inside their hidden layers to force the networks to identify a reduced representation $\bm{z}$ of the training examples \citep{hinton1994autoencoders,gehring2013extracting}. We will discuss these linear and nonlinear compression approaches in detail in the following sections. \par

Next, for the fourth step (purple), for each snapshot, we map $t$ and $\bm{\mu}$ to its representation in the linear and nonlinear manifolds equivalently. This operation produces mappings between each pair $(t, \bm{\mu})$, with $t \in \{t^0, \hdots, t^{N^t}\}$ ($N^t$ is different for each $\mu$) and $\bm{\mu}$ in the training set, and the best approximation in the reduced space $\mathcal{U}_\mathrm{N}$, $\mathcal{P}_\mathrm{N}$, $\mathcal{T}_\mathrm{N}$ for $\bm{u}_h(\bm{\mu})$, ${p}_h(\bm{\mu})$, and ${T}_h(\bm{\mu})$ at time $t$, respectively. We note that the notion of \emph{best approximation} is different between linear subspaces and nonlinear manifolds, and it will be discussed in the following sections. Finally, during the online phase (red), for given values of the parameter time instance $t$ and $\bm{\mu}$, we aim to recover the online approximation to our primary variables by querying the corresponding reduced subspace and reconstructing the resulting finite element representation. \par


\subsection{Linear approach} \label{sec:linear_com}

\begin{figure}[!ht]
   \centering
    \includegraphics[width=17.0cm,keepaspectratio]{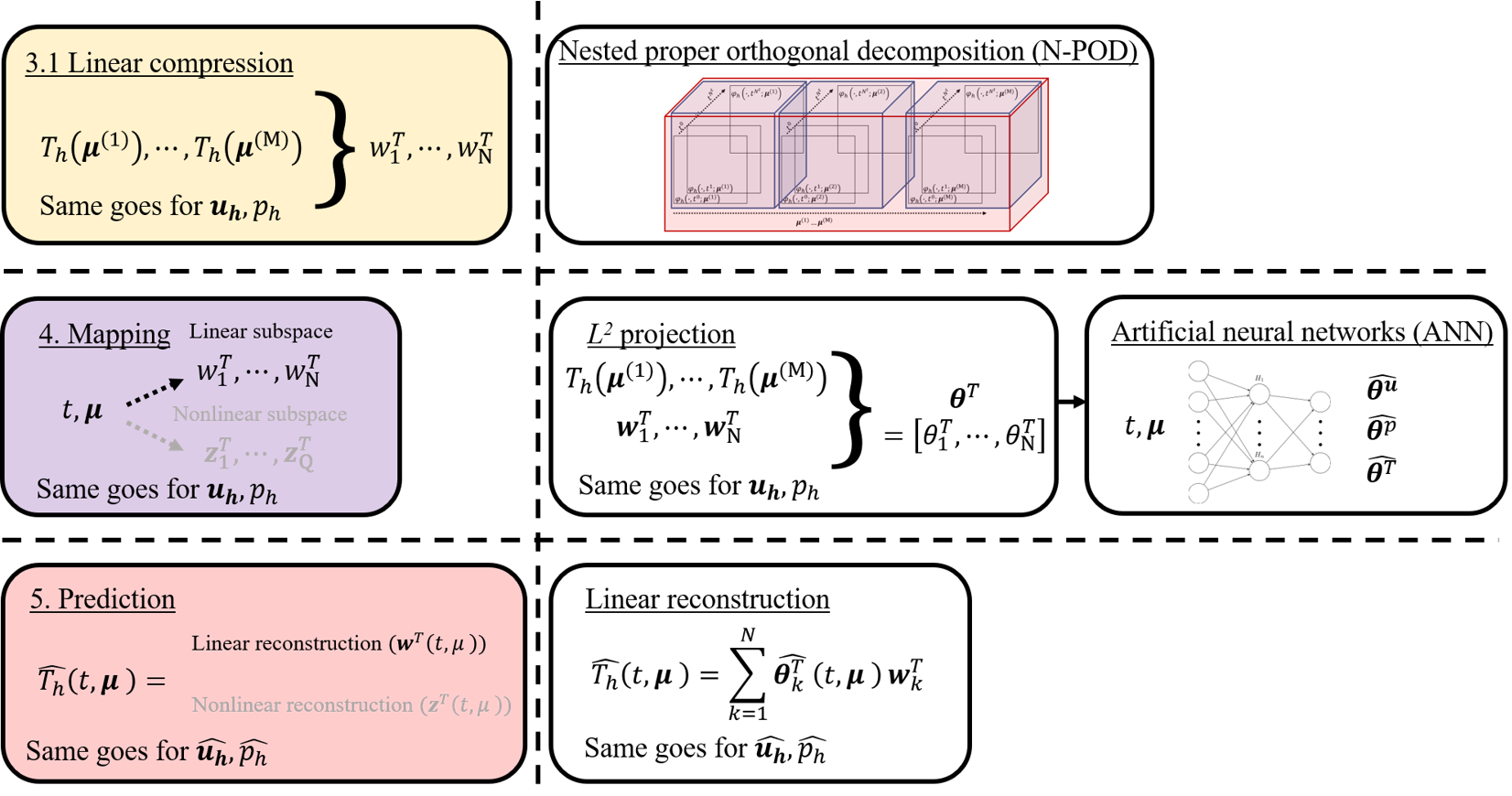} 
   \caption{Procedures taken in linear compression approach. Note that step 1 and 2 are presented in Figure 
   \ref{fig:data_driven}. We note that the nested POD technique compresses $t$ (blue cubes) and $\bm{\mu}$ (red cube), consecutively.}
   \label{fig:data_driven_linear_com}
\end{figure}

We provide a graphical illustration of procedures taken in the linear compression approach in Figure \ref{fig:data_driven_linear_com}. We note that as steps 1 and 2 of our data-driven framework shown in Figure \ref{fig:data_driven} are similar for both linear and nonlinear compression approaches, we here begin our procedure at the third step. Moreover, for the sake of compactness, we will focus on an expression and operation carried out for the temperature field ${T}_{h}$ for the rest of this section. A very similar procedure is similarly carried out for the velocity field $\bm{u}_h$ and the pressure field $p_h$ as well. \par

We have two main approaches to perform POD: (1) the \emph{standard} POD method, in which all domains, $t$ and $\bm{\mu}$, are compressed simultaneously (2) the \emph{nested} POD technique, in which $t$ and $\bm{\mu}$ domains are compressed consecutively. In this work, we employ the latter since our $N^t$ is large (i.e., small-time-steps to satisfy Courant–Friedrichs–Lewy (CFL) condition, see Equation  \eqref{eq:time_mult}). Consequently, the SVD computation of the standard POD case may become unfeasible because it requires a large amount of resources, both in terms of CPU time and memory storage (i.e., the bottleneck is due to the simultaneous compression in $t$ and $\bm{\mu}$.). Moreover, from \cite{kadeethum2021non}, the nested POD could provide comparable accuracy to the standard POD. We note that, in the reduced order modeling community, the nested POD can be also referred to as two-level POD or hierarchical approximate POD \citep{Audouze1,rapun2010reduced,Audouze2,ballarin2016fast,himpe2018hierarchical,wang2019non,jacquier2020non}. The detail of \emph{nested} POD could be found in Appendix \ref{sec:sec_nested_pod}. \par

We then, in the fourth step, obtain the optimal representation of each snapshot in the reduced basis spaces by means of an $L^2$ projection. 
This operation defines a map between each pair $(t, \bm{\mu})$, with $t \in \{t^0, \hdots, t^{N^t}\}$ and $\bm{\mu}$ in the training set, and a vector of coefficients $\bm{\theta}^T(t, \bm{\mu})$ that characterize the best approximation in the reduced space $\mathcal{T}_\mathrm{N}$ for the temperature field ${T}_h(\bm{\mu})$ at time $t$.

Let $\left\{\mathbf{w}_{1}, \cdots, \mathbf{w}_{\mathrm{N}}\right\}$ denote the basis functions spanning $\mathcal{T}_{\mathrm{N}}$. Given a time $t^{n}$ in the discretization of the time interval $\mathbb{T}$ and a parameter instance $\bm{\mu}^{(i)}$ in the training set we can define the best approximation $\widetilde{{T}}_{h}\left(\cdot; t^{n}, \bm{\mu}^{(i)}\right)$ to ${T}_{h}\left(\cdot; t^{n}, \bm{\mu}^{(i)}\right)$ in $\mathcal{T}_{\mathrm{N}}$ as
\begin{equation}
\widetilde{{T}}_{h}\left(\cdot; t^{n}, \bm{\mu}^{(i)}\right) = \sum_{k=1}^{\mathrm{N}} {\theta}_k^T(t^n, \bm{\mu}^{(i)}) \mathbf{w}_{k}
\end{equation}
where the coefficients $\theta_j^T$ are solutions to the $L^2$ projection problem, which can be stated as: Given ${T}_{h}\left(\cdot; t^{n}, \bm{\mu}^{(i)}\right)$, find $\bm{\theta}^T(t^{n}, \bm{\mu}^{(i)})=\left[\theta_{1}^T(t^{n}, \bm{\mu}^{(i)}), \cdots, \theta_{\mathrm{N}}^T(t^{n}, \bm{\mu}^{(i)})\right] \in \mathbb{R}^{\mathrm{N}}$ such that: $$\sum_{j=1}^{\mathrm{N}} {\theta}_j^T(t, \bm{\mu}) (\mathbf{w}_{j}, \mathbf{w}_{k})_T = \left({T}_{h}\left(\cdot; t^{n}, \bm{\mu}^{(i)}\right), \mathbf{w}_{k}\right)_T,  k = 1, \cdots, \mathrm{N}.$$

\noindent
We note that this results in a linear system, which left-hand side $(\mathbf{w}_{j}, \mathbf{w}_{k})_T$ can be easily precomputed and stored in a $\mathrm{N} \times \mathrm{N}$ matrix. However, the right-hand side can only be computed once the FOM solutions are available for the training set and corresponding time-steps. The next step is to generalize the computation of the coefficients of the ROM expansion for any (time, parameter) pair using artificial neural networks trained on the available data $\bm{\theta}^T(t^{n}, \bm{\mu}^{(i)})$.

We now define a map $\widehat{\bm{\theta}}^T(t, \bm{\mu})$ for any time $t$ in the time interval $\mathbb{T}$ and any value of the parameter $\bm{\mu}$ by training artificial neural networks (ANN) to approximate $\bm{\theta}^T(t, \bm{\mu})$ based on the training data points obtained by $L^2$ projection \citep{hesthaven2018non}. The neural networks are built on the PyTorch platform \citep{NEURIPS2019_9015}. For simplicity, our ANN has five hidden layers, and each layer has seven neurons. We use \emph{tanh} as our activation function. A detailed discussion on these hyperparameters could be found in \citep{kadeethum2021non}. Here, we use a mean squared error ($\mathrm{MSE}^{\theta, T}$) as the metric of our network loss function, defined as follows

\begin{equation}\label{eq:loss_mse_linear}
{\mathrm{MSE}^{\theta, T}}=\frac{1}{\mathrm{M} N^t} \sum_{i=1}^{\mathrm{M}}\sum_{k=0}^{N^t}\left|\widehat{\bm{\theta}}^T\left(t^k, \bm{\mu}^{(i)}\right)-{\bm{\theta}^T}\left(t^k, \bm{\mu}^{(i)}\right)\right|^{2}.
\end{equation}

\noindent
To minimize Equation \eqref{eq:loss_mse_linear}, we train the neural network by adjusting each neuron weight ($\mathrm{W}$) and bias ($\mathrm{b}$) using the adaptive moment estimation (ADAM) algorithm \citep{kingma2014adam}. Throughout this study, we use a batch size of 32, a learning rate of 0.001, a number of epoch of 10,000, and we normalize both our input and output to $[0, 1]$. To prevent our networks from overfitting behavior, we follow early stopping and generalized cross-validation criteria \citep{hesthaven2018non,prechelt1998early,prechelt1998automatic}. Note that instead of literally stopping our training cycle, we only save the set of trained weight and bias to be used in the online phase when the current validation loss is lower than the lowest validation from all the previous training cycle. We note that we do not use the RBF regression in linear approach because we follow procedures proposed in \citep{hesthaven2018non}.

During the online or prediction phase (fifth step), for each inquiry (i.e., a value of $\bm{\mu}$), we evaluate the ANN to obtain $\widehat{\bm{\theta}}^T(t, \bm{\mu})$ for each $t \in \{t^0, \cdots, t^{{N^t}}\}$). Subsequently, we reconstruct  $\widehat{{T}}_{h}\left(\cdot; t, \bm{\mu}\right)$ as

\begin{equation}
\widehat{{T}}_{h}\left(\cdot; t, \bm{\mu}\right) = \sum_{k=1}^{\mathrm{N}} \widehat{\theta}_k^T(t, \bm{\mu}) \mathbf{w}_{k},
\label{eq:online_solution}
\end{equation}

\noindent
We note that the reduced basis $\{\mathbf{w}_{k}\}_{k=1}^{\mathrm{N}}$ is already constructed during the POD phase; hence, recovering the online solutions, requires to evaluate $\widehat{\theta}_k^T(t, \bm{\mu})$ from the trained ANN (which is typically extremely fast), and subsequently, perform a reconstruction using Equation \eqref{eq:online_solution} (which only requires a linear combination of finite element functions). As a result, one typically enjoys an inexpensive online phase for each inquiry.

\subsection{Nonlinear approach} \label{sec:nonlinear_com}

\begin{figure}[!ht]
   \centering
    \includegraphics[width=17.0cm,keepaspectratio]{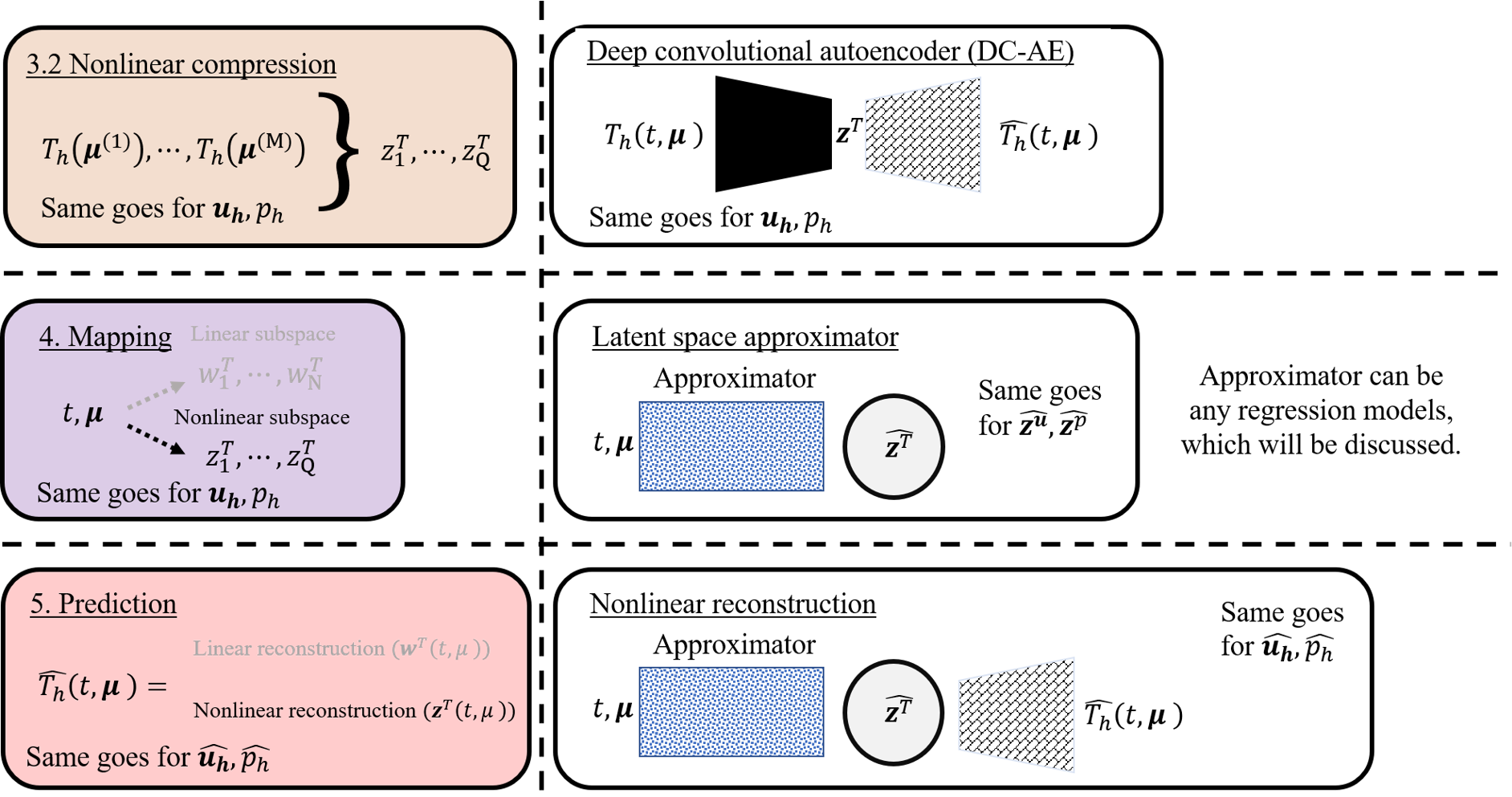} 
   \caption{Procedures taken in nonlinear compression approach. Note that step 1 and 2 are presented in Figure 
   \ref{fig:data_driven}.}
   \label{fig:data_driven_nonlinear_com}
\end{figure}

In Figure \ref{fig:data_driven_nonlinear_com}, we provide a graphical illustration of procedures considered in the nonlinear approach that we propose here. Again, as steps 1 and 2 of our data-driven framework shown in Figure \ref{fig:data_driven} are similar for both the linear and nonlinear approaches, we here begin our procedure at the third step. The main difference between the linear and nonlinear approaches lie within the compression as well as the reconstruction tool. For linear compression we use POD and for the reconstruction we use the $L_2$ projection. For the nonlinear compression (of  $T_h \left(\bm{\mu}^{ \left(1\right)}\right) , \cdots, T_h \left(\bm{\mu}^{\left(\mathrm{M}\right)}\right)$) and reconstruction we utilize AE. More specifically, for the compression we utilize the encoder, whereas for the reconstruction we utilize the decoder of the AE. To reiterate, we note that $T_h \left(\bm{\mu}^{ \left(i\right)}\right)$ represents $T_h$ given $\bm{\mu}^{\left(i\right)}$, belonging to the training set, evaluated at all space and time. The AE could be considered as a nonlinear variation of POD \citep{baldi2012autoencoders,hinton1994autoencoders,vincent2008extracting}, and it uses the so-called \emph{bottleneck} inside its hidden layers to force the networks to identify a nonlinear manifolds (${z}_1^T, \cdots, {z}_Q^T$) of the training examples \citep{hinton1994autoencoders,gehring2013extracting}.   \par


The AE is composed of three main components: encoder, bottleneck, and decoder. The encoder task is to produce the nonlinear manifolds $\bm{z}^T$ with given $T_h$, i.e.

\begin{equation}
    \bm{z}^T\left(t, \bm{\mu}\right) = \mathrm{encoder}\left(  {T}_h\left(t, \bm{\mu}\right)  \right), \quad \forall t \in \mathbb{T} \: \text{and} \: \forall\bm{\mu} \in \mathbb{P}.
\end{equation}

This $\bm{z}^T$ lies within the bottleneck layers. The decoder then attempts to reconstruct $\widehat{T}_h$ with given $\bm{z}^T$, i.e.

\begin{equation}
   \widehat{T}_h\left(t, \bm{\mu}\right) = \mathrm{decoder}\left(  \bm{z}^T\left(t, \bm{\mu}\right)  \right), \quad \forall t \in \mathbb{T} \: \text{and} \: \forall\bm{\mu} \in \mathbb{P}.
\end{equation}

The encoder utilizes a contracting block that performs two convolutions followed by a max pool operation. The bottleneck employs two linear layers. The decoder uses expanding block in which it performs an upsampling and convolution of its two inputs. We note that each contracting block uses LeakyReLU with a negative slope of 0.2 as its activation function, and each expanding block uses ReLU as its activation function. Our model, in total, has seven contracting blocks and seven expanding blocks. We provide the detail of our AE in Appendix \ref{sec:dc_used}. \par


We train our AE by minimizing

\begin{equation} \label{eq:loss_ae}
    {\mathrm{MSE}^{T}} = 
    \frac{1}{\mathrm{M} N^t} \sum_{i=1}^{\mathrm{M}}\sum_{k=0}^{N^t}\left|\widehat{T}_h\left(t^k, \bm{\mu}^{(i)}\right)-T_h\left(t^k, \bm{\mu}^{(i)}\right)\right|^{2}.
\end{equation}

\noindent
Here, we follow the training used in \cite{kadeethum2021framework}. We use the ADAM algorithm to adjust learnable parameters ($\mathrm{W}$ and $\mathrm{b}$) \citep{kingma2014adam}. Throughout this study, we use $\mathrm{B} = 32$ and a number of the epoch of 50. The learning rate ($\eta$) is calculated as \citep{loshchilov2016sgdr}

\begin{equation}
\eta_{c}=\eta_{\min }+\frac{1}{2}\left(\eta_{\max }-\eta_{\min }\right)\left(1+\cos \left(\frac{\mathrm{step_c}}{\mathrm{step_f}} \pi\right)\right)
\end{equation}

\noindent
where $\eta_{c}$ is a learning rate at step $\mathrm{step_c}$, $\eta_{\min }$ is the minimum learning rate, which is set as $1 \times 10^{-16}$, $\eta_{\max }$ is the maximum or initial learning rate, which is selected as $1 \times 10^{-5}$, $\mathrm{step_c}$ is the current step, and $\mathrm{step_f}$ is the final step. We note that each step refers to each time we perform back-propagation, including updating both encoder and decoder's parameters. \par

Following the training of the AE, we now establish the manifold $\bm{z}^T\left(t, \bm{\mu}\right), \quad \forall t \in \mathbb{T} \: \text{and} \: \forall\bm{\mu} \in \mathbb{P}$ during the fourth step shown in Figure \ref{fig:data_driven_nonlinear_com}. The data available for this task are the pairs $(t, \bm{\mu})$ in the training set. We achieve this through the training of a so-called \emph{approximator}. In this study, we test three types of the approximators. The first one is ANN as also used in Section \ref{sec:linear_com}. To reiterate, our ANN has five hidden layers, and each layer has seven neurons. We use \emph{tanh} as our activation function. Its loss function is

\begin{equation}\label{eq:loss_ann_nonlinear}
{\mathrm{MSE}^{\bm{z}^T}}=\frac{1}{\mathrm{M} N^t} \sum_{i=1}^{\mathrm{M}}\sum_{k=0}^{N^t}\left|\widehat{\bm{z}}^T\left(t^k, \bm{\mu}^{(i)}\right)-{\bm{z}^T}\left(t^k, \bm{\mu}^{(i)}\right)\right|^{2}.
\end{equation}

\noindent
Similar to Section \ref{sec:linear_com}, we use the ADAM algorithm to adjust each neuron $\mathrm{W}$ and $\mathrm{b}$ \citep{kingma2014adam}, a batch size of 32, a learning rate of 0.001, a number of epoch of 10,000, and we normalize both our input ($t, \bm{\mu}$) and output ($\bm{z}^T$) to $[0, 1]$. We follow the same procedures presented in Section \ref{sec:linear_com} to avoid over-fitting. The second and the third approximators are based on radial basis function (RBF) interpolation \citep{wright2003radial}. The second approximator uses a linear RBF, and the third one employs a cubic RBF. Similar to the ANN approximator, our input is $t, \bm{\mu}$, and our output is $\bm{z}^T(t, \bm{\mu})$. Please refer to \cite{Jones2001} for the detailed implementation. \par

During the online phase (the fifth step shown in Figure \ref{fig:data_driven_nonlinear_com}), we utilize the \emph{trained} approximator and the \emph{trained} decoder to approximate $\widehat{{T}}_{h}\left(\cdot; t, \bm{\mu}\right)$ for each inquiry (i.e., a value of $\bm{\mu}$) through

\begin{equation}
\widehat{\bm{z}}^T\left(\cdot; t, \bm{\mu}\right) = \operatorname{approximator} \left( t, \bm{\mu} \right),
\label{eq:online_solution_z_nonlinear}
\end{equation}

\noindent
and, subsequently, 

\begin{equation}
\widehat{{T}}_{h}\left(\cdot; t, \bm{\mu}\right) = \operatorname{decoder} \left(\widehat{\bm{z}}^{T}\left(\cdot; t, \bm{\mu}\right) \right).
\label{eq:online_solution_nonlinear}
\end{equation}

\noindent
We note that each evaluation of approximator and decoder is extremely fast. Hence, one typically enjoys an inexpensive online phase for each inquiry.

\begin{remark} \label{remark:2}
Our ROM does not require a large number of training parameters $\bm{\mu}$ to achieve the desired accuracy and avoid overfitting. We achieve this by stacking our time domain and parameter ($\bm{\mu}$) domain together during the training of AE. Consequently, the input to AE corresponds to not only each FEM results with given $\bm{\mu}$, but also each time-step. To elaborate, we refer to our Example 2; we have only 40 training values of $\bm{\mu}$, but our total input size to AE is 36110. By doing this, during our online inquiry, we simply interpolate at the time of interest within the time domain provided during the training phase.
\end{remark}

\begin{remark} \label{remark:4}
We present the performance of the model that uses autoencoder without convolutional layers in appendix \ref{sec:non_conv}. In summary, we observe that the model with convolutional layers (deep convolutional autoencoder) performs better.
\end{remark}

\section{Numerical examples} \label{sec:numer_examples}

We examine the performance of the proposed non-intrusive reduced order model following the nonlinear compression and reconstruction approach compared to the linear approach through three numerical examples. The information of each example is presented in Table \ref{tab:main_info}. Throughout this section, for compactness, we will focus on the results of the temperature field. The velocity field and the pressure field results are similar to those of the temperature field. We note that $\mathrm{M}_\mathrm{v}$ and $\mathrm{M}_\mathrm{t}$ represent number of validation and testing sets, respectively. The parameters $\bm{\mu}$ belonging to validation set and testing set are named $\bm{\mu}_\mathrm{v}^{(i)}$ and $\bm{\mu}_\mathrm{t}^{(i)}$, respectively. \par

As mentioned previously in Section \ref{sec:data_driven}, our FOM, see Appendix \ref{sec:fem}, relies on unstructured grids. Our nonlinear compression and reconstruction, however, are applicable to a structured data set because they utilize convolutional layers. Hence, we interpolate the finite element result $T_h$ to structured grids. We then replace the FOM dimension ${N}_h^T$, associated with the unstructured grid, with a pair $(\widetilde{N}_h^T, \widetilde{N}_h^T)$, associated to the structured grid. As we present in Table \ref{tab:main_info}, the degrees of freedom or DOFs of the linear approach are less than those of the nonlinear approach. \par

\begin{table}[!ht]
\centering
\caption{Summary of main information for each example.}
\begin{tabular}{|l|c|c|c|l|}
\hline
                            & \textbf{Example 1}                      & \textbf{Example 2}                        & \textbf{Example 3}           & \multicolumn{1}{c|}{\textbf{remark}}               \\ \hline
$\mathrm{M}$                & 40                                      & 40                                        & 100                          & training set                                       \\ \hline
$\mathrm{M}_\mathrm{v}$              & 10                                      & 10                                        & 20                           & validation set                                     \\ \hline
$\mathrm{M}_\mathrm{t}$              & 10                                      & 10                                        & 20                           & test set                                           \\ \hline

$\mathrm{M} N^t$            & 16802                                   & 36110                                     & 86086                        & total training data                 \\ \hline
$N^t$ range                 & {[}226, 477{]}                          & {[}790, 1010{]}                           & {[}740, 985{]}               & for training, validation, and test sets            \\ \hline
$N_{h}^T$                   & 7110                                    & 9600                                      & 9600                         & DOFs: $T_h$                                        \\ \hline
$\widetilde{N}_h^T$                   & $128 \times 128$                                    & $128 \times 128$                                        & $128 \times 128$                           & interpolated $T_h$ on structured mesh                                       \\ \hline
$t$ range                 & {[}0.0, 0.1{]}                          & {[}0.0, 0.1{]}                           & {[}0.0, 0.1{]}               & $\left[t^{0}, t^{N}\right]$          \\ \hline

\multirow{2}{*}{$\bm{\mu}$} & \multirow{2}{*}{$\mathrm{Ra}=[40, 80]$} & \multirow{2}{*}{$\mathrm{Ra}=[350, 450]$} & $\mathrm{Ra_1}=[350, 450]$   & \multirow{2}{*}{only Example 3 has two parameters} \\ \cline{4-4}
                            &                                         &                                           & $\mathrm{Ra_2}=[0.001, 100]$ &                                                    \\ \hline
\end{tabular}
\label{tab:main_info}
\end{table}

\begin{figure}[!ht]
   \centering
    \includegraphics[width=16.0cm,keepaspectratio]{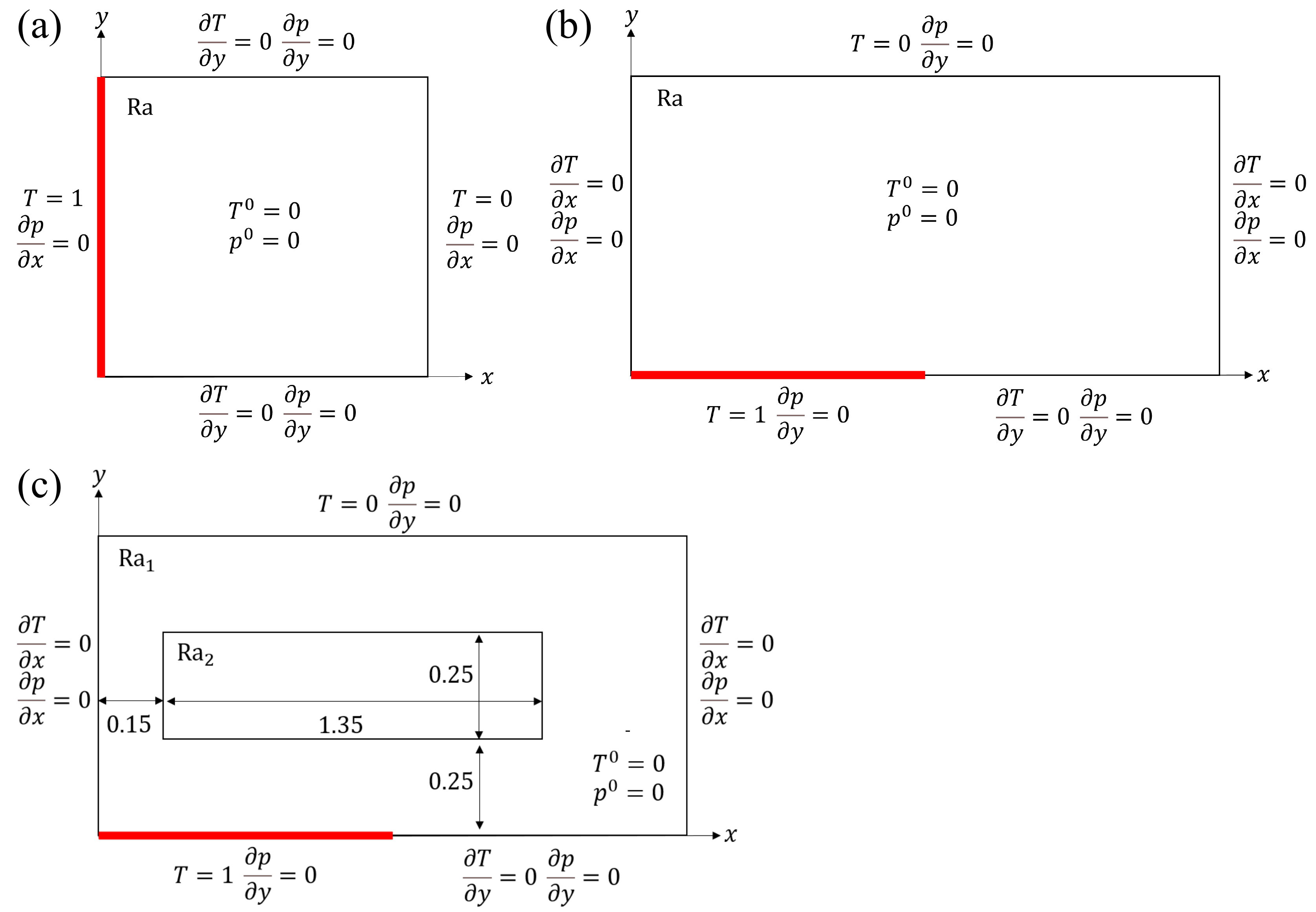} 
   \caption{Domain and boundary conditions for (a) Example 1, (b) Example 2, and (c) Example 3. The red line indicates the region of the boundary where the temperature is elevated.}
   \label{fig:geo1}
\end{figure}

We present a summary of all geometries and boundary conditions we use in Figure \ref{fig:geo1}. In short, Examples 1 and 2 represent cases that we have only one $\bm{\mu}$ - $\mathrm{Ra}$ while Example 3 illustrates a case where we have two $\bm{\mu}$ - $\mathrm{Ra_1}$ and $\mathrm{Ra_2}$ The detailed specification of each example will be discussed in the following subsections.

The summary of naming for each model is summarized in Table \ref{tab:naming}. They will be used in figures and discussion throughout the rest of this paper. Please refer to Sections \ref{sec:data_driven}, \ref{sec:linear_com}, and \ref{sec:nonlinear_com} for more detail on how we develop each model. In short, \emph{linear} compression refers to the models that uses POD as a compression tool (see Section \ref{sec:linear_com}), and \emph{nonlinear} compression refers to the models with AE as its compression tool (see Section \ref{sec:nonlinear_com}). For the \emph{linear} compression, subspace dimension refers to the number of reduced basis or $\mathrm{N}$ as well as the number of intermediate reduced basis or $\mathrm{N_{int}}$. We assume $\mathrm{N} = \mathrm{N_{int}}$ for all models for simplicity. The subspace dimension is the number of latent space or $\mathrm{Q}$ for the \emph{nonlinear} compression. \par

\begin{table}[!ht]
\centering
\caption{Summary of naming for each model.}
\begin{tabular}{|c|c|c|c|}
\hline
\textbf{model name} & \textbf{compression} & \textbf{subspace dimension} & \textbf{approximator} \\ \hline
POD 16 RB           & linear               & 16                          & ANN                   \\ \hline
POD 50 RB           & linear               & 50                          & ANN                   \\ \hline
POD 100 RB          & linear               & 100                         & ANN                   \\ \hline
POD 500 RB          & linear               & 500                         & ANN                   \\ \hline
AE 4 z: L-RBF       & nonlinear            & 4                           & linear RBF            \\ \hline
AE 4 z: C-RBF       & nonlinear            & 4                           & cubic RBF             \\ \hline
AE 4 z: ANN         & nonlinear            & 4                           & ANN                   \\ \hline
AE 16 z: L-RBF      & nonlinear            & 16                          & linear RBF            \\ \hline
AE 16 z: C-RBF      & nonlinear            & 16                          & cubic RBF             \\ \hline
AE 16 z: ANN        & nonlinear            & 16                          & ANN                   \\ \hline
AE 256 z: L-RBF     & nonlinear            & 256                         & linear RBF            \\ \hline
AE 256 z: C-RBF     & nonlinear            & 256                         & cubic RBF             \\ \hline
AE 256 z: ANN       & nonlinear            & 256                         & ANN                   \\ \hline
\end{tabular}
\label{tab:naming}
\end{table}

We note that FOM, POD, $L_2$ projection, RBF, and linear reconstruction are computed using AMD Ryzen Threadripper 3970X. The training of AE and ANN and nonlinear reconstruction are done by a single Quadro RTX 6000. We note that without using GPU, the training time of our AE is impractical. The training of ANN takes longer with CPU, but it is still able to finish within a reasonable time frame. A discussion of computational time will be provided in the discussion section (see Section \ref{sec:discussion}).

\subsection{Example 1: Heating from side problem}

\begin{figure}[!ht]
   \centering

         \includegraphics[keepaspectratio, height=4.5cm]{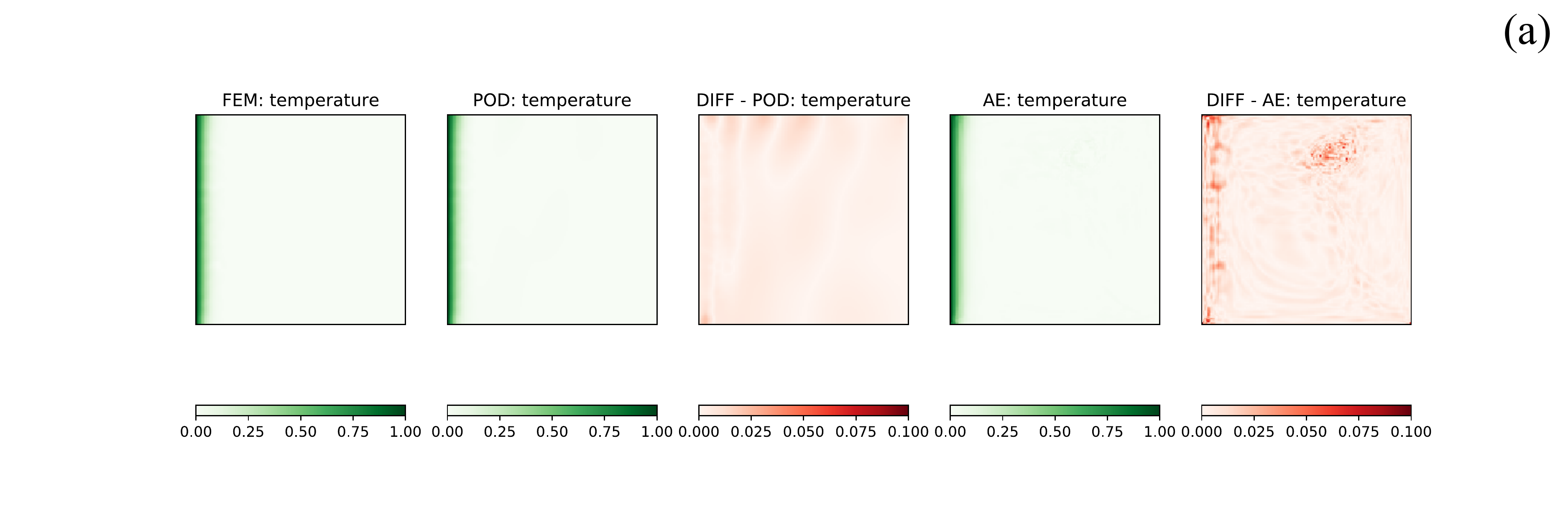}
         \includegraphics[keepaspectratio, height=4.5cm]{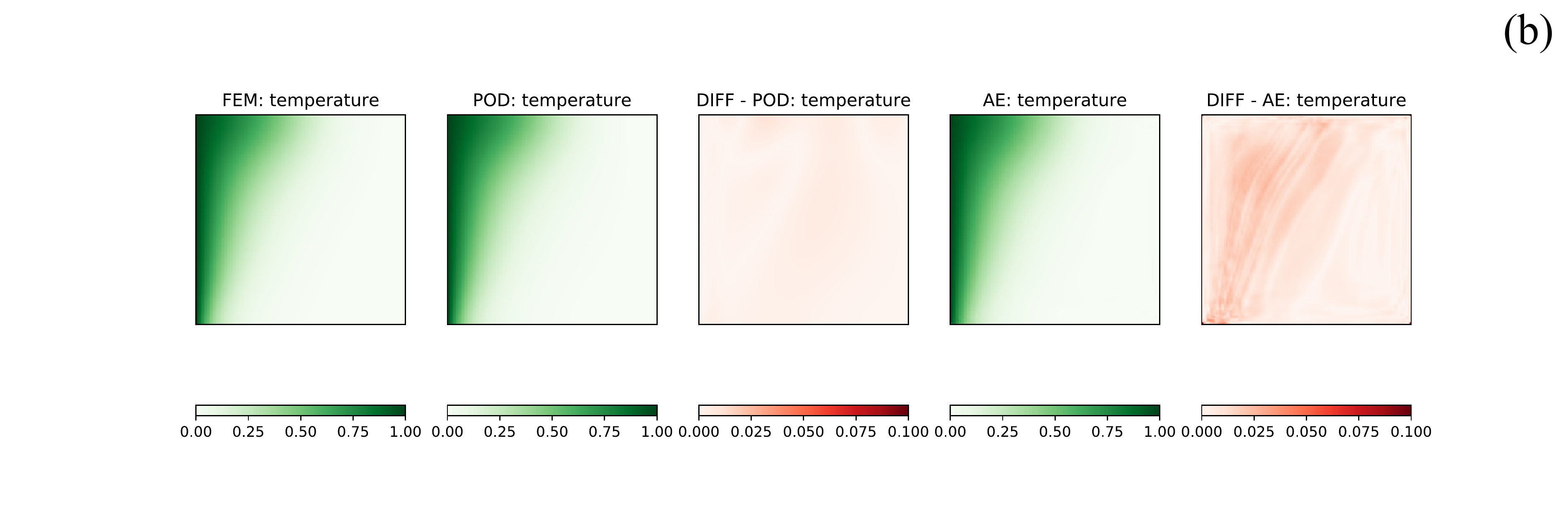}
         \includegraphics[keepaspectratio, height=4.5cm]{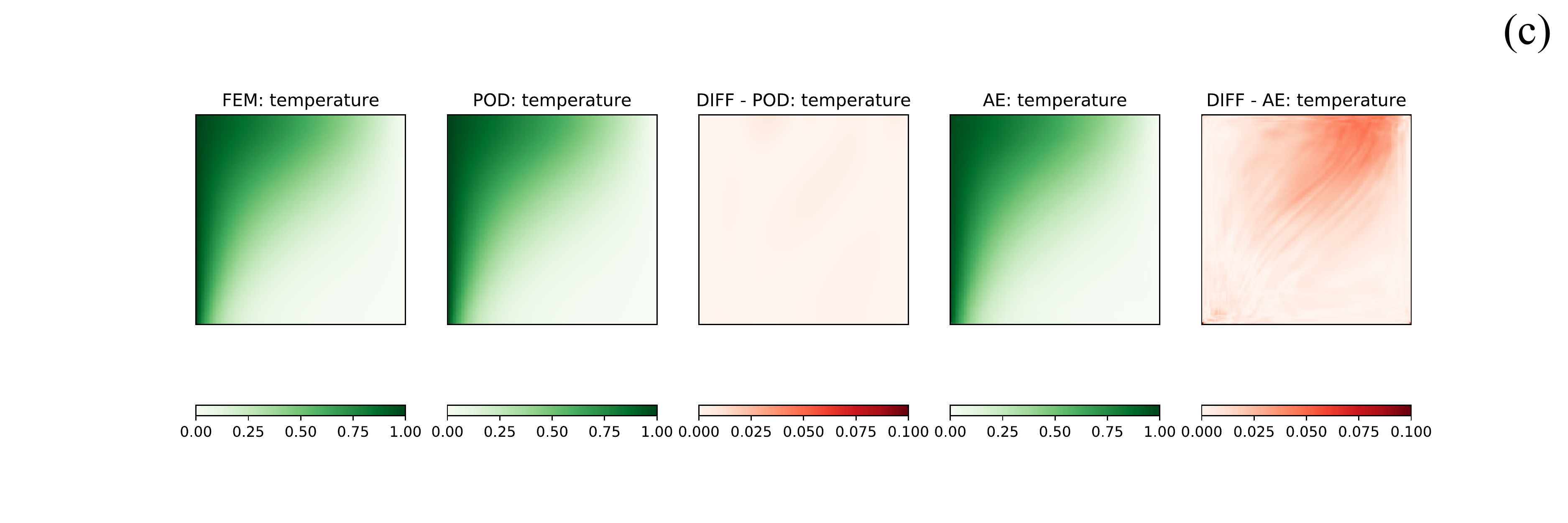}
   \caption{Example 1: test case ($\mathrm{Ra} = 72.82$) results. (a) $t = 0.001$, (b) $t = 0.03$, and (c) $t = 0.07$. These results are produced by POD 16 RB and AE 16 z: ANN for POD: temperature and AE: temperature, respectively.}
   \label{fig:ex1_pic}

\end{figure}

We adapt this example from \cite{zhang2016mixed}. The geometry and boundary conditions are shown in Figure \ref{fig:geo1}a. The flow is driven by buoyancy as the fluid is heated on the left side of the domain. The fluid then flows upwards and rotates to the right side of the domain (see Figure \ref{fig:ex1_pic}a-c, first column). As presented in Table \ref{tab:main_info}, we set $\bm{\mu} = (\mathrm{Ra})$, and its admissible range of variation is $[40.0, 80.0]$. We use $\mathrm{M} = 40$, $\mathrm{M}_\mathrm{v} = 10$, $\mathrm{M}_\mathrm{t} = 10$ where $\mathrm{M}_\mathrm{v}$ is a number of cases in validation set and  $\mathrm{M}_\mathrm{t}$ is a number of cases in test set. We have in total $\mathrm{M} N^t = 16802$ training data points. Moreover, as our FOM utilizes the adaptive time-stepping, see Equation  \eqref{eq:time_mult}, the minimum and maximum $N^t$ is 226 and 477, respectively.  \par

\begin{figure}[!ht]
   \centering
         \includegraphics[keepaspectratio, height=6.0cm]{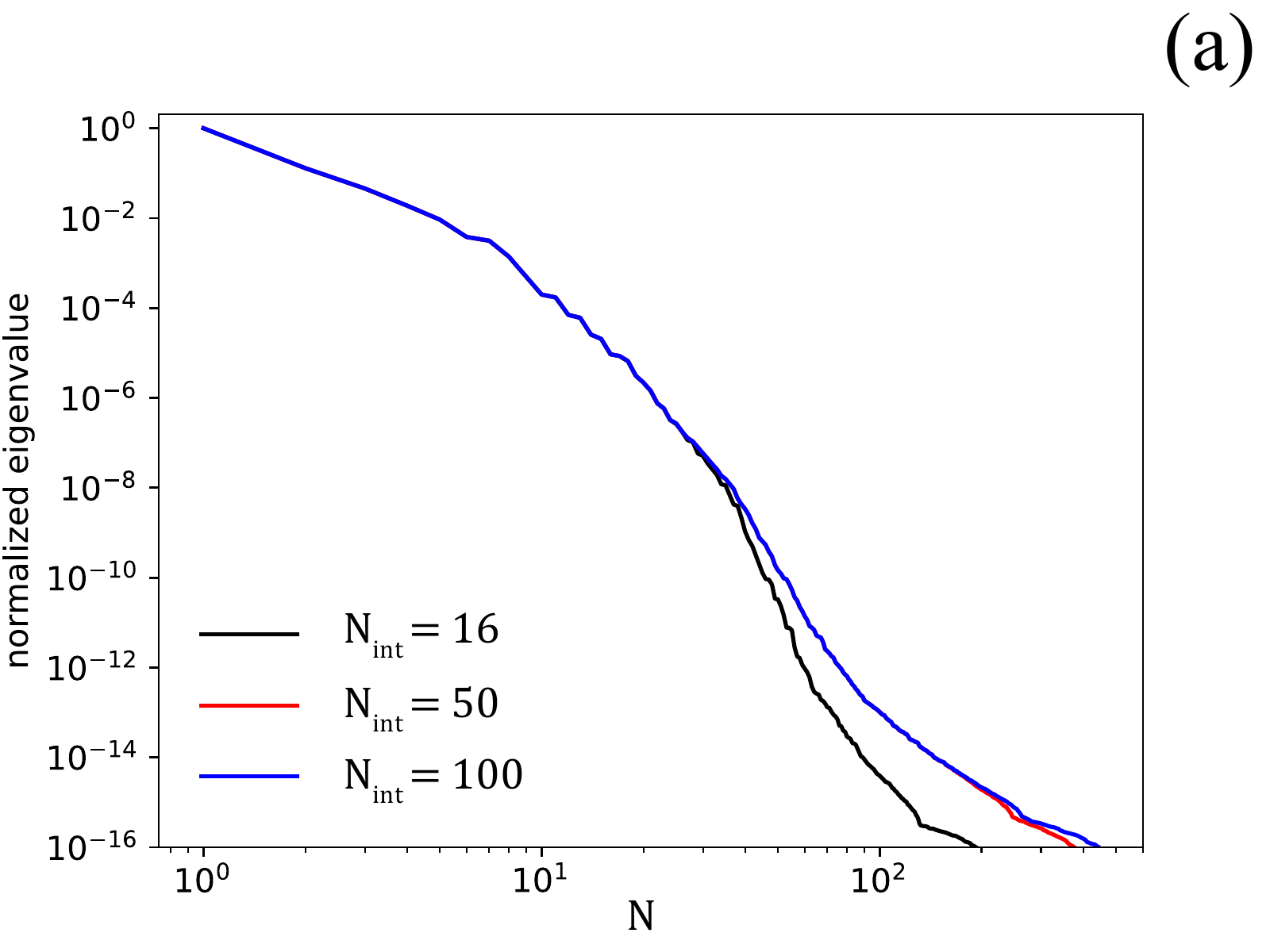}
         \includegraphics[keepaspectratio, height=6.0cm]{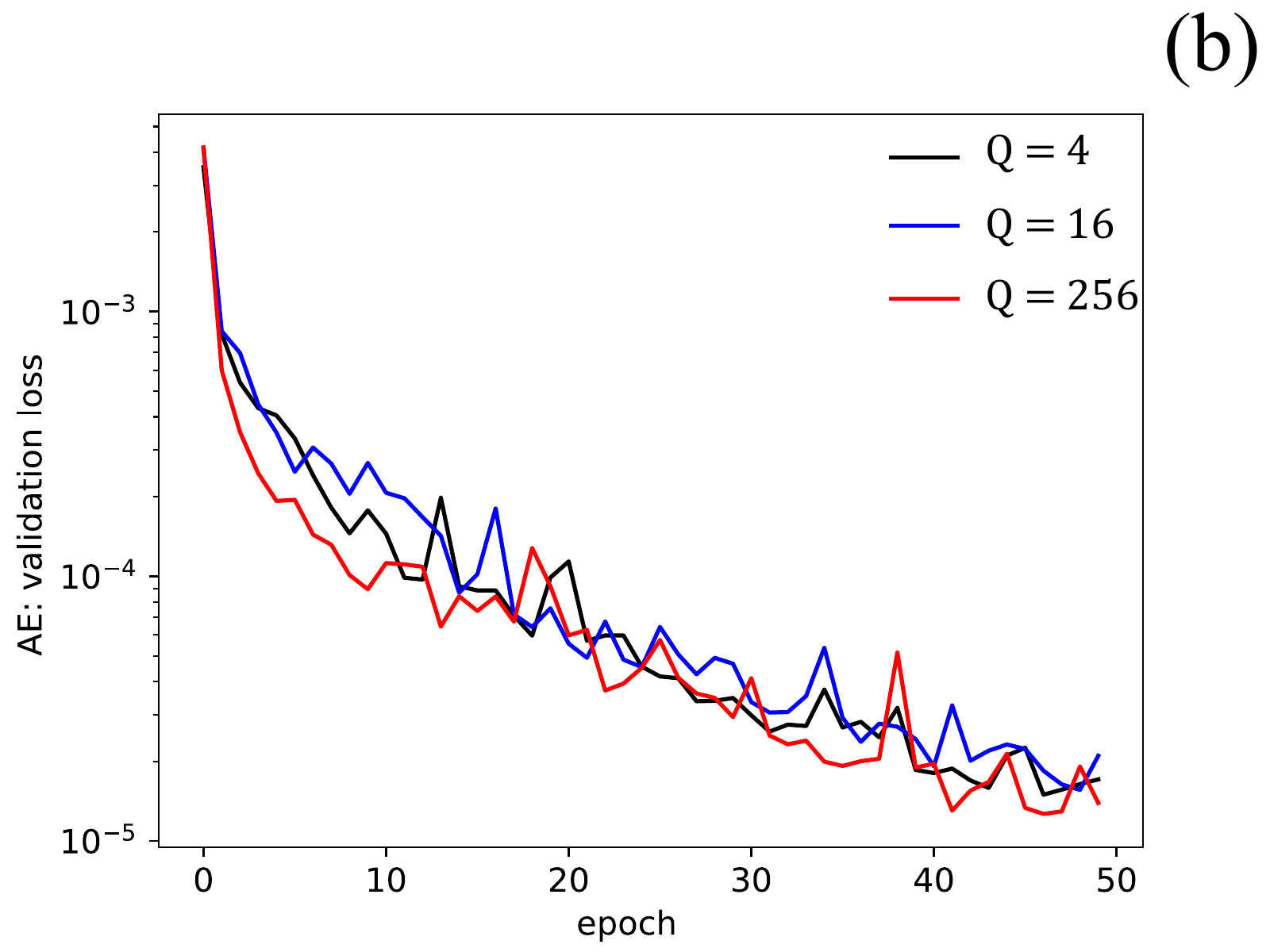}
   \caption{Example 1: (a) normalized eigenvalue as a function of reduced basis and (b) AE validation loss during the training.}
   \label{fig:ex1_train}
\end{figure}

We present the normalized eigenvalue as a function of the reduced basis for linear compression and AE validation loss during the training for nonlinear compression in Figure \ref{fig:ex1_train}. We observe that by using $\mathrm{N_{int}} = 50$, the POD could capture information similar to a case where we use $\mathrm{N_{int}} = 100$. Moreover, the normalized eigenvalues of all cases decay reasonably fast, which may imply that the POD could capture most of the information with a small number of $\mathrm{N_{int}}$ and $\mathrm{N}$. For the nonlinear compression, Figure \ref{fig:ex1_train}b, the validation loss of all $\mathrm{Q}$ values is approximately $2 \times 10^{-5}$. This behavior implies that using only $\mathrm{Q}=4$; the AE could capture most of the information. 
 \par

For the testing, we focus on a fixed realization of the uncertain parameter $\bm{\mu}_\mathrm{t}$ belonging to $\mathrm{M_t}$, which is outside of the training set. To reiterate, we use $\mathrm{M_t} = 10$. One example (out of 10) of the test cases is presented in Figure \ref{fig:ex1_pic} for three different $t$ values. In this figure, the difference between solutions produced by the FOM and ROM for both linear and nonlinear compression (further referred to as DIFF) is calculated by

\begin{equation}\label{eq:diff}
\operatorname{DIFF}_\varphi(t^k, \bm{\mu}_\mathrm{t}^{(i)})= \left|\varphi_h(\cdot; t^k, \bm{\mu}_\mathrm{t}^{(i)}) - \widehat{\varphi}_h(\cdot; t^k, \bm{\mu}_\mathrm{t}^{(i)})\right|,
\end{equation}

\noindent
where $\varphi_h$ is a finite-dimensional approximation of the set of primary variables corresponding to velocity, pressure, and temperature fields in this study (see appendix \ref{sec:fem}). $\widehat{\varphi_h}$ is an approximation of $\varphi_h$ produced by the ROM (either linear and nonlinear compression approaches). Thus, $\varphi_h(\cdot; t^k, \bm{\mu}_\mathrm{t}^{(i)})$ and $\widehat{\varphi}_h(\cdot; t^k, \bm{\mu}_\mathrm{t}^{(i)})$ represent $\varphi_h$ and $\widehat{\varphi_h}$ at all space coordinates (i.e., evaluations at each DOF) at time $t^k$ with input parameter $\bm{\mu}_\mathrm{t}^{(i)}$, respectively. We note that, as mentioned at the beginning of Section \ref{sec:numer_examples}, we here only present the results of the temperature field. Hence, $\varphi_h$ and $\widehat{\varphi_h}$ represent $T_h$ and $\widehat{T_h}$, respectively. From Figure \ref{fig:ex1_pic}, we observe that both linear and nonlinear compression (POD and AE, respectively) provide a reasonable approximation of the temperature field. \par

We then evaluate the performance of linear and nonlinear approaches by evaluating the mean square error ($\mathrm{MSE}_\varphi(:, \bm{\mu}_\mathrm{t}^{(i)})$) and maximum DIFF (or max(DIFF)) of the test cases defined as follows

\begin{equation}\label{eq:validation_mse}
    {\mathrm{MSE}_\varphi(:, \bm{\mu}_\mathrm{t}^{(i)}) :=\frac{1}{N^t} { \sum_{k=0}^{N^t} \left| \varphi_h(\cdot; t^k, \bm{\mu}_\mathrm{t}^{(i)}) - \widehat{\varphi}_h(\cdot; t^k, \bm{\mu}_\mathrm{t}^{(i)})\right|_{\varphi}^2}},
\end{equation}


\noindent
and

\begin{equation}
    {\mathrm{max(DIFF)}_\varphi(:, \bm{\mu}_{\mathrm{t}}^{(i)}) := \left\|\operatorname{DIFF}_\varphi(:, \bm{\mu}_{\mathrm{t}}^{(i)})\right\|_{\varphi}^{\infty}.}
\end{equation}

\noindent
We note that $\mathrm{MSE}_\varphi(:, \bm{\mu}_\mathrm{t}^{(i)})$ and $\mathrm{max(DIFF)}_\varphi(:, \bm{\mu}_\mathrm{t}^{(i)})$ represent the MSE and max(DIFF) of all $t$ for each $\bm{\mu}_\mathrm{t}^{(i)}$. To reiterate, each $\bm{\mu}_\mathrm{t}^{(i)}$ belongs to the testing set. \par

\begin{figure}[!ht]
   \centering
         \includegraphics[keepaspectratio, height=8.0cm]{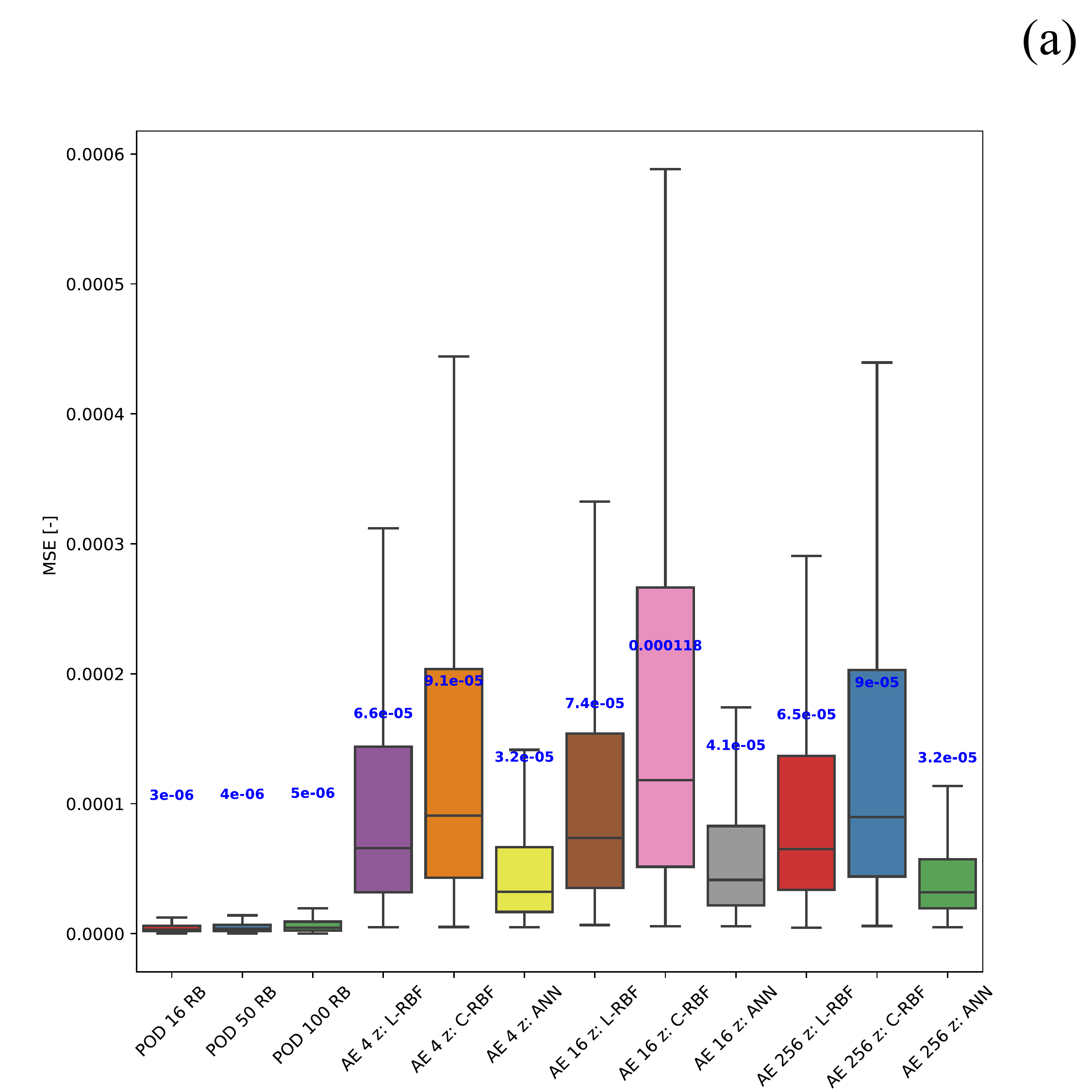}
         \includegraphics[keepaspectratio, height=8.0cm]{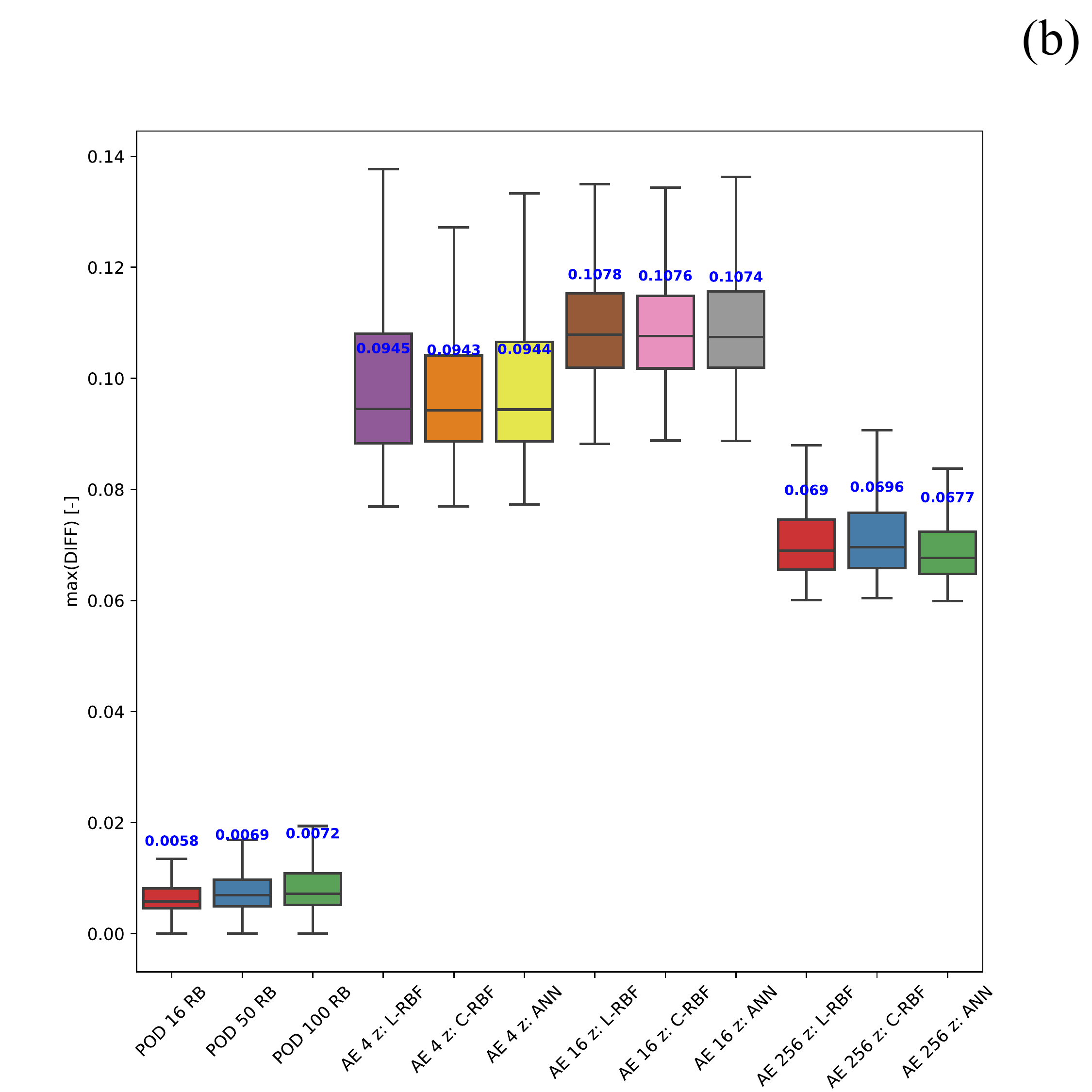}
   \caption{Example 1: Each model's performance on a test set (a) mean square error (MSE) and (b) maximum DIFF (max(DIFF)). Blue text represents a mean value. 16 RB, 50 RB, and 100 RB represent linear compression models with $\left[\mathrm{N_{int}} = 16, \mathrm{N} = 16\right]$, $\left[\mathrm{N_{int}} = 50, \mathrm{N} = 50\right]$, and $\left[\mathrm{N_{int}} = 100, \mathrm{N} = 100\right]$, respectively. The L-RBF and C-RBF are the nonlinear compression models with RBF with linear function and RBF with cubic function as its approximator, respectively. }
   \label{fig:ex1_test}
\end{figure}

The box-plots of the MSE and max(DIFF) for all $\bm{\mu}_\mathrm{t}^{(i)}$ (testing set), are shown in Figure \ref{fig:ex1_test}. From this figure, we observe that the linear approach (POD - as a compression tool) delivers a better accuracy than the proposed nonlinear approach. Moreover, as expected from the decay of eigenvalue shown in Figure \ref{fig:ex1_train}, the model accuracy is not much different among the models with 16 RB, 50 RB, or 100 RB ($\left[\mathrm{N_{int}} = 16, \mathrm{N} = 16\right]$, $\left[\mathrm{N_{int}} = 50, \mathrm{N} = 50\right]$, or $\left[\mathrm{N_{int}} = 100, \mathrm{N} = 100\right]$, respectively). One observation is, for the linear approach, the error increases as the dimension of POD increases. This behavior occurs because there are two sources of errors. The first one is from the truncation of reduced basis (i.e., $\mathrm{N} \le \mathrm{M}$), and the second one is the mapping between $\left( t, \bm{\mu} \right)$ and linear subspace ($\bm{w}$). Further detail and investigation could be found in \cite{kadeethum2021non}. \par

Among the nonlinear compression models, there is no obvious best choice. All models have the MSE results in the order of magnitude of $10^{-5}$. The models that use ANN as the approximator to map ($t$, $\bm{\mu}$) to $\bm{z}$ (see Section \ref{sec:nonlinear_com} for more detail) seem to have a better performance on the MSE results than those that use RBF as their approximator. However, for the max(DIFF) results, the model with ANN approximator seems to be less accurate. Moreover, it is hard to conclude that by increasing the dimension of $\bm{z}$, the framework's accuracy will be improved. We provide RBF training loss in Figure \ref{fig:si_training_loss_rbf_hfs} and ANN validation loss during the training in Figure \ref{fig:si_training_loss_ann}a. \par

\begin{figure}[!ht]
   \centering

         \includegraphics[keepaspectratio, height=7.5cm]{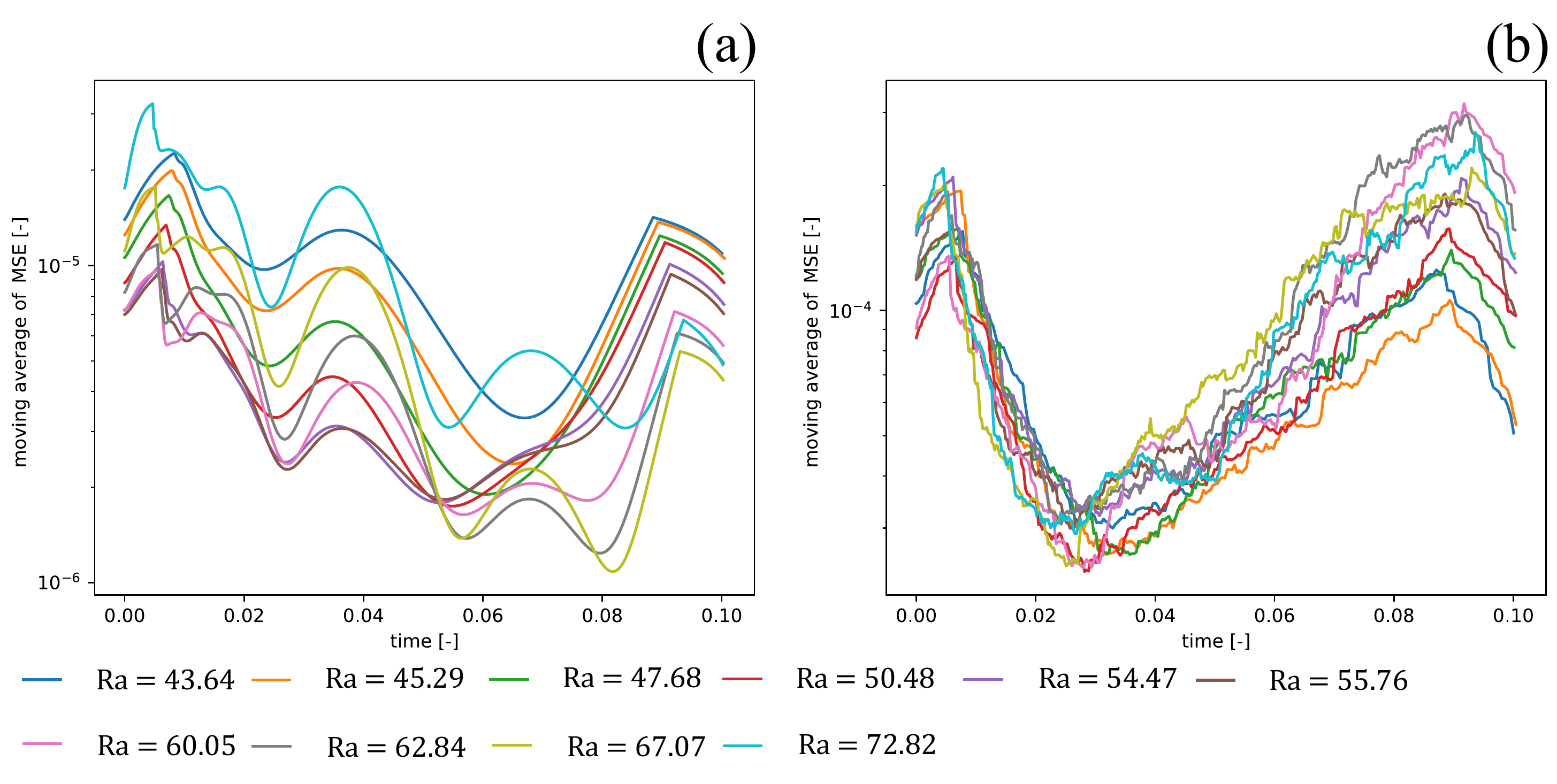}
   \caption{Example 1: $50^{th}$ moving average of MSE on a test set of (a) POD 16 RB and (b) AE 16 z: ANN.}
   \label{fig:ex1_time}

\end{figure}

The $50^{th}$ moving average of MSE as a function of $t$ for Example 1 test set is presented in Figure \ref{fig:ex1_time}. The MSE moving average of the linear compression, Figure \ref{fig:ex1_time}a, gradually decreases and then abruptly increases toward the end of the simulation. The MSE moving average of the nonlinear compression, on the other hand, drops sharply at the beginning, then gradually picks up, and finally decreases again at the end of the simulation. We observe no clear relationship between the MSE values and $\mathrm{Ra}$ values.  \par

\subsection{Example 2: Elder problem}

\begin{figure}[!ht]
   \centering

         \includegraphics[keepaspectratio, height=4.5cm]{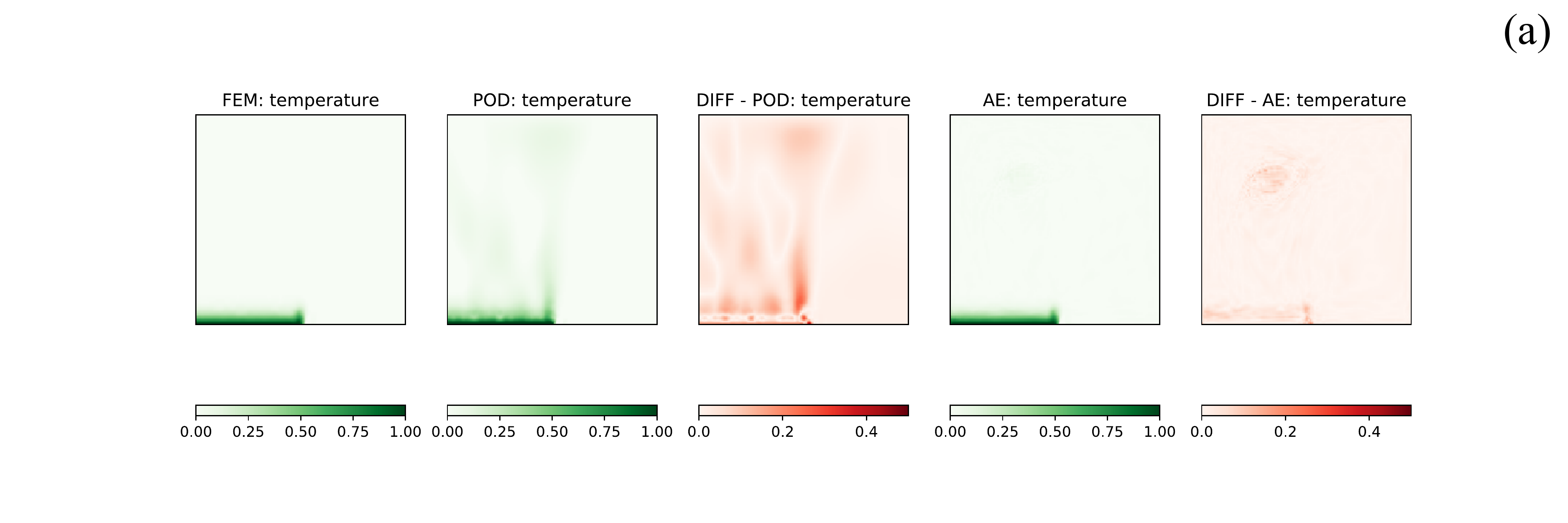}
         \includegraphics[keepaspectratio, height=4.5cm]{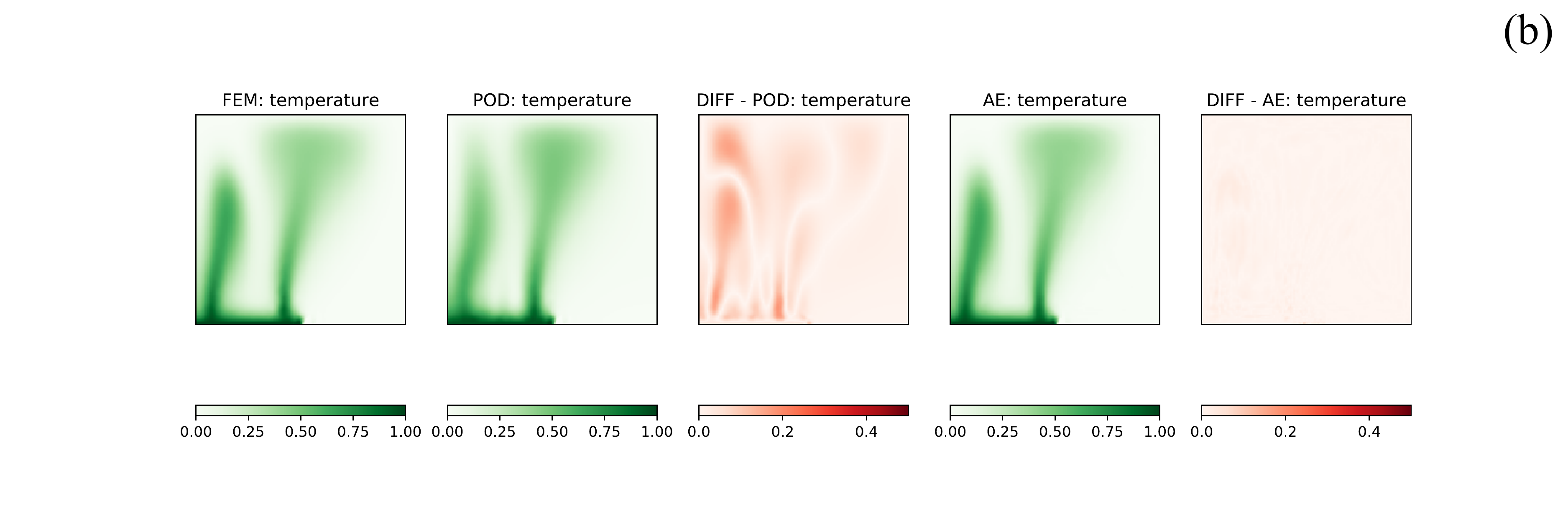}
         \includegraphics[keepaspectratio, height=4.5cm]{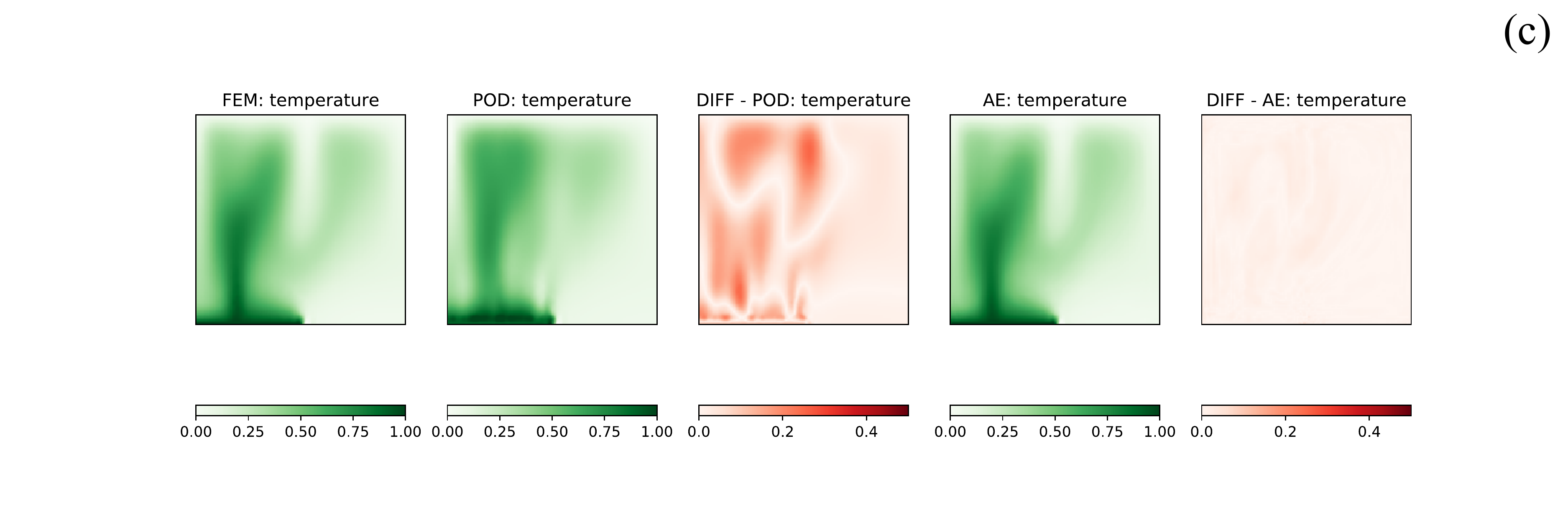}
   \caption{Example 2: test case ($\mathrm{Ra} = 380.16$) results. (a) $t = 0.001$, (b) $t = 0.03$, and (c) $t = 0.07$. These results are produced by POD 16 RB and AE 16 z: ANN for POD: temperature and AE: temperature, respectively.}
   \label{fig:ex2_pic}

\end{figure}

We progress to the Elder problem \citep{elder1967transient}, which could be considered as a benchmark in the natural convection literature \citep{simpson2003theoretical}. The problem's domain and boundary conditions are presented in Figure \ref{fig:geo1}b \citep{diersch2002variable, zhang2016mixed}. As the medium is heated at its lower part, the flow is driven upward by buoyancy. Again, see Table \ref{tab:main_info}, we set $\bm{\mu} = (\mathrm{Ra})$, and its admissible range of variation is $[350.0, 400.0]$. We note that as the $\mathrm{Ra}$ number considered in this case is relatively high, the flow instability may cause fingering behavior, see Figure \ref{fig:ex2_pic}a-c, first column,  \citep{simpson2003theoretical,zhang2016mixed}. Moreover, the higher $\mathrm{Ra}$ value also affect the minimum and maximum $N^t$ as its ranges is increased (compared with Example 1) to $[790, 1010]$. We use $\mathrm{M} = 40$, $\mathrm{M}_\mathrm{v} = 10$, and $\mathrm{M}_\mathrm{t} = 10$. We have in total $\mathrm{M} N^t = 36110$ training data points. \par



\begin{figure}[!ht]
   \centering
         \includegraphics[keepaspectratio, height=6.0cm]{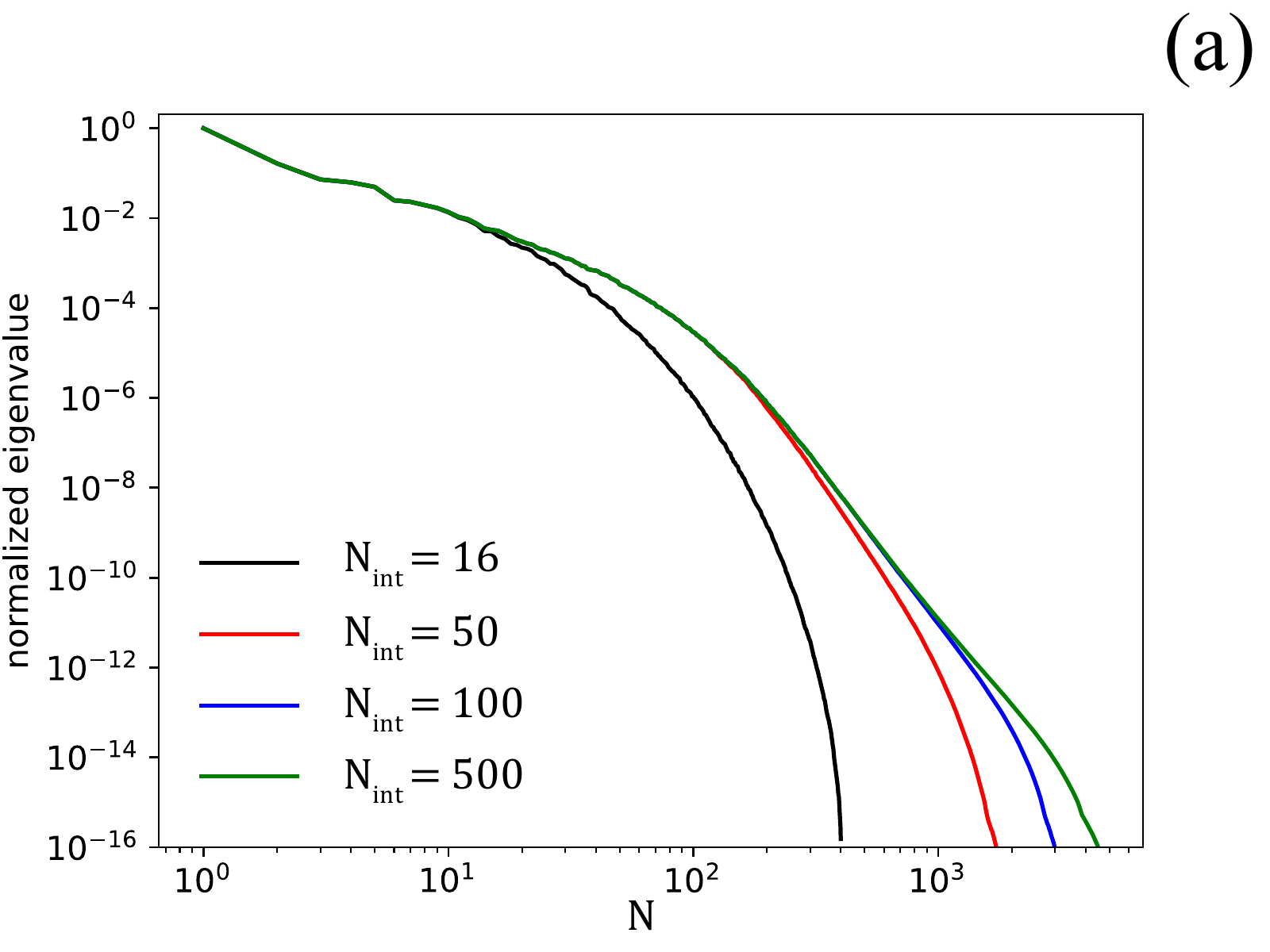}
         \includegraphics[keepaspectratio, height=6.0cm]{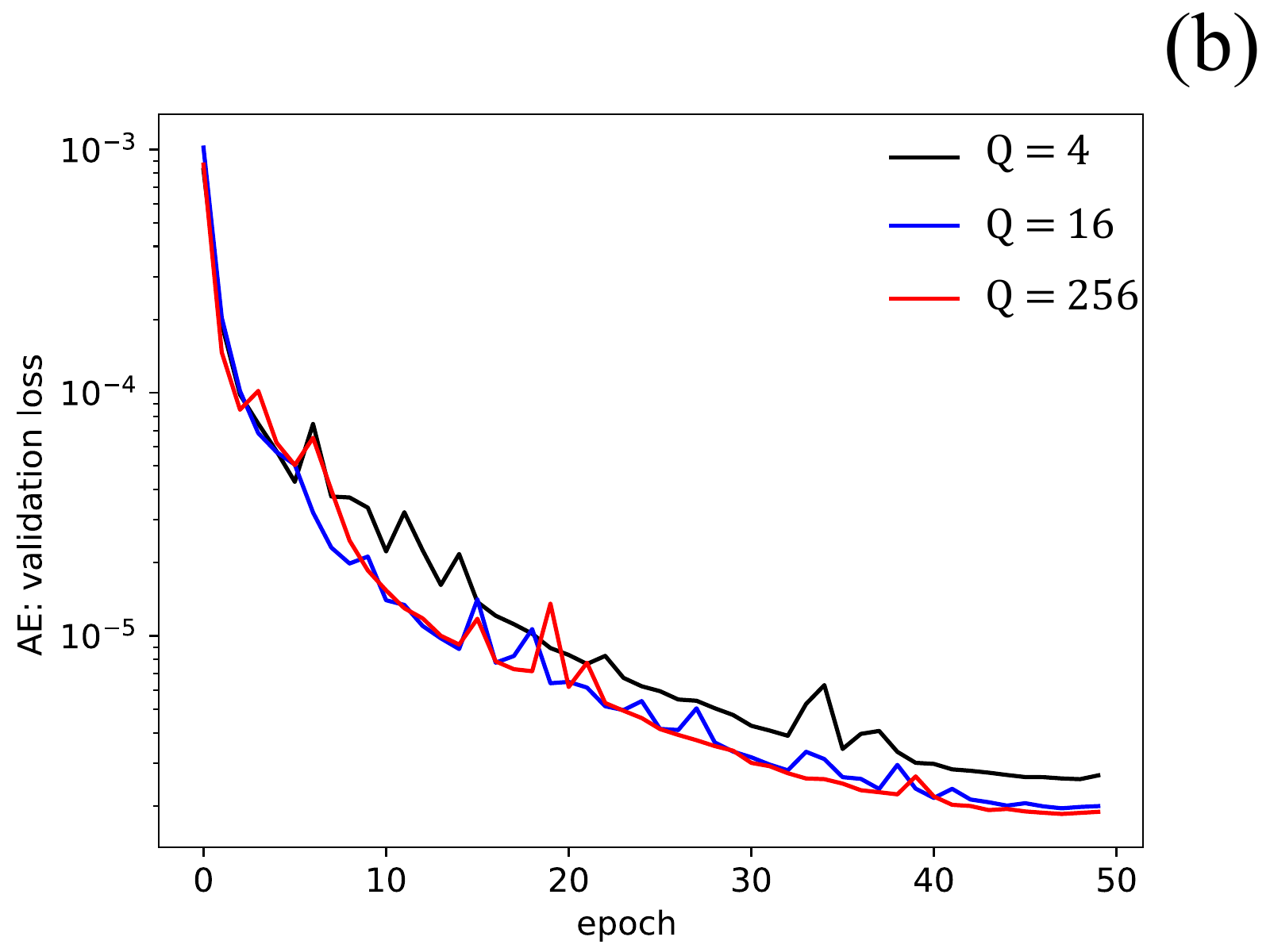}
   \caption{Example 2: (a) normalized eigenvalue as a function of reduced basis and (b) autoencoder validation loss during the training.}
   \label{fig:ex2_train}

\end{figure}

The normalized eigenvalue as a function of the reduced basis for the linear compression and AE validation loss for the nonlinear compression are presented in Figure \ref{fig:ex2_train}. The eigenvalue decay behavior is different from Example 1. We observe that as $\mathrm{N_{int}}$ is increased, the POD requires more modes or $\mathrm{N}$ to capture information (i.e., the decay becomes slower as $\mathrm{N_{int}}$ grows). This behavior reflects that Example 2 is more challenging than Example 1 for the linear compression. For the nonlinear compression, Figure \ref{fig:ex2_train}b, the validation loss of models with $\mathrm{Q} \geq 16$ is approximately $2 \times 10^{-6}$, while the model with $\mathrm{Q}=4$ has the validation loss of $4 \times 10^{-6}$. This behavior implies that we may need to use the AE with at least 16 latent spaces to obtain better accuracy. We note that the AE validation loss of Example 2 is lower than that of Example 1 because we have more available training data points, see Table \ref{tab:main_info}. \par

Similar to Example 1, we use $\mathrm{M_t} = 10$. Again, we want to emphasize that our testing ($\bm{\mu}_{\mathrm{t}}^{(i)}$) set does not contain any $\bm{\mu}$ values used in the training or validation set. One example (out of 10) of the test cases is presented in Figure \ref{fig:ex2_pic} for three different $t$ values. As expected and dissimilar to Example 1, we observe that the nonlinear compression provides a much better result than the linear compression, see $\operatorname{DIFF}$ values. \par

\begin{figure}[!ht]
   \centering
         \includegraphics[keepaspectratio, height=8.0cm]{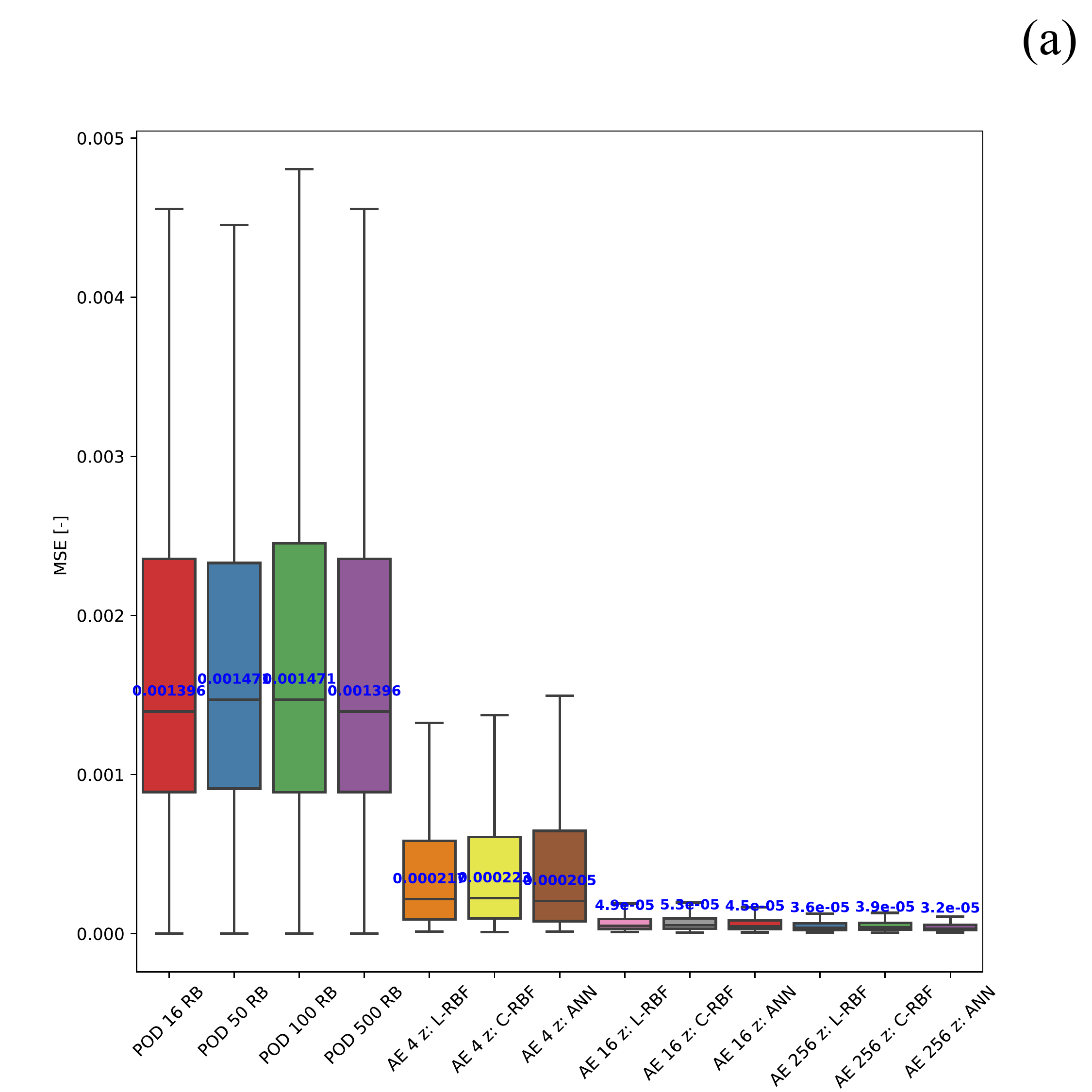}
         \includegraphics[keepaspectratio, height=8.0cm]{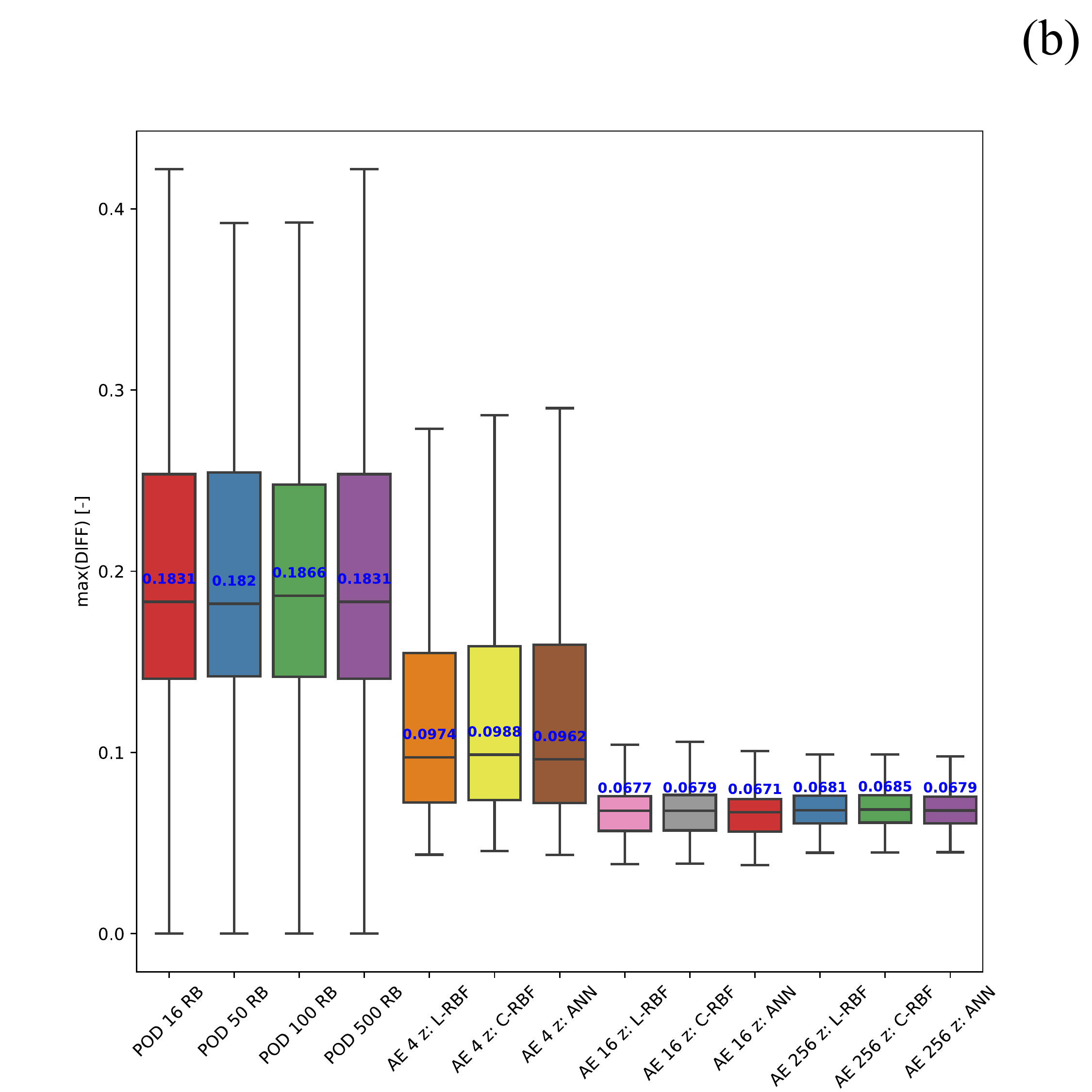}
   \caption{Example 2: Each model's performance on a test set (a) mean square error (MSE) and (b) maximum DIFF (max(DIFF)). Blue text represents a mean value. 16 RB, 50 RB, 100 RB, and 500 RB represent linear compression models with $\left[\mathrm{N_{int}} = 16, \mathrm{N} = 16\right]$, $\left[\mathrm{N_{int}} = 50, \mathrm{N} = 50\right]$, $\left[\mathrm{N_{int}} = 100, \mathrm{N} = 100\right]$, and $\left[\mathrm{N_{int}} = 500, \mathrm{N} = 500\right]$, respectively. The L-RBF and C-RBF are the nonlinear compression models with RBF with linear function and RBF with cubic function as its approximator, respectively.}
   \label{fig:ex2_test}
\end{figure}

The MSE and max(DIFF) results of Example 2 are illustrated in Figure \ref{fig:ex2_test}. From Figure \ref{fig:ex2_test}a, the MSE values of the POD approach is approximately one order of magnitude higher than the AE with $\mathrm{Q}=4$, and two orders of magnitude higher than the nonlinear compression with $\mathrm{Q} \geq 16$. The max(DIFF) results follow the same behavior of MSE results, see Figure \ref{fig:ex2_test}b. We observe that there is no accuracy difference between the nonlinear compression with $\mathrm{Q}=16$ or $\mathrm{Q}=256$. Moreover, the nonlinear compression models with L-RBF, C-RBF, or ANN as its approximator have approximately the same accuracy. We provide RBF training loss in Figure \ref{fig:si_training_loss_rbf_elder} and ANN validation loss during the training in Figure \ref{fig:si_training_loss_ann}b. \par

\begin{figure}[!ht]
   \centering
         \includegraphics[keepaspectratio, height=8.0cm]{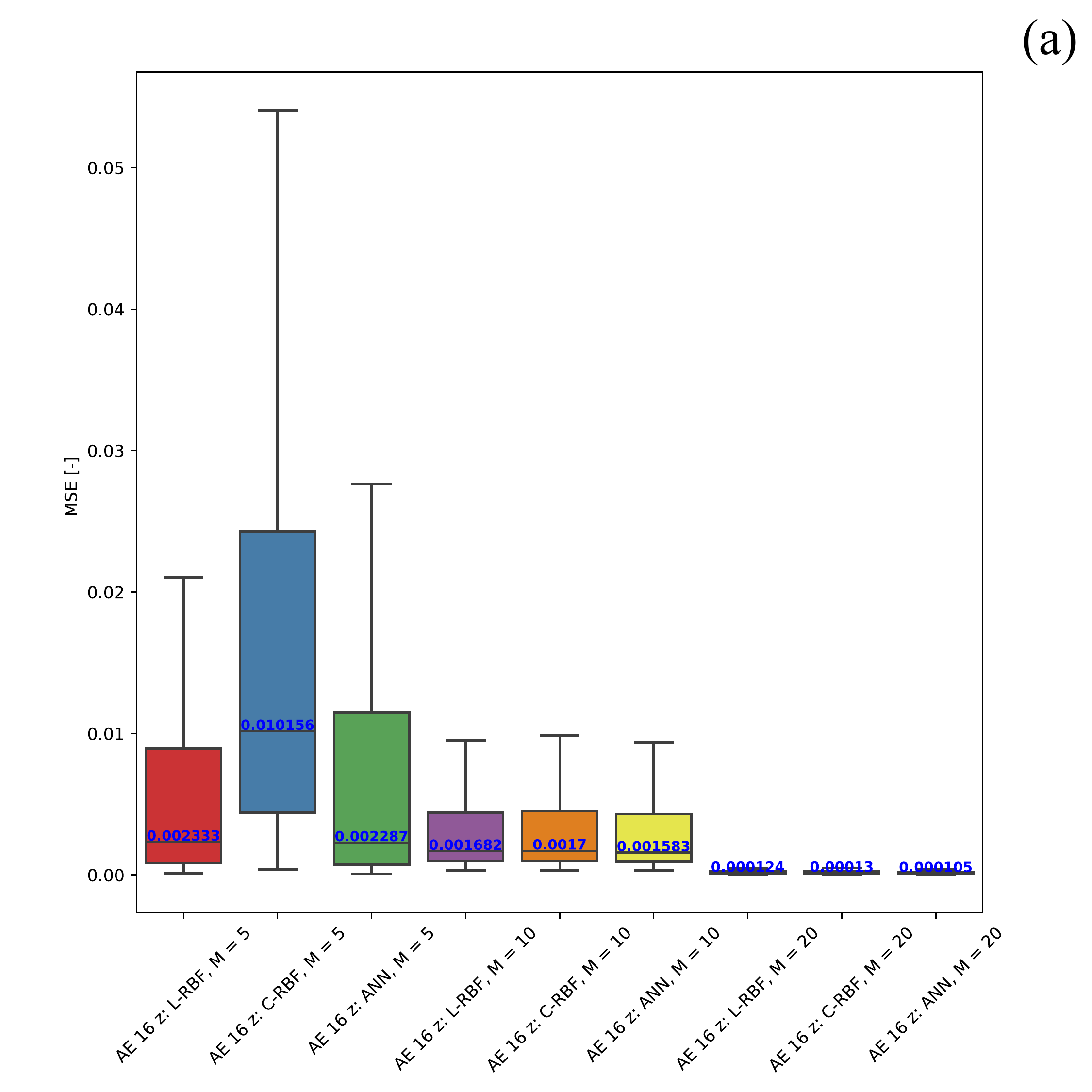}
         \includegraphics[keepaspectratio, height=8.0cm]{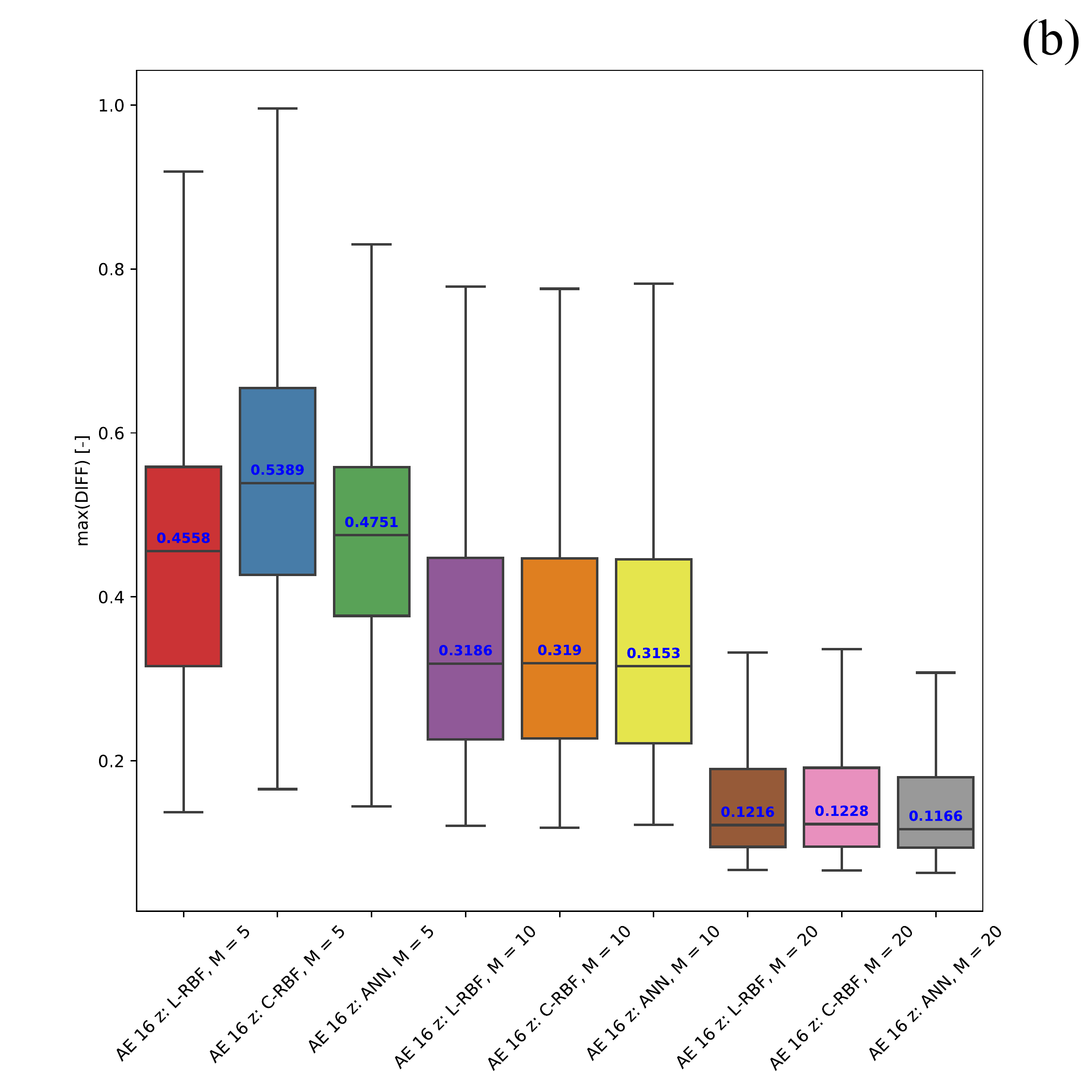}
   \caption{Example 2: Each model's performance on a test set with different number of training data, $\mathrm{M} = 5$, $\mathrm{M} = 10$, and $\mathrm{M} = 20$, (a) mean square error (MSE) and (b) maximum DIFF (max(DIFF)). Blue text represents a mean value. We here only present results of nonlinear approach using $\mathrm{Q} = 16$ with L-RBF, C-RBF, or ANN as its approximator.}
   \label{fig:ex2_test_sen_M}
\end{figure}

We also present the model's accuracy as a function of $\mathrm{M}$ in Figure \ref{fig:ex2_test_sen_M}, for $\mathrm{M} = 5$, $\mathrm{M} = 10$, and $\mathrm{M} = 20$. For $\mathrm{M} = 40$, please refer to Figure \ref{fig:ex2_test}. Note that we only present results of nonlinear approach using $\mathrm{Q} = 16$ with L-RBF, C-RBF, or ANN as its approximator. As expected, the model with higher $\mathrm{M}$ has a better accuracy, see Figures \ref{fig:ex2_test} and \ref{fig:ex2_test_sen_M}. The nonlinear models with $\mathrm{M} \geq 10$ can deliver a better accuracy than those of the linear approach with $\mathrm{M} = 40$. \par

\begin{figure}[!ht]
   \centering
         \includegraphics[keepaspectratio, height=7.5cm]{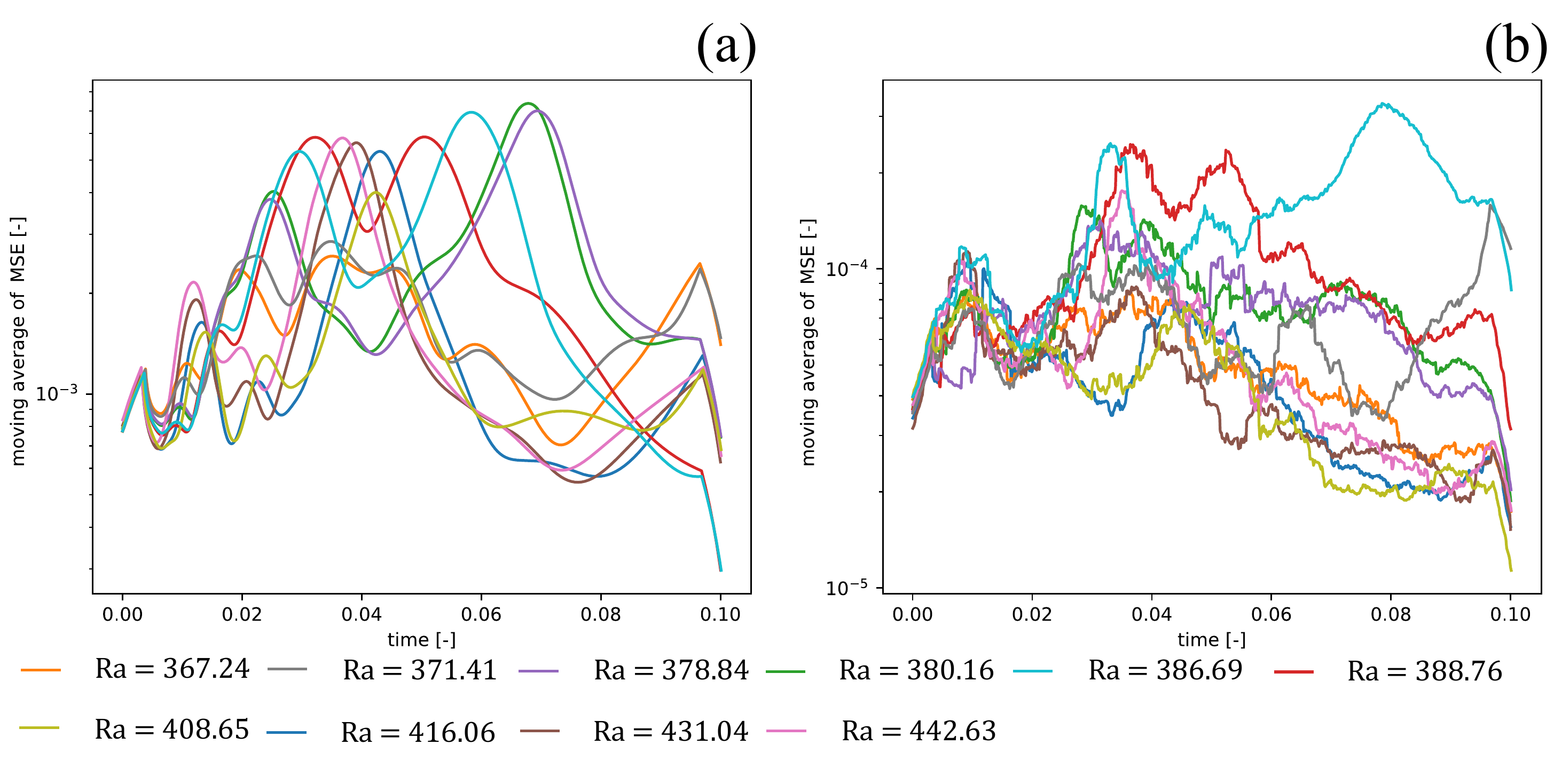}
   \caption{Example 2: $50^{th}$ moving average of MSE on a test set of (a) POD 16 RB and (b) AE 16 z: ANN.}
   \label{fig:ex2_time}

\end{figure}

We plot the $50^{th}$ moving average of MSE as a function of $t$ in Figure \ref{fig:ex2_time}. Unlike the previous example, see Figure \ref{fig:ex1_time}, we could not find any patterns of MSE as a function of $t$ in this plot. Moreover, as we also observe in Example 1, there is no clear relationship between the MSE values and $\mathrm{Ra}$ values. \par

\subsection{Example 3: Modified Elder problem}

\begin{figure}[!ht]
   \centering

         \includegraphics[keepaspectratio, height=4.5cm]{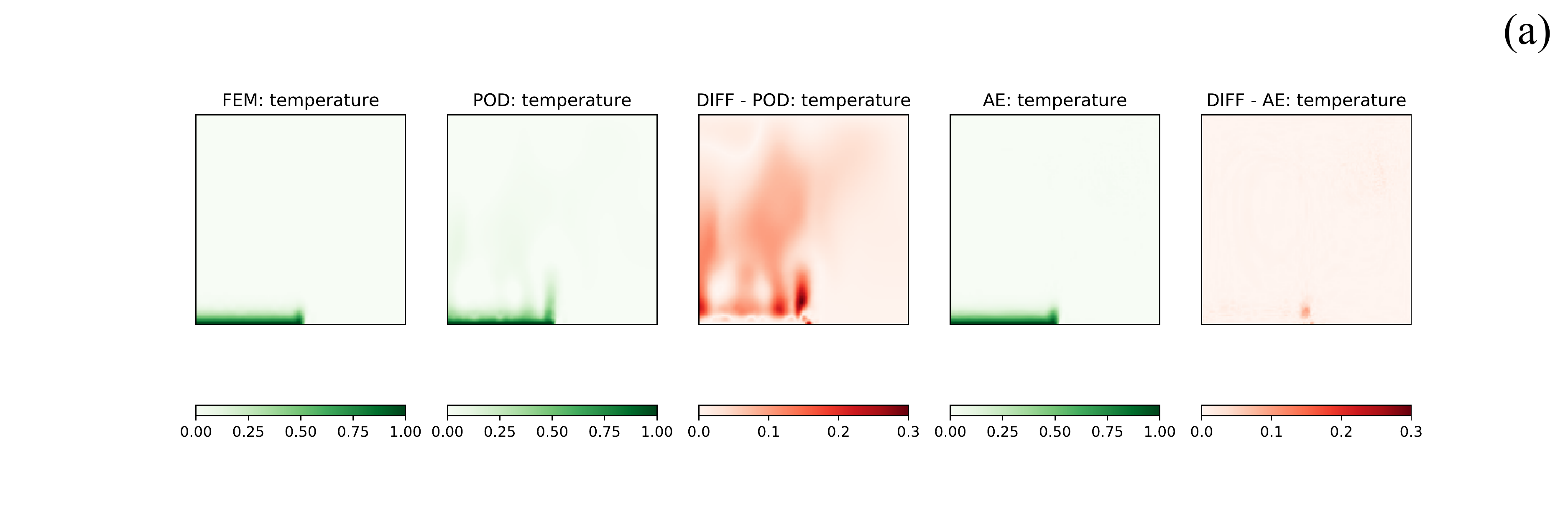}
         \includegraphics[keepaspectratio, height=4.5cm]{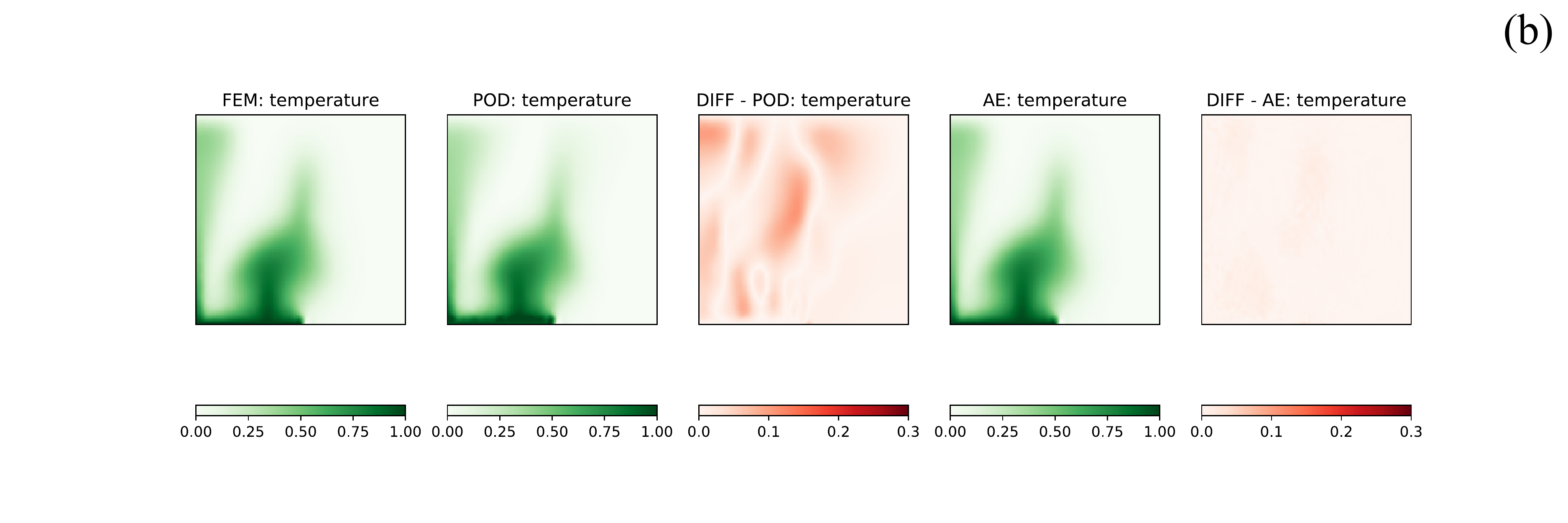}
         \includegraphics[keepaspectratio, height=4.5cm]{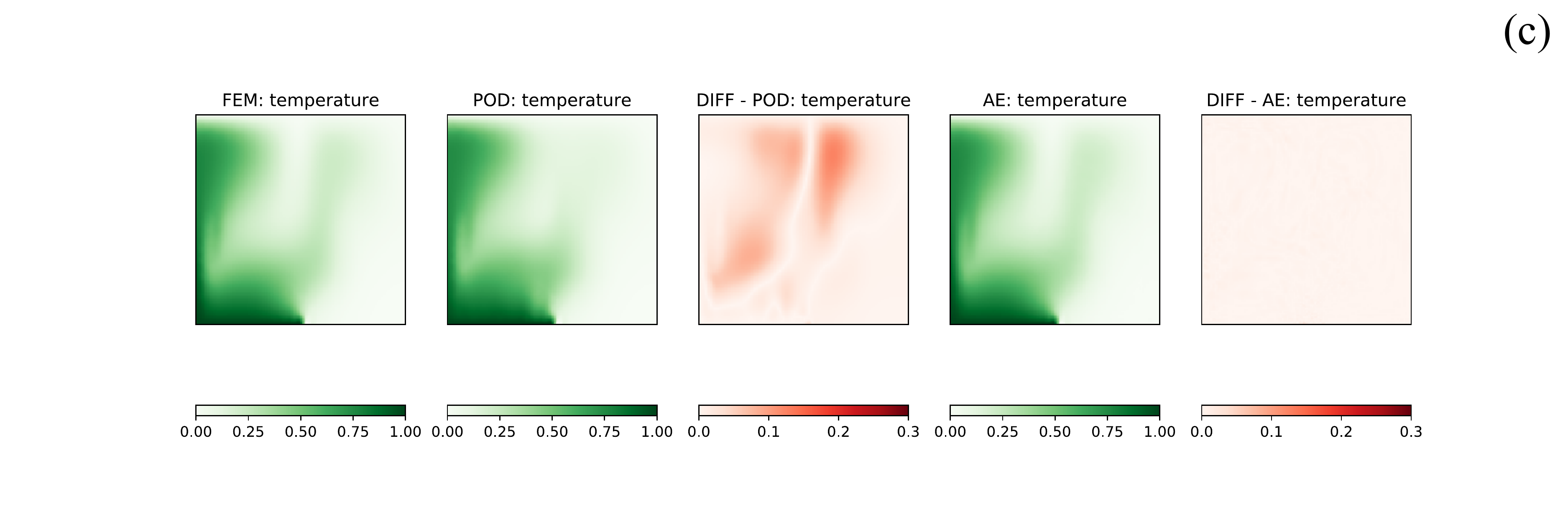}
   \caption{Example 3: test case ($\mathrm{Ra_1} = 357.39$, $\mathrm{Ra_2} = 0.004$) results. (a) $t = 0.001$, (b) $t = 0.03$, and (c) $t = 0.07$. These results are produced by POD 16 RB and AE 16 z: ANN for POD: temperature and AE: temperature, respectively.}
   \label{fig:ex3_pic}

\end{figure}

So far, we have tested our framework on problems with only one parameter (i.e. $\bm{\mu}$ is a $1-$vector). In this example, we modify the Elder problem by inserting one subdomain as presented in Figure \ref{fig:geo1}c. We add one more parameter and set $\bm{\mu} = (\mathrm{Ra}_1, \mathrm{Ra}_2)$. Their admissible range of variation are $[350.0, 400.0]$ and $[0.001, 100.0]$ for $\mathrm{Ra}_1$ and $\mathrm{Ra}_2$, respectively. Similar to the previous example, the range of $\mathrm{Ra}_1$ may cause fingering behavior, see Figure \ref{fig:ex3_pic}a-c, first column. The minimum and maximum $N^t$ is $[740, 985]$, which is not much different from Example 2. We use $\mathrm{M} = 100$, $\mathrm{M}_\mathrm{v} = 20$, and $\mathrm{M}_\mathrm{t} = 20$. We have in total $\mathrm{M} N^t = 86086$ training data points. Please refer to Table \ref{tab:main_info} for the summary. \par



\begin{figure}[!ht]
   \centering
         \includegraphics[keepaspectratio, height=6.0cm]{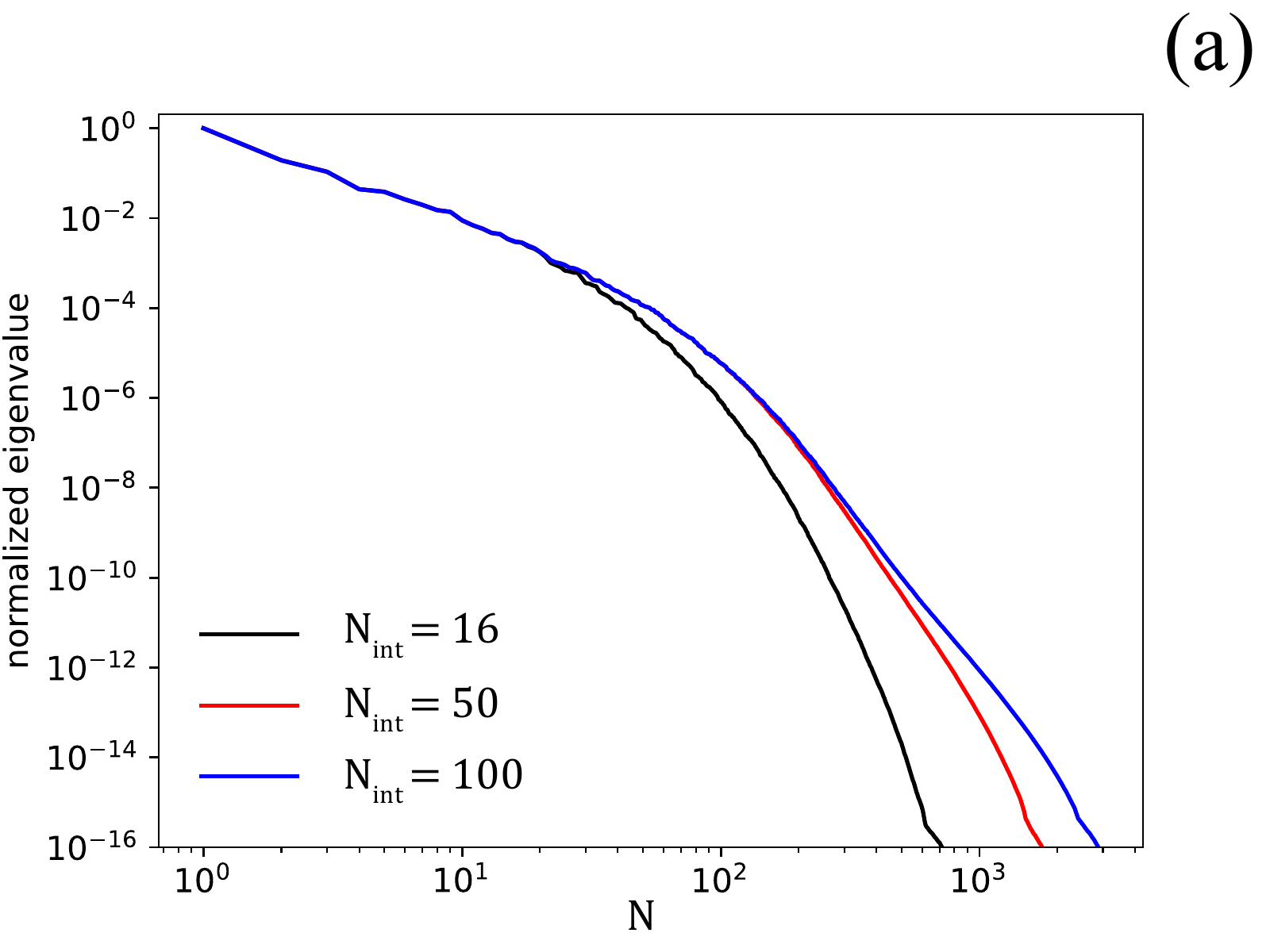}
         \includegraphics[keepaspectratio, height=6.0cm]{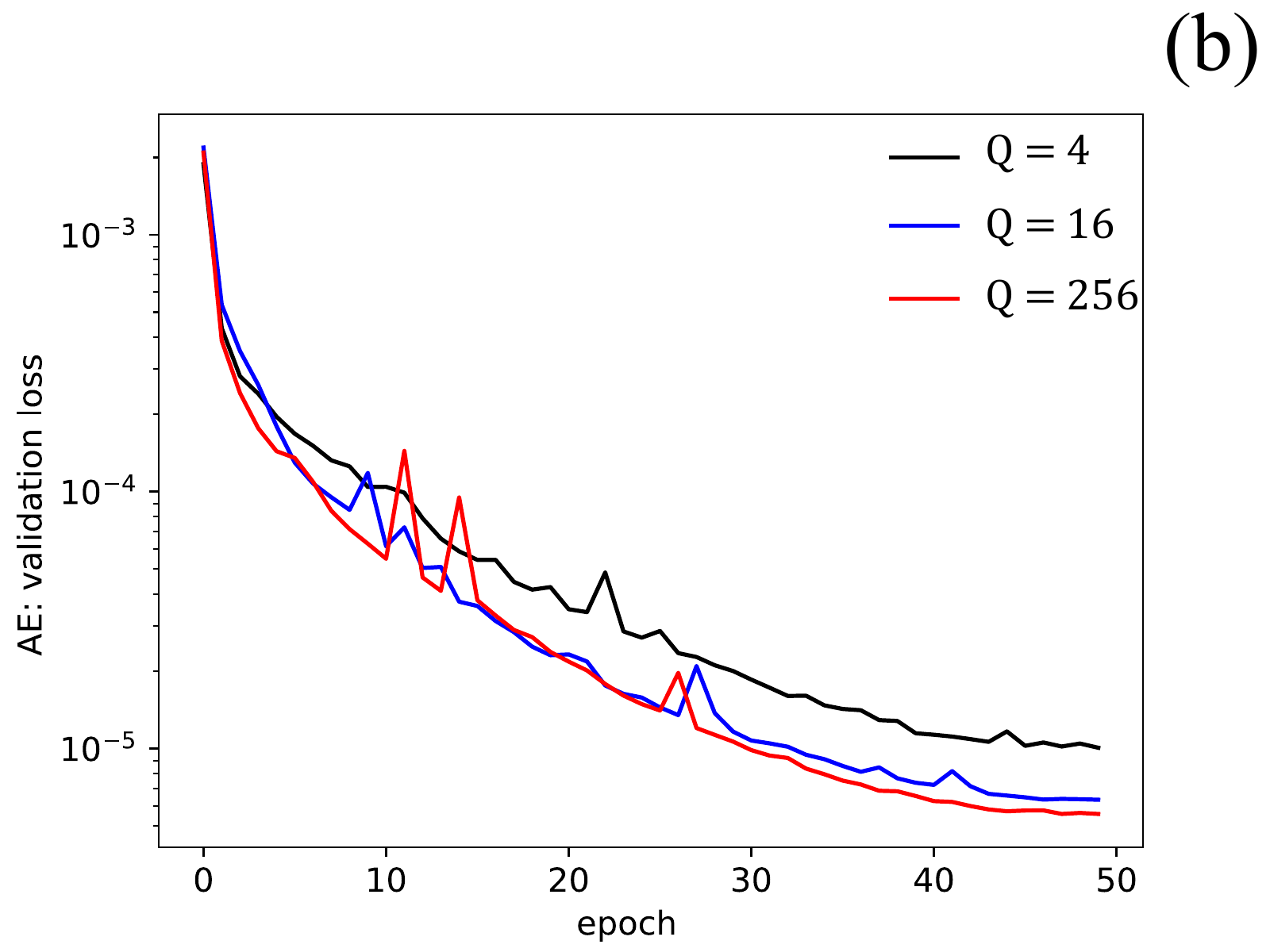}
   \caption{Example 3: (a) normalized eigenvalue as a function of reduced basis and (b) autoencoder validation loss during the training.}
   \label{fig:ex3_train}

\end{figure}

We present the normalized eigenvalue as a function of reduced basis and AE validation loss during the training in Figure \ref{fig:ex3_train}. As expected, the decay behavior of the eigenvalue (Figure \ref{fig:ex3_train}a) is similar to that of Example 2, which means the decay becomes slower as $\mathrm{N_{int}}$ grows. This behavior shows that Examples 2 and 3 are more complex than Example 1. From Figure \ref{fig:ex3_train}b, the nonlinear compression, the validation loss of model with $\mathrm{Q} = 4$ is approximately $2 \times 10^{-5}$, model with $\mathrm{Q} = 16$ is approximately $7 \times 10^{-6}$, and model with $\mathrm{Q} = 256$ is approximately $6 \times 10^{-6}$. We note that the AE validation loss of Example 3 is higher than that of Example 1 even though we have more available training data points, see Table \ref{tab:main_info}. However, the trend of the validation loss is still decreasing, which means it may require more epochs to achieve the same level of the validation loss. \par

In Example 3, we use $\mathrm{M_t} = 20$, and we want to emphasize that our testing set ($\bm{\mu}_{\mathrm{t}}^{(i)}$) does not contain any $\bm{\mu}$ values used in the training or validation set. One example (out of 20) of the test cases is presented in Figures \ref{fig:ex3_pic} for three different $t$ values. Similar to Example 2, we observe that the nonlinear compression provides a much better result than the linear compression, see $\operatorname{DIFF}$ values.

\begin{figure}[!ht]
   \centering

         \includegraphics[keepaspectratio, height=8.0cm]{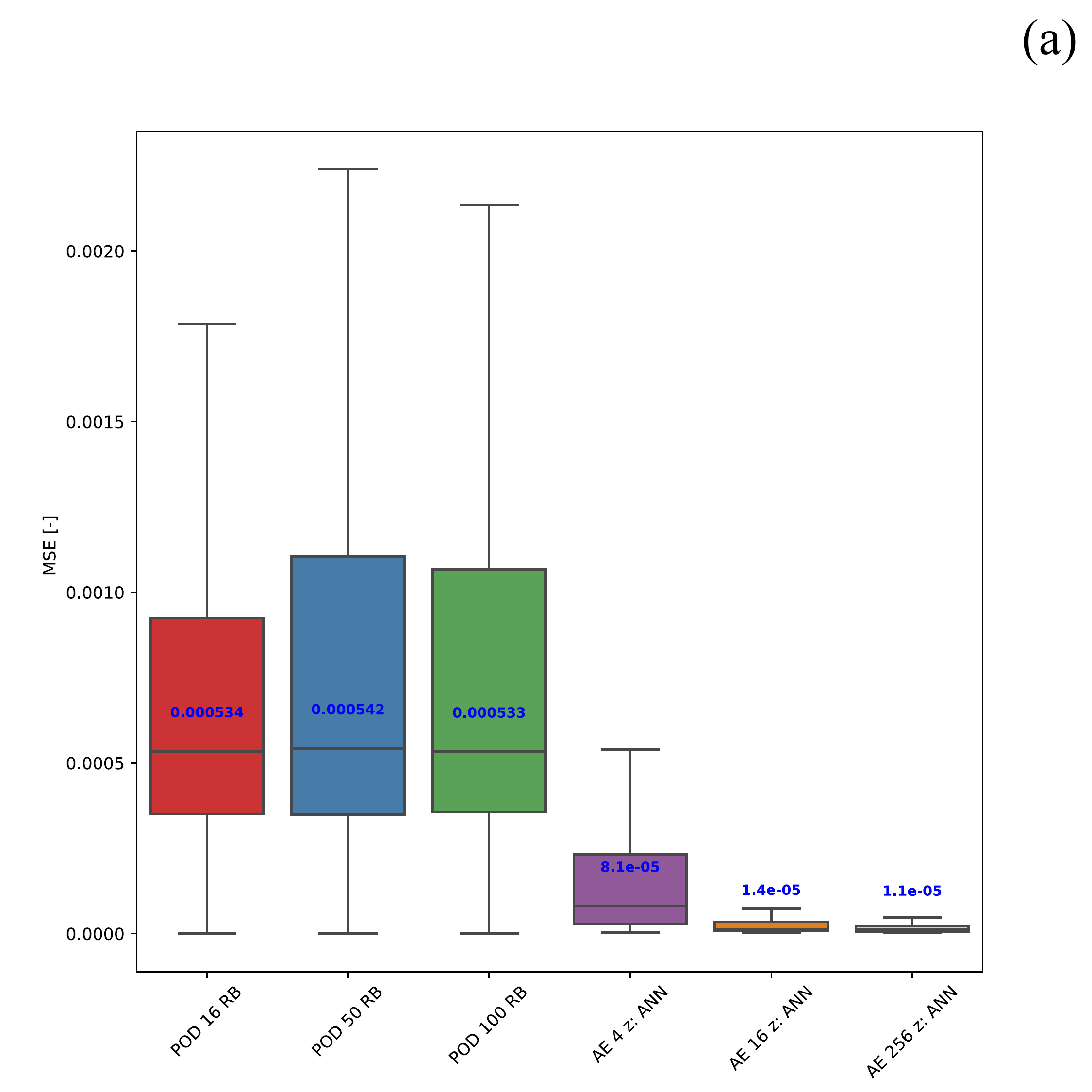}
         \includegraphics[keepaspectratio, height=8.0cm]{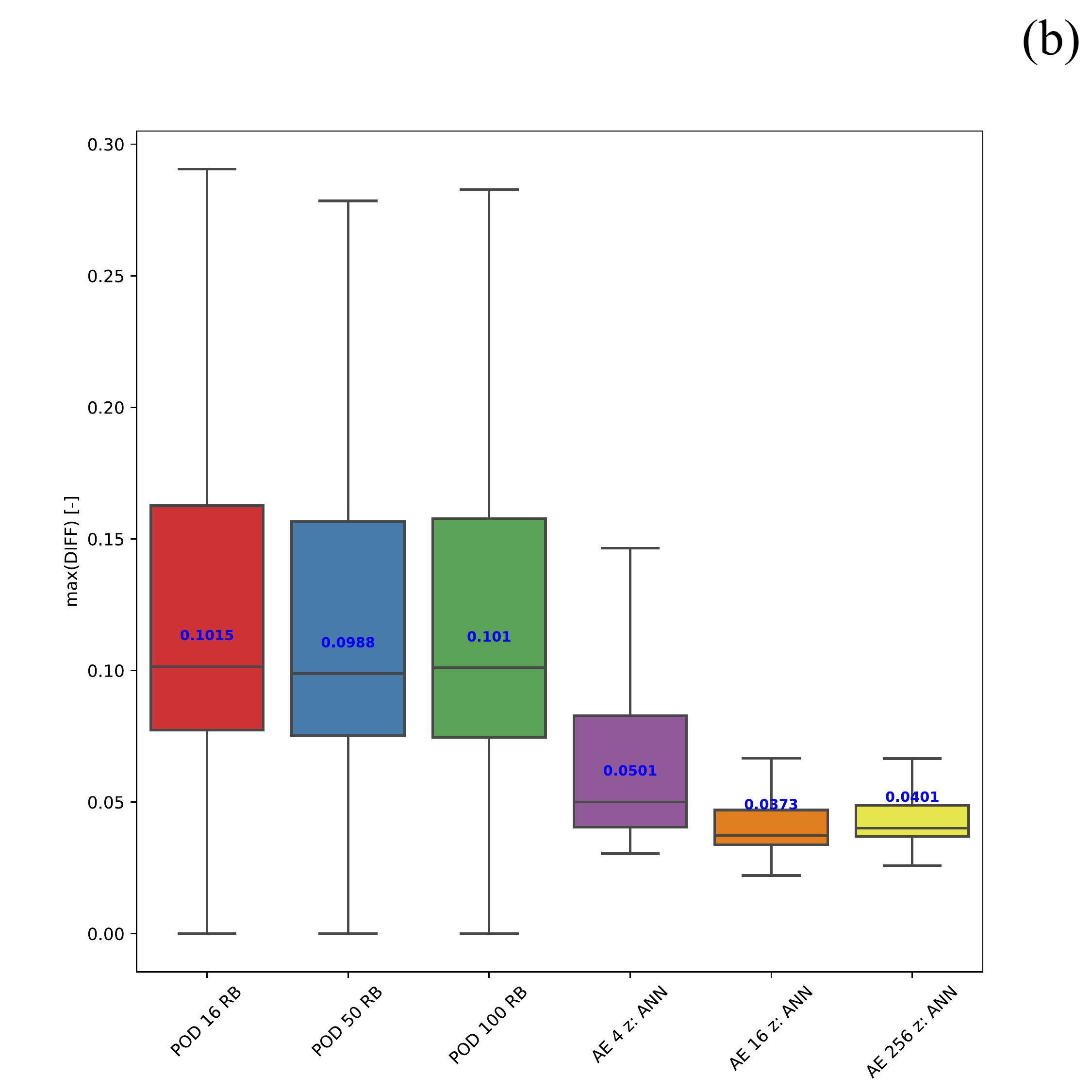}
   \caption{Example 3: Each model's performance on a test set (a) mean square error (MSE) and (b) maximum DIFF (max(DIFF)). Blue text represents a mean value.}
   \label{fig:ex3_test}

\end{figure}

The MSE and max(DIFF) results of Example 3 are presented in Figure \ref{fig:ex3_test}. We note that as we could not see much difference of the nonlinear compression models with RBF or ANN as their approximator (see Figures \ref{fig:ex1_test} and \ref{fig:ex2_test}), we here present the results of the AE models with ANN approximator for the sake of compactness. The MSE values of the POD approach are approximately one order of magnitude higher than the AE with $\mathrm{Q}=4$ and almost two orders of magnitude higher than the nonlinear compression with $\mathrm{Q} \geq 16$. From Figure \ref{fig:ex3_test}b, the max(DIFF) results follow the same trend of the MSE results. Similar to Example 2, we observe no accuracy difference between the nonlinear compression with $\mathrm{Q}=16$ or $\mathrm{Q}=256$. We provide ANN validation loss during the training in Figure \ref{fig:si_training_loss_ann}c. \par

\begin{figure}[!ht]
   \centering

         \includegraphics[keepaspectratio, height=9.5cm]{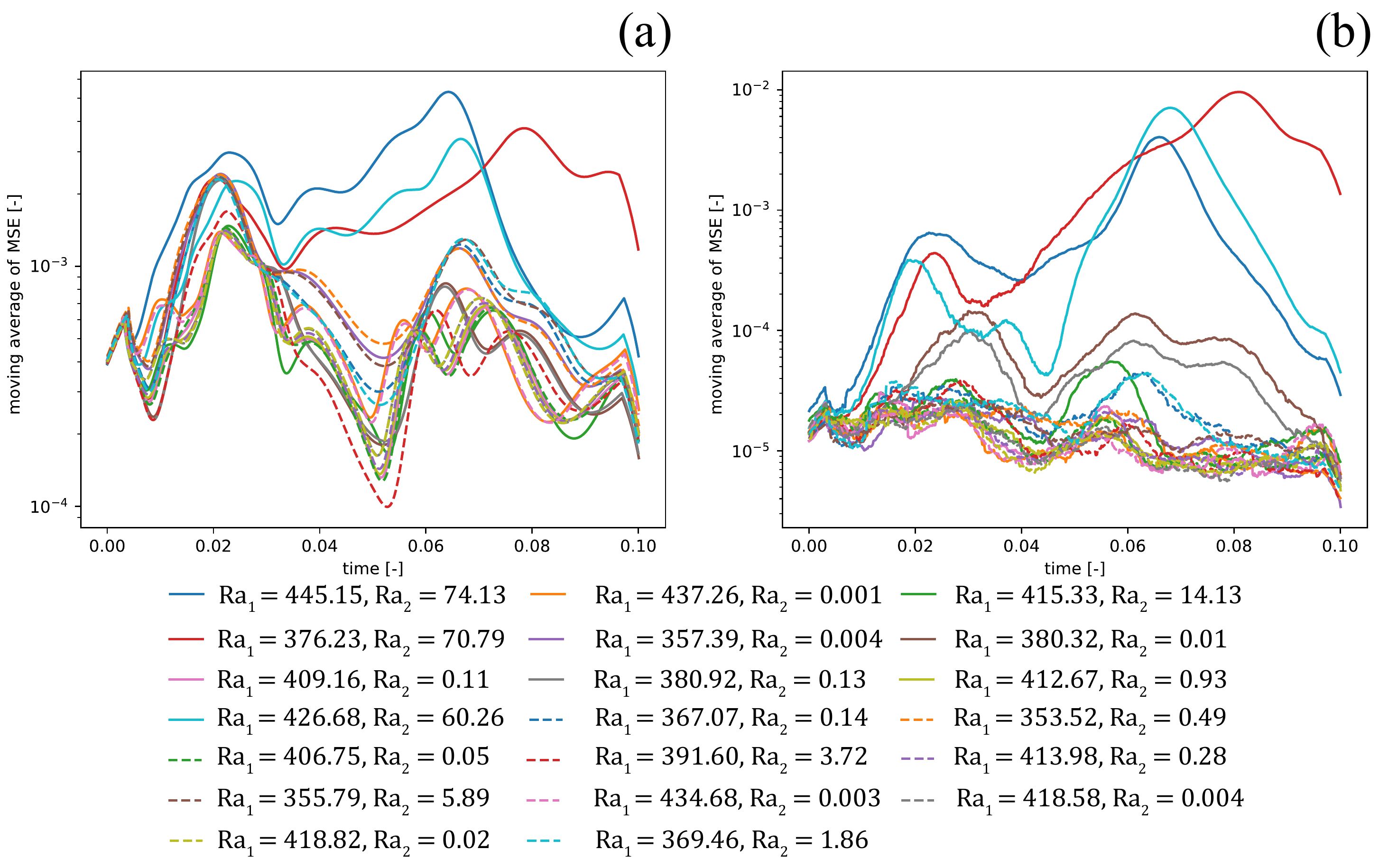}
   \caption{Example 3: $50^{th}$ moving average of MSE on a test set of (a) POD 16 RB and (b) AE 16 z: ANN. 16 RB, 50 RB, and 100 RB represent linear compression models with $\left[\mathrm{N_{int}} = 16, \mathrm{N} = 16\right]$, $\left[\mathrm{N_{int}} = 50, \mathrm{N} = 50\right]$, and $\left[\mathrm{N_{int}} = 100, \mathrm{N} = 100\right]$, respectively.}
   \label{fig:ex3_time}

\end{figure}

We present the $50^{th}$ moving average of MSE as a function of $t$ in Figure \ref{fig:ex3_time} for all the samples in our test set. Unlike Examples 1 and 2, in which we could not observe any relationships between the MSE values and $\mathrm{Ra}$ values, here we could see that three cases have significantly higher errors than those of the rest. These three cases share the same characteristic in which their $\mathrm{Ra_2}$ values are relatively high, $\mathrm{Ra_2} = 60.26$, $\mathrm{Ra_2} = 74.13$, and $\mathrm{Ra_2} = 74.13$.  \par

\begin{figure}[!ht]
   \centering
         \includegraphics[keepaspectratio, height=6.0cm]{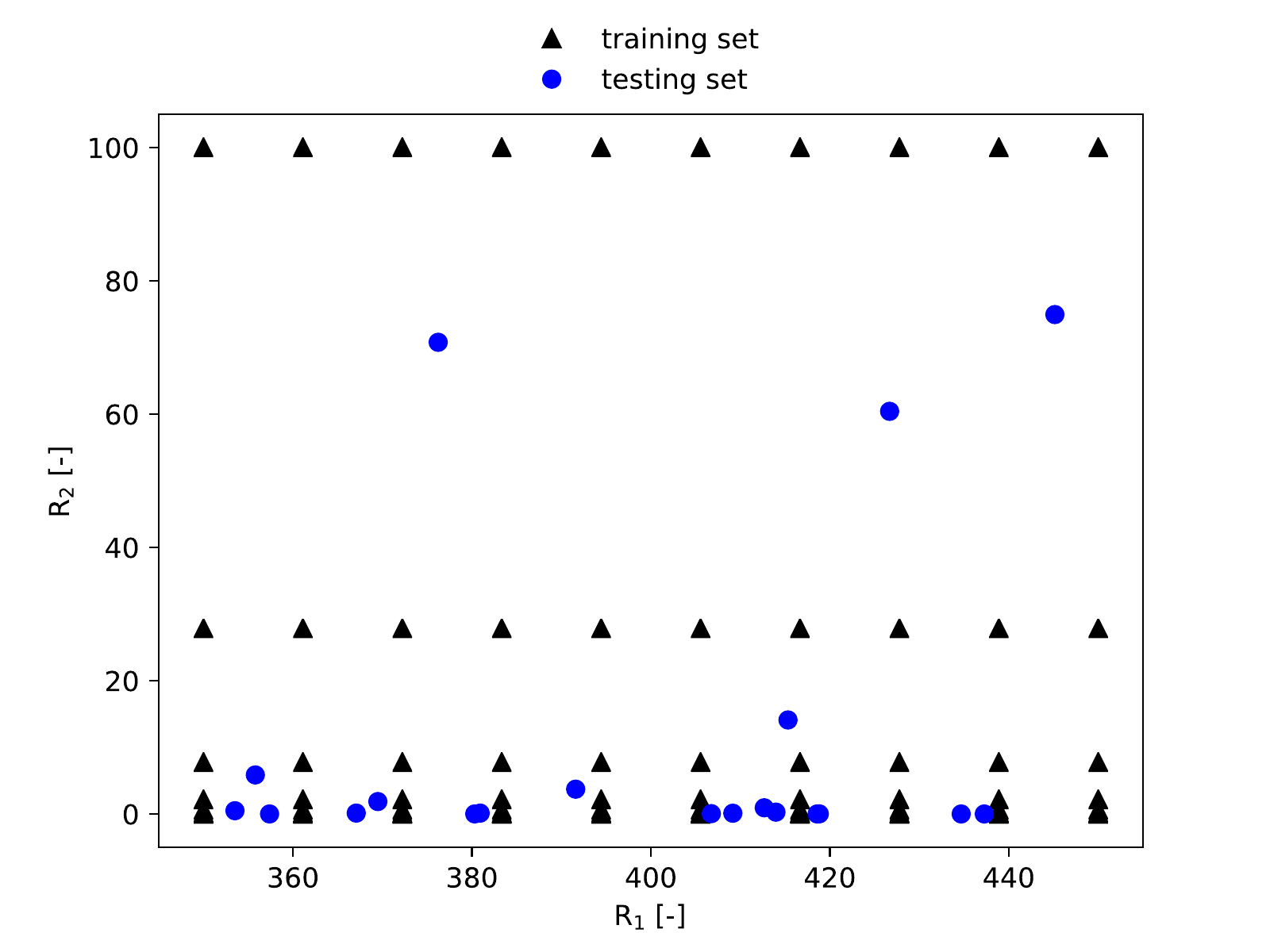}
   \caption{Example 3: training and testing samples: linear scale.}
   \label{fig:ex3_train_test}
\end{figure}

The reasons that these three cases have considerably higher errors than the rest of the test set stem from how we sample our training and testing sets. We plot our training and testing samples in Figure \ref{fig:ex3_train_test}. As the range of $\mathrm{Ra_2}$ spans several orders of magnitude (i.e., $\mathrm{Ra_2}=[0.001, 100.0]$), we sample $\mathrm{Ra_2}$ through its log-scale (i.e., $\log_{10}(\mathrm{Ra_2})=[-3, 2]$) using equispaced distribution for the training set. Consequently, there are not many training samples that lie in a range where $\mathrm{Ra_2}$ values are relatively high (see Figure \ref{fig:ex3_train_test}).  \par

The nonlinear compression approach seems to suffer from this sampling technique more than the linear one. To elaborate, for the POD approach, we observe that the MSE values of these three cases are approximately one order of magnitude higher than those of the group. However, MSE values become much higher (about two orders of magnitude) than the rest of the testing cases for the AE models. Therefore, one may want to consider adaptive sampling approaches \citep{paul2015adaptive,vasile2013adaptive,choi2020gradient} when dealing with nonlinear compression models.   \par

\section{Discussion} \label{sec:discussion}

In this section, we will discuss three main aspects of our findings: (1) model accuracy, (2) computational cost, and (3) a suggestion to select between linear and nonlinear non-intrusive reduced order modeling approaches. \par

\subsection{Model accuracy}

From the numerical experiments, we observe that the linear and nonlinear approaches can each perform better depending on the problem at hand. The linear approach has a better performance than its nonlinear counterpart, for Example 1. The linear compression average of MSE is about one order of magnitude less than that of the nonlinear compression.  In Example 2, the AE model with at least 16 latent spaces has MSE of two orders of magnitude less than the POD approach. The same goes for Example 3, where we have more parameters than Examples 1 and 2 (i.e., we have two parameters; $\mathrm{Ra_1}$ and $\mathrm{Ra_2}$). \par

For all examples, there is no significant accuracy difference among POD models with a different number of reduced bases $\mathrm{N}$ (as well as the number of intermediate reduced basis $\mathrm{N_{int}}$). As discussed extensively in \cite{kadeethum2021non}, the errors not only arise in the POD compression process but also during the mapping between the pair ($t$, $\bm{\mu}$) and the solution in the reduced linear subspace by ANN. Another observation is, for Examples 2 and 3, as we increase $\mathrm{N_{int}}$, the decay of eigenvalue as a function of $\mathrm{N}$ grows slower. This behavior indicates that the model evolution in time requires a large number of $\mathrm{N_{int}}$, which implies that the linear compression approach may not be suitable for Examples 2 and 3. For the nonlinear compression approach, we observe that using three different types of approximators, linear RBF, cubic RBF, and ANN, to map between the pair ($t$, $\bm{\mu}$) and the solution in the reduced nonlinear manifold ($\bm{z}$) results in approximately the same accuracy. Moreover, the training time and prediction time of each model are not significantly different.  \par

Another interesting observation is that using $\mathrm{M} = 40$ the nonlinear approach achieves around five to ten $\%$ of maximum relative errors for Examples 1, 2, and 3. On the other hand, the linear approach delivers a wide range of relative errors across Examples 1, 2, and 3. We also test our nonlinear model using different $\mathrm{M}$. As expected, the model's accuracy increases as $\mathrm{M}$ increases.

In Example 3, we illustrate that the sampling technique used to generate our training set could affect our framework's accuracy. Especially, the performance of the nonlinear approach is substantially influenced by regions of sparse training data (i.e., test cases that lie within the sparse training data zone have MSE values about two orders of magnitude higher than the rest of the testing cases). Hence, we recommend a future study of the application of adaptive sampling approaches \citep{paul2015adaptive,vasile2013adaptive} applied to the nonlinear compression model. \par

We want to emphasize that the linear approach could handle data from unstructured mesh, which is beneficial because it is applicable to complex geometries. Our nonlinear approach; however, relies on convolutional layers (without convolutional layers, its accuracy decreased as shown in Appendix \ref{sec:non_conv}); hence, to apply this model to more complex geometries, one may require additional techniques such as graph convolutional networks \citep{kipf2016semi} or space-filling curves \citep{heaney2020applying}. \par

\subsection{Computational cost}

We perform FOM, POD, $L_2$ projection, RBF, and linear reconstruction on AMD Ryzen Threadripper 3970X. The training of AE and ANN and nonlinear reconstruction are done by a single Quadro RTX 6000. We note that without using Quadro RTX 6000, our AE model requires impractical training time on AMD Ryzen Threadripper 3970X. The FOM is done using  multiphenics (\url{https://github.com/multiphenics/multiphenics}) and FEniCS (\url{https://fenicsproject.org/}) frameworks. The POD is done using RBniCS (\url{https://github.com/RBniCS/RBniCS/}). The ANN and AE models are developed based on PyTorch (\url{https://pytorch.org/}), and the RBF is performed using SciPy (\url{https://www.scipy.org/}). \par

The computational time used to develop our ROM can be broken down into three main parts: (i) generation of training data through FOM - the second step in Figure \ref{fig:data_driven}, (ii) data compression - the third step in Figure \ref{fig:data_driven}, and (iii) mapping of $t$ and $\bm{\mu}$ to reduced subspace - the fourth step in Figure \ref{fig:data_driven}. We note that initialization of $\bm{\mu}$ - the first step in Figure \ref{fig:data_driven} and prediction or online phase - the fifth step in Figure \ref{fig:data_driven} are relatively much cheaper and insignificant compared to steps 2, 3, and 4. \par

The generation of training data through FOM - the second step in Figure \ref{fig:data_driven}, requires the highest computational resources. On average, each FOM snapshot, each $\bm{\mu}^{(i)}$ for all $t$, takes about two hours on AMD Ryzen Threadripper 3970X (4 threads). Again, we note that our FOM utilizes the adaptive time-stepping, see Equation  \eqref{eq:time_mult}; hence, each $\bm{\mu}^{(i)}$ may require a substantially different computational time. To elaborate, cases that have higher $\mathrm{Ra}$ usually have a smaller time-step ($N^t$ becomes larger); and subsequently, they require more time to complete. \par  


For the data compression step - the third step in Figure \ref{fig:data_driven}, the linear compression technique usually takes a longer time than its nonlinear counterpart. However, we note that this may not be a fair comparison as we perform POD and AE using different platforms and machines. Moreover, we want to emphasize that without using \emph{nested} POD technique, our linear compression approach would not be able to complete its operation in a timely manner. On average, the POD operation takes about five to eight hours, while the AE model takes approximately four to six hours. We note that this comparison is made using different machines. As mentioned previously, the POD is computed using CPU while the AE is trained using GPU. We want to emphasize that training our AE model using CPU would require an impractical training time. Moreover, using incremental SVD or randomized SVD may reduce our computational time of POD significantly \citep{choi2019librom}.  \par

The mapping of $t$ and $\bm{\mu}$ to reduced subspace - the fourth step in Figure \ref{fig:data_driven} is the cheapest among the three main parts. For linear compression, we only use the ANN model as a mapping tool between $t$ and $\bm{\mu}$ and  linear subspace $\bm{w}$. The training of the ANN model used throughout this study takes about one and a half to two hours. For nonlinear compression, as mentioned previously, using linear RBF, cubic RBF, and ANN, to map the time/parameters pair ($t$ and $\bm{\mu}$) to the approximation of the snapshot in the  nonlinear manifolds ($\bm{z}$) leads to approximately the same accuracy. The linear RBF is slightly cheaper than the cubic RBF and ANN. Again, we note that this may not be a fair comparison because even though we use the same machine for these calculations, we build our RBF models on SciPy but the ANN model on PyTorch. These three processes, on average, take about one to two and a half hours depending on the value of $\mathrm{Q}$.  \par

The computational time used during the prediction or the online phase - the fifth step in Figure \ref{fig:data_driven} -, for both linear and nonlinear approaches, is a completely different story. Once the ROM is trained, it could be used in an extremely fast fashion. One inquiry - a pair of $t^{k}$ and $\bm{\mu}^{(i)}$ takes only several milliseconds. Hence, one typically enjoys an inexpensive online phase compared to FOM for each inquiry. To elaborate, one FOM simulation, i.e., $\bm{\mu}^{(i)}$ for all $t$, takes two hours, as mentioned previously, but one ROM simulation would take approximately one second given that you require an evaluation of your quantities of interest at all timestamps used in FOM, $0=: t^{0}<t^{1}<\cdots<t^{N} := \tau$. In practice, however, the ROM is not bound by CFL condition; hence, it can deliver the quantities of interest at any specific timestamps without going through any timestamps in between in order to satisfy CFL condition. Assuming we only require our quantities of interest at the final time, for Example 2, our ROM could provide a speed-up of $7 \times 10^{6}$. In the worst-case scenario where an evaluation of your quantities of interest at all timestamps used in FOM is required, our framework could still provide a speed-up of $7 \times 10^{3}$. In short, our framework could provide speed-up of $7 \times 10^{3}$ to $7 \times 10^{6}$. \par


\subsection{Linear vs. nonlinear approach?}

\begin{figure}[!ht]
   \centering
         \includegraphics[keepaspectratio, height=12.0cm]{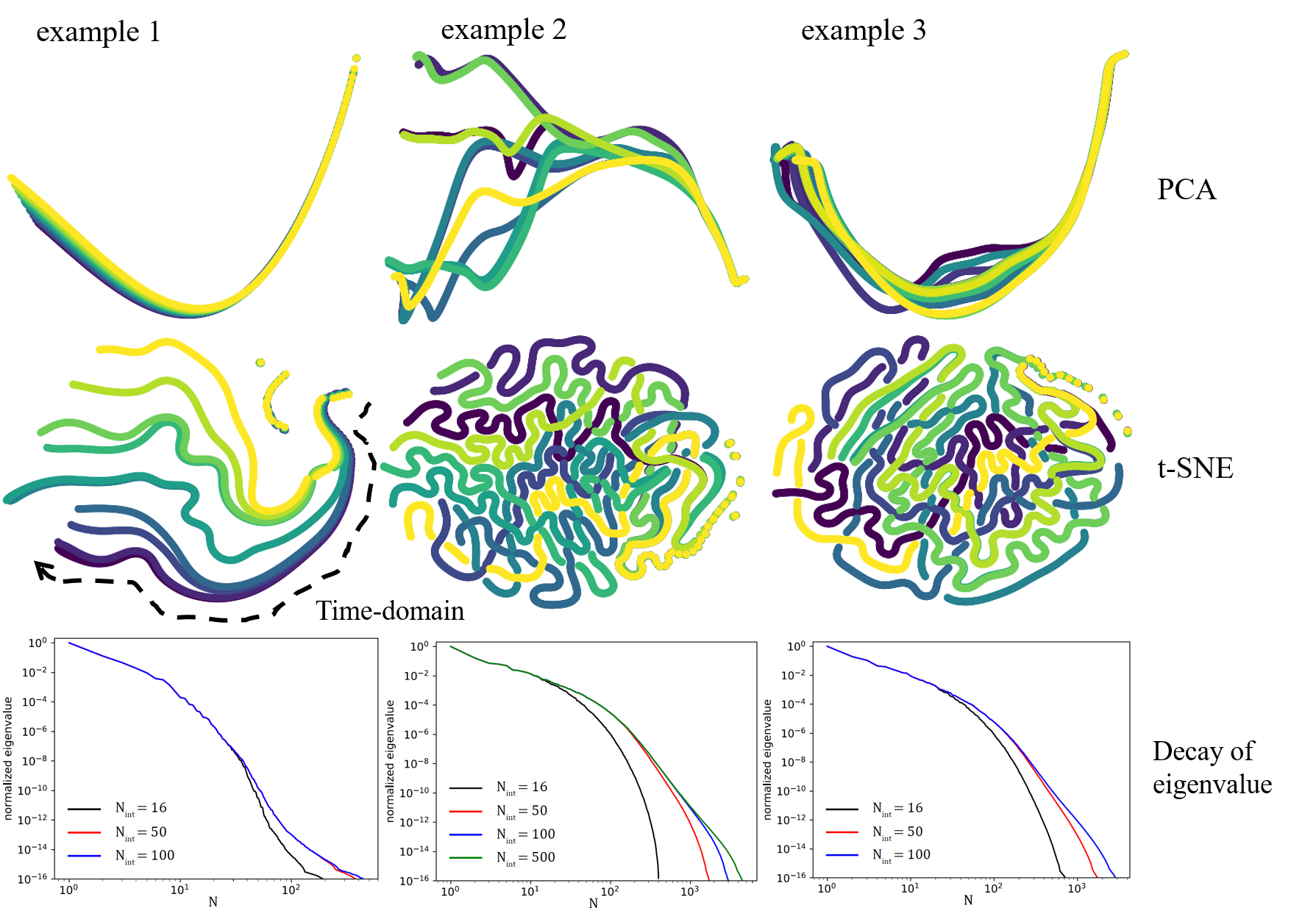}
    \caption{Subspace of $T_h$ visualization using principal component analysis (PCA) and t-Distributed Stochastic Neighbor Embedding (t-SNE). Each color represents each $\bm{\mu}^{(i)}$. The bottom row shows a decay of eigenvalue as a function of a number of reduced linear basis ($\mathrm{N}$). These plots are created using data from our test set.}
   \label{fig:pca_tsne}
\end{figure}

As discussed in previous sections, the linear approach performs better in Example 1, while the proposed nonlinear approach delivers superior performance in Examples 2 and 3. A popular indicator to determine if the linear approach will work well or not is to see the singular value decay (or the eigenvalue decay). If the eigenvalue decays fast, which is the case for Example 1 (see the bottom row and first column of Figure~\ref{fig:pca_tsne}), then the linear approach will work well. Alternatively, we here propose a figurative way to estimate which method will perform better without performing an actual compression and reconstruction. We visualize the linear and nonlinear manifolds  of each example using principal component analysis (PCA) \citep{abdi2010principal} and t-Distributed Stochastic Neighbor Embedding (t-SNE) \citep{van2008visualizing} in Figure \ref{fig:pca_tsne}. We note that this visualization is done directly on the data produced from FOM, $T_h$ in this case; hence, this can be done before the construction of ROM. To be precise, this visualization could be done between steps 2 and 3 presented in Figure \ref{fig:data_driven}. These plots are done using SciPy (\url{https://www.scipy.org/}) and  created using data from our test set, and each color represents a specific $\bm{\mu}^{(i)}$      \par

The PCA could be considered a linear subspace visualization, while the t-SNE is a nonlinear visualization manifold. From Figure \ref{fig:pca_tsne}, we observe that the PCA and t-SNE plots are not much different, for Example 1. However, the PCA and t-SNE plots are substantially different for Examples 2 and 3. This behavior may imply that Example 1 is less complex, and its reduced representation could be fitted into linear subspace. Examples 2 and 3, on the other hand, exhibit more complex behavior, which leads to substantially different behavior between linear and nonlinear approaches. Hence, a qualitative comparison between PCA and t-SNE plots may indicate the performance of linear and nonlinear non-intrusive ROM approaches. We note that this observation requires further investigation before more definitive conclusions can be reached. \par

\section{Conclusion} \label{sec:conclusion}

We present a non-intrusive reduced order model of natural convection in porous media. Our reduced order model is developed based on deep convolutional autoencoder as data compression and reconstruction tools and a regressor, radial basis function regression, or artificial neural networks for mapping our parameter space to nonlinear manifolds. The framework could handle data from a finite element model with adaptive time-stepping, commonly in the natural convection problem since Courant–Friedrichs–Lewy (CFL) condition controls the time-step size. During an online phase, our framework is not bound by CFL condition; hence, it could deliver quantities of interest at any given time without going through timestamps in between. Therefore, our reduced order model is much faster than the finite element model (speed-up of up to $7 \times 10^{6}$), while its accuracy, in the worst-case scenario, still lies within a mean squared error of 0.07. We also provide comprehensive comparisons between the linear and nonlinear compression approaches. We illustrate that in specific settings, the nonlinear compression outperforms its linear counterpart and vice versa. We suggest that it might be possible to use a quantitative comparison of subspace visualization using principal component analysis (PCA) and t-Distributed Stochastic Neighbor Embedding (t-SNE) to indicate which method will perform better.

\section{Acknowledgments}
FB thanks Horizon 2020 Program for Grant H2020 ERC CoG 2015 AROMA-CFD project 681447 (PI Prof. Gianluigi Rozza) that supported the development of RBniCS and multiphenics, and the project ``Numerical modeling of flows in porous media'' organized at the Catholic University of the Sacred Heart.
YC acknowledges LDRD funds (21-FS-042) from Lawrence Livermore National Laboratory (LLNL-JRNL-824706). Lawrence Livermore National Laboratory is operated by Lawrence Livermore National Security, LLC, for the U.S. Department of Energy, National Nuclear Security Administration under Contract DE-AC52-07NA27344.
DO acknowledges support from Los Alamos National Laboratory's Laboratory Directed Research and Development Early Career Award (20200575ECR). 
HY was supported by the Laboratory Directed Research and Development program at Sandia National Laboratories. Sandia National Laboratories is a multimission laboratory managed and operated by National Technology and Engineering Solutions of Sandia LLC, a wholly owned subsidiary of Honeywell International Inc. for the U.S. Department of Energy’s National Nuclear Security Administration under contract DE-NA0003525. 
NB acknowledges startup  support from the Sibley School of Mechanical and Aerospace Engineering, Cornell University.

\newpage

\begin{appendices}

\beginapp

\section{Finite element discretization and solution}\label{sec:fem}

In this section, we describe the numerical methods for solving the governing system described in Section \ref{sec:governing_equations}. Here, we utilize a combination of a mixed finite element method for spatial discretization, and employ a backward differentiation formula for temporal discretization.

\subsection{Domain discretization and geometrical operators}

We follow the framework proposed in \cite{kadeethum2021locally}. We here briefly introduce the notations used throughout this section.
Let $\mathcal{T}_h$ be a shape-regular triangulation obtained by a partition of $\Omega$ into $d$-simplices (triangles in $d=2$, for example). For each cell $\mathcal{T} \in \mathcal{T}_h$, we denote by $h_{\mathcal{T}}$ the diameter of $\mathcal{T}$, and we set $h_{\mathrm{max}}=\max_{\mathcal{T} \in \mathcal{T}_h} h_{\mathcal{T}}$ and $h_{\mathrm{min}}=\min_{\mathcal{T} \in \mathcal{T}_h} h_{\mathcal{T}}$.
We further denote by $\mathcal{E}_h$ the set of all faces (i.e., $d - 1$ dimensional entities connected to at least a $T \in \mathcal{T}_h$) and by $\mathcal{E}_h^{I}$ and $\mathcal{E}_h^{\partial}$ the collection of all interior and boundary facets, respectively.
The boundary set $\mathcal{E}_h^{\partial}$ is decomposed as $\mathcal{E}_{h}^{D,m} \cup \mathcal{E}_{h}^{N,m}$, where $\mathcal{E}_{h}^{D,m}$ and $\mathcal{E}_{h}^{N,m}$ are two disjoint subsets associated with the Dirichlet boundary faces on $\partial \Omega_p$ and the Neumann boundary faces on $\partial \Omega_{q}$ as defined in Equation  \eqref{eq:mass_dimless}. For Equation  \eqref{eq:temp_dimless}, $\mathcal{E}_h^{\partial}$ is decomposed as $\mathcal{E}_{h}^{D,T} \cup\mathcal{E}_{h}^{\mathrm{In}}\cup\mathcal{E}_{h}^{\mathrm{Out}}$ defined based on $\partial \Omega_{{T}}$ $\partial \Omega_{\mathrm{In}}$ $\partial \Omega_{\mathrm{Out}}$, respectively. \par

Next, 
$$
e = \partial \mathcal{T}^{+}\cap \partial \mathcal{T}^{-}, \ \ e \in \mathcal{E}_h^I,
$$
\noindent
where  $\mathcal{T}^{+}$ and $\mathcal{T}^{-}$ are the two neighboring elements to $e$. We denote by $h_e$ the characteristic length of $e$ calculated as
\begin{equation}
h_{e} :=\frac{\operatorname{meas}\left(\mathcal{T}^{+}\right)+\operatorname{meas}\left(\mathcal{T}^{-}\right)}{2 \operatorname{meas}(e)},
\end{equation}

\noindent
depending on the argument, meas($\cdot$) represents the measure of a cell ($d=3$) or of a facet ($d=2$).

Let $\mathbf{n}^{+}$ and $\mathbf{n}^{-}$ be the outward unit normal vectors to  $\partial \mathcal{T}^+$ and $\partial \mathcal{T}^-$, respectively.
For any given scalar function $\zeta: \mathcal{T}_h \to \mathbb{R}$ and vector function $\bm{\sigma}: \mathcal{T}_h \to \mathbb{R}^d$, we denote by $\zeta^{\pm}$ and $\bm{\sigma}^{\pm}$ the restrictions of $\zeta$ and $\bm{\sigma}$ to $\mathcal{T}^\pm$, respectively.
Subsequently, we define the weighted average operator as

\begin{equation}
\{\zeta\}_{\delta e}=\delta_{e} \zeta^{+}+\left(1-\delta_{e}\right) \zeta^{-}, \ \text{ on } e \in \mathcal{E}_h^I,
\end{equation}

\noindent
and

\begin{equation}
\{\bm{\sigma}\}_{\delta e}=\delta_{e} \bm{\sigma}^{+}+\left(1-\delta_{e}\right) \bm{\sigma}^{-},
\ \text{ on } e \in \mathcal{E}_h^I,
\end{equation}

\noindent
where $\delta_{e}$ is weighted average, which is 0.5 throughout this manuscript. The jump across an interior edge will be defined as
\begin{align*}
\jump{\zeta} = \zeta^+\mathbf{n}^++\zeta^-\mathbf{n}^- \quad \mbox{ and } \quad \jump{\bm{\sigma}} = \bm{\sigma}^+\cdot\mathbf{n}^+ + \bm{\sigma}^-\cdot\mathbf{n}^- \quad \mbox{on } e\in \mathcal{E}_h^I.
\end{align*}

Finally, for $e \in \mathcal{E}^{\partial}_h$, we set $\av{\zeta}_{\delta_e} :=   \zeta$ and $\av{\bm{\sigma}}_{\delta_e} :=  \bm{\sigma}$ for what concerns the definition of the weighted average operator, and $\jump{\zeta} :=  \zeta \mathbf{n}$ and $\jump{\bm{\sigma}} :=  \bm{\sigma} \cdot \mathbf{n}$ as definition of the jump operator.

\subsection{Temporal discretization} \label{sec_time_discre}

The time domain $\mathbb{T} = \left(0, \tau\right]$ is partitioned into $N^t$ subintervals such that $0=: t^{0}<t^{1}<\cdots<t^{N} := \tau$. The length of each subinterval $\Delta t^{n-1}$ is defined as $\Delta t^{n-1}=t^{n}-t^{n-1}$ where $n$ represents the current time-step and $n-1$ is previous time-step. We assume that the user provides the initial $\Delta t^0$, while an adaptive procedure is carried out to choose $\Delta t^{n-1}$, $n > 1$, as follows

\begin{equation} \label{eq:time_mult}
\Delta t^{n-1} :=
\begin{cases}
\mathrm{CFL} \frac{h_l}{\left\|\bm{u}^{n-1}\right\|_{\infty}} & \text{if} \ \Delta t^{n-1} \le \Delta t_{\max} \ \\
 \Delta t_{\max}  & \text{if} \ \Delta t^{n-1} > \Delta t_{\max},
\end{cases}
\end{equation}

\noindent
where $\mathrm{CFL}$ is a constant that the user can provide according to the Courant-Friedrichs-Lewy condition \citep{courant1967partial}, $\left\|\cdot\right\|_{\infty}$ is the maximum norm of a vector function, and $\Delta t_{\max}$ is a maximum allowed time-step.
Note that we use $\Delta t_{\max}$ to control $\Delta t^n$ when the model approaches a steady-state condition since $\left\|\bm{u}^{n-1}\right\|_{\infty}$ may approach zero, which would lead to a very large ratio $\frac{h_l}{\left\|\bm{u}^{n-1}\right\|_{\infty}}$.

Let $\varphi(\cdot, t)$ be a scalar function and $\varphi^{n}$ be its approximation at time $t^n$, i.e. $\varphi^{n} \approx \varphi\left(t^{n}\right)$. We employ the following backward differentiation formula \citep{ibrahim2007implicit,akinfenwa2013continuous,lee2018phase}

\begin{equation} \label{eq:bdf_gen}
\mathrm{BDF}_{m}\left(\varphi^{n}\right):=\left\{\begin{array}{ll}
\frac{1}{\Delta t^n}\left(\varphi^{n}-\varphi^{n-1}\right) & m=1 \\ 
\frac{1}{2\Delta t^n}\left(3 \varphi^{n}-4 \varphi^{n-1}+\varphi^{n-2}\right) & m=2 \\
\frac{1}{6\Delta t^n}\left(11 \varphi^{n} -18 \varphi^{n-1}+9 \varphi^{n-2}-2\varphi^{n-3}\right) & m=3 \\
\frac{1}{12\Delta t^n}\left(25 \varphi^{n} -48 \varphi^{n-1}+36 \varphi^{n-2}-16\varphi^{n-3}+3\varphi^{n-4}\right) & m=4
\end{array}\right.
\end{equation}
for the discretization of the time derivative of $\varphi(\cdot, t)$ at time $t^n$. 

\subsection{Spatial discretization}\label{sec:space_discretize}

Folowing \cite{kadeethum2021locally}, the fluid velocity and pressure fields in Equation \eqref{eq:mass_dimless} are discretized by
the Brezzi-Douglas-Marini (BDM) element \citep{brezzi2012mixed}
and the piecewise constants element, respectively, to ensure local mass conservation.
The temperature field in Equation \eqref{eq:temp_dimless} is discretized by the enriched Galerkin (EG) method \citep{lee2016locally,sun2009locally}.

We begin with defining the BDM function space as follows \citep{brezzi2012mixed}

\begin{equation}
\mathcal{U}_{h}^{\mathrm{BDM}_{k}}\left(\mathcal{T}_{h}\right) :=\left\{\bm{\psi_u} \in H(\operatorname{div}, \Omega):\left.\bm{\psi_u}\right|_{\mathcal{T}} \in \mathrm{BDM}(\mathcal{T}), \forall \mathcal{T} \in \mathcal{T}_{h}\right\}. 
\label{eq:BDM_V}
\end{equation}
\noindent
Here, $\bm{\psi_u}$ denotes a generic function of $\mathcal{U}_{h}^{\mathrm{BDM}_{k}}\left(\mathcal{T}_{h}\right)$ and $\mathrm{BDM}(\mathcal{T})$ is defined according to \citep{brezzi2012mixed}.

Next, the discontinuous Galerkin (DG) function space is 

\begin{equation}\label{eq:DG_P}
\mathcal{P}_{h}^{\mathrm{DG}_{k}}\left(\mathcal{T}_{h}\right) :=\left\{\psi_p \in L^{2}(\Omega) :\left.\psi_p\right|_{\mathcal{T}} \in \mathbb{P}_{k}(T), \forall \mathcal{T} \in \mathcal{T}_{h}\right\},
\end{equation}

\noindent
where $L^{2}(\Omega)$ is the space of square-integrable scalar functions, and $\psi_p$ a generic function of $\mathcal{P}_{h}^{\mathrm{DG}_{k}}\left(\mathcal{T}_{h}\right)$. We note that $\mathcal{P}_{h}^{\mathrm{DG}_{0}}\left(\mathcal{T}_{h}\right)$ represents the piecewise constants function. We then define the finite element space for the continuous Galerkin (CG) function space for a scalar-valued function

\begin{equation}
\mathcal{P}_{h}^{\mathrm{CG}_{k}}\left(\mathcal{T}_{h}\right) :=\left\{\psi_T \in \mathbb{C}^{0}(\Omega) :\left.\psi_T\right|_{\mathcal{T}} \in \mathbb{P}_{k}(\mathcal{T}), \forall \mathcal{T} \in \mathcal{T}_{h}\right\}.
\label{eq:CG_space_s}
\end{equation}

\noindent
Here, $\mathbb{C}^{0}(\Omega) := \mathbb{C}^{0}(\Omega; \mathbb{R})$ denotes the space of scalar-valued piecewise continuous polynomials, and $\mathbb{P}_{k}(\mathcal{T}) := \mathbb{P}_{k}(\mathcal{T}{; \mathbb{R}})$ is the space of polynomials of degree at most $k$ over each element $\mathcal{T}$. Finally, the EG finite element space with polynomial order $k$ is
\begin{equation}\label{eq:EG_P}
\mathcal{P}_{h}^{\mathrm{EG}_{k}}\left(\mathcal{T}_{h}\right) :=\mathcal{P}_{h}^{\mathrm{CG}_{k}}\left(\mathcal{T}_{h}\right) \oplus \mathcal{P}_{h}^{\mathrm{DG}_{0}}\left(\mathcal{T}_{h}\right),
\end{equation}

\noindent
where ${\psi_T}$ denotes a generic function of $\mathcal{P}_{h}^{\mathrm{EG}_{k}}\left(\mathcal{T}_{h}\right)$ (as well as $\mathcal{P}_{h}^{\mathrm{CG}_{k}}\left(\mathcal{T}_{h}\right)$).

\subsection{Full discretization}

We now present the fully discrete form of Equations \eqref{eq:mass_dimless} and \eqref{eq:temp_dimless} using the above-described combination of finite element spaces. In particular, we seek the approximated velocity $\bm{u}_h \in \mathcal{U}_{h}^{\mathrm{BDM}_{1}}\left(\mathcal{T}_{h}\right)$,
approximated fluid pressure $p_h \in \mathcal{P}_{h}^{\mathrm{DG}_{0}}\left(\mathcal{T}_{h}\right)$,
and approximated $T_h \in \mathcal{P}_{h}^{\mathrm{EG}_{1}}\left(\mathcal{T}_{h}\right)$. The weak form of the first part of mass balance Equation  \eqref{eq:mass_dimless} is obtained multiplying by $\bm{\psi_u} \in \mathcal{U}_{h}^{\mathrm{BDM}_{1}}\left(\mathcal{T}_{h}\right)$ and integrating by parts, resulting in

\begin{equation}
\mathcal{N}_{\bm{u}}\left(\bm{\psi_u};  \bm{u}_{h}^{n}, p_{h}^{n},T_{h}^{n}  \right) = 0
, \quad\forall \bm{\psi_u} \in \mathcal{U}_{h}^{\mathrm{BDM}_{1}}\left(\mathcal{T}_{h}\right),
\label{eq:N_p_3f}
\end{equation}

\noindent
for each time-step $t^n$, where

\begin{equation*}
\begin{split}
\mathcal{N}_{\bm{u}}\left(\bm{\psi_u};  \bm{u}_{h}^{n}, p_{h}^{n},T_{h}^{n}  \right) & =
\sum_{\mathcal{T} \in \mathcal{T}_{h}} \int_{\mathcal{T}}  \bm{u}_{h}^{n} \cdot  \bm{\psi_u} \: d V 
+ \sum_{\mathcal{T} \in \mathcal{T}_{h}} \int_{\mathcal{T}}   p_{h}^{n} \nabla \cdot  \bm{\psi_u} \: d V  - \sum_{\mathcal{T} \in \mathcal{T}_{h}} \int_{\mathcal{T}} \mathbf{z} \mathrm{Ra} T_{h}^{n} \cdot  \bm{\psi_u} \: d V \\
& + \sum_{e \in \mathcal{E}_{h}^{D,m}} \int_{e} p_D \bm{\psi_u} \cdot \mathbf{n} \: d S
, \quad \forall \bm{\psi_u} \in \mathcal{U}_{h}^{\mathrm{BDM}_{1}}\left(\mathcal{T}_{h}\right).
\end{split}
\end{equation*}

\noindent
Here, $\int_{\mathcal{T}} \cdot\  d V$ and $\int_{e} \cdot\  d S$ refer to volume and surface integrals, respectively. Furthermore, the notation for $\mathcal{N}_{\bm{u}}\left(\bm{\psi_u};  \bm{u}_{h}^{n}, p_{h}^{n},T_{h}^{n}  \right)$ in Equation \eqref{eq:N_p_3f} highlights before the semicolon the test function, and after the semicolon the nonlinear dependence on discrete solutions to the coupled problem. The same notation will be used hereafter for the remaining equations.

Then we multiply the $\nabla \cdot \bm{u}=0$ of Equation \eqref{eq:mass_dimless} by $\psi_p \in \mathcal{P}_{h}^{\mathrm{DG}_{0}}\left(\mathcal{T}_{h}\right)$ and integrating by parts as follows

\begin{equation}
\mathcal{N}_p\left(\psi_p;  \bm{u}_{h}^{n}, p_{h}^{n},T_{h}^{n}  \right) = 0, \quad\forall \psi_p \in \mathcal{P}_{h}^{\mathrm{DG}_{0}}\left(\mathcal{T}_{h}\right),
\label{eq:N_u_3f}
\end{equation}

\noindent
for each time-step $t^n$, where

\begin{equation*}
\begin{split}
\mathcal{N}_p\left(\psi_p;  \bm{u}_{h}^{n}, p_{h}^{n},T_{h}^{n}  \right) & =
\sum_{\mathcal{T} \in \mathcal{T}_{h}} \int_{\mathcal{T}} \nabla \cdot \bm{u}_{h}^{n}   \psi_p \: d V, \quad\forall \psi_p \in \mathcal{P}_{h}^{\mathrm{DG}_{0}}\left(\mathcal{T}_{h}\right).
\end{split}
\end{equation*}

Next, we discretize Equation \eqref{eq:temp_dimless}

\begin{equation}
\mathcal{N}_T\left(\psi_T ; \bm{u}_{h}^{n}, p_{h}^{n},T_{h}^{n} \right) = 0, \quad\forall \psi_T \in \mathcal{P}_{h}^{\mathrm{EG}_{1}}\left(\mathcal{T}_{h}\right)
\label{eq:N_c}
\end{equation}

\noindent
for each time-step $t^n$, where

\begin{equation*}
\begin{split}
\mathcal{N}_T\left(\psi_T ; \bm{u}_{h}^{n}, p_{h}^{n},T_{h}^{n} \right) & = 
\sum_{\mathcal{T} \in \mathcal{T}_{h}} \int_{\mathcal{T}} \mathrm{BDF}_{4}\left(T_{h}^{n} \right) \psi_T \: d V
+ \sum_{\mathcal{T} \in \mathcal{T}_{h}} \int_{\mathcal{T}} \nabla  T_{h}^{n} \cdot \nabla \psi_{T} \: d V \\ & - \sum_{e \in \mathcal{E}_h^{I} \cup \mathcal{E}_{h}^{D,T}}  \int_{e}\left\{\nabla T_{h}^{n} \right\}_{\delta_{e}} \cdot \llbracket \psi_T \rrbracket \: d S + \theta\sum_{e \in \mathcal{E}_h^{I} \cup \mathcal{E}_{h}^{D,T}}  \int_{e}\left\{  \nabla \psi_{T}\right\}_{\delta_{e}} \cdot \llbracket T_{h}^{n} \rrbracket \: d S \\ & +  \sum_{e \in \mathcal{E}_h^{I} \cup \mathcal{E}_{h}^{D,T}}  \int_{e} \frac{\beta}{h_{e}}   \llbracket T_{h}^{n} \rrbracket \cdot \llbracket \psi_T \rrbracket \: d S 
-\sum_{\mathcal{T} \in \mathcal{T}_{h}} \int_{\mathcal{T}} \bm{u}_h^n T_{h}^{n} \cdot \nabla \psi_T \: d V \\
& +\sum_{e \in \mathcal{E}_h^{I} }  \int_{e} \bm{u}_h^n \cdot \mathbf{n} T^{\mathrm{up}}_h\llbracket \psi_T\rrbracket \: d S  
+ \sum_{e \in \mathcal{E}_{h}^{D,T}} \int_{e} \nabla \psi_{T} \cdot T_{D} \mathbf{n}  \: d S 
- \sum_{e \in \mathcal{E}_{h}^{D, T}} \int_{e} \frac{\beta}{h_{e}} \llbracket \psi_T \rrbracket \cdot T_D \mathbf{n} \: d S \\
&+\sum_{e \in \mathcal{E}_{h}^{\mathrm{Out}}} \int_{e} \bm{u}_h^n  \cdot \mathbf{n}T_{h}^{n} \psi_T \: d S 
+\sum_{e \in \mathcal{E}_{h}^{\mathrm{In}}} \int_{e} \bm{u}_h^n \cdot \mathbf{n} T_{\mathrm{in}} \psi_T \: d S-\sum_{T \in \mathcal{T}_{h}} \int_{T} f_c \psi_{T} \: d V, \quad\forall \psi_T \in \mathcal{P}_{h}^{\mathrm{EG}_{1}}\left(\mathcal{T}_{h}\right).
\end{split}
\end{equation*}

\noindent
Two parameters $\theta$ and $\beta$ define corresponding interior penalty methods.
The discretization becomes the symmetric interior penalty Galerkin method (SIPG) when $\theta = -1$,
the incomplete interior penalty Galerkin method (IIPG)  when $\theta = 0$,
and the non-symmetric interior penalty Galerkin method (NIPG) when $\theta = 1$ \citep{riviere2008discontinuous}. In this study, we set $\theta = 0$ and $\beta =1.1$ throughout this paper \citep{kadeethum2021locally}. Also, $T^{\mathrm{up}}_h$ is an upwind value of $T_{h}^{n}$ defined as \citep{riviere2000discontinuous,riviere2008discontinuous}:
\begin{equation}
T^{\mathrm{up}}_h=\left\{\begin{array}{ll}
T_{h}^{n+} & \text { if } \quad \bm{u}_h^n \cdot \mathbf{n} \geq 0 \\
T_{h}^{n-} & \text { if } \quad \bm{u}_h^n \cdot \mathbf{n}<0
\end{array} \quad \forall e=\partial \mathcal{T}^+ \cap \partial \mathcal{T}^-\right.
\end{equation}
\noindent
where $T_{h}^{n+}$ and $T_{h}^{n-}$ correspond to $T_{h}^{n}$ of $\mathcal{T}^+$ and $\mathcal{T}^-$, respectively.

\section{Information on nested proper orthogonal decomposition (POD)}\label{sec:sec_nested_pod}

Let $\bm{\mu}^{(i)}$ be a parameter instance in the training set, $i = 1, \hdots, \mathrm{M}$. The corresponding ${T}_{h}$ snapshot associated to $\bm{\mu}^{(i)}$ contains

\begin{equation}
{\mathbb{S}_T^{(i)}}=\left[{{T}}_{h}\left({\cdot; t^{0}, \bm{\mu}^{(i)}}\right), \cdots, {{T}}_{h}\left({\cdot; t^{{N^t}}, \bm{\mu}^{(i)}}\right)\right] \in \mathbb{R}^{{N_{h}^T} \times N^t}. 
\end{equation}

\noindent
where ${{T}}_{h}\left({\cdot; t^{n}, \bm{\mu}^{(i)}}\right)$ represent the temperature field at time $t^{n}$ and parameter instance $\bm{\mu}^{(i)}$. We recall that $N_{h}^T$ is the number of DOFs in the temperature finite element space, and $N^t$ is the total number of time-steps corresponding to specific $\bm{\mu}$.
In this study, $N_{h}^T$ is constant (i.e., the mesh and finite element function space remain the same), but $N^t$ varies because our finite element solver utilizes an adaptive time-stepping (see Equation  \eqref{eq:time_mult}). Therefore, each snapshot has a different $N^t$ even though each snapshot utilizes the same initial and final time (i.e., the same starting and endpoints but a different number of intervals). \par

The nested POD algorithm can be summarized in the two following sequential stages:
\begin{enumerate}
\item[1)] \emph{compression on the temporal evolution}: for each parameter instance $\bm{\mu}^{(i)}$ in the training set compress the temporal evolution stored in $\mathbb{S}_T^{(i)} \in \mathbb{R}^{N_{h}^T \times N^t}$ by means of a POD, retaining only the first $\mathrm{N}_{\mathrm{int}} \ll N^t$ modes. A compressed matrix $\widetilde{\mathbb{S}}_T^{(i)} \in \mathbb{R}^{N_{h}^T \times \mathrm{N}_{\mathrm{int}}}$ is then assembled by storing by column the first $\mathrm{N}_{\mathrm{int}}$ modes, scaled by the respective singular values. The value of $\mathrm{N}_{\mathrm{int}}$ can be chosen according to energy criteria (and thus it will be, in general, depending on the index $i$) or can be fixed a priori (as we do in this study), and is typically considerably smaller than the number of time-steps $N^t$.
\item[2)] \emph{compression on the parameter space}: after the temporal evolution of each parameter instance has been compressed, one can assemble the following matrix
\begin{equation}
{\widetilde{\mathbb{S}}_T}=\left[{\widetilde{\mathbb{S}}_T^{(1)}, \cdots, \widetilde{\mathbb{S}}_T^{(\mathrm{M})}}\right] \in \mathbb{R}^{{N_{h}^T} \times \mathrm{N}_{\mathrm{int}}\mathrm{M}}.
\end{equation}
by horizontally stacking all matrices $\widetilde{\mathbb{S}}_T^{(i)}$, $i = 1, \cdots, \mathrm{M}$. We then perform the singular value decomposition (SVD) of ${\widetilde{\mathbb{S}}_T}$ as

\begin{equation}
{\widetilde{\mathbb{S}}_T}=\mathbb{W}\left[\begin{array}{cc}
\mathbb{D} & 0 \\
0 & 0
\end{array}\right] \mathbb{B}^{\top}
\end{equation}

\noindent
where $\mathbb{W}=\left[\mathbf{w}_{1},\cdots, \mathbf{w}_{{N_{h}^T}}\right] \in \mathbb{R}^{{N_{h}^T} \times {N_{h}^T}}$ and $\mathbb{B}=\left[\mathbf{b}_{1},\cdots, \mathbf{b}_{\mathrm{N}_{\mathrm{int}}\mathrm{M} }\right] \in \mathbb{R}^{\mathrm{N}_{\mathrm{int}}\mathrm{M} \times\mathrm{N}_{\mathrm{int}}\mathrm{M} }$ are orthogonal matrices, $\mathbb{D}=\operatorname{diag}\left(d_{1}, \cdots, d_{r}\right) \in \mathbb{R}^{r \times r}$
is a diagonal matrix, with singular values $d_{1} \geq d_{2} \geq \cdots \geq d_{r}>0 .$ Here, $r$ is the number of non-zero singular values and $r \leq \min \left\{{N_{h}^T}, \mathrm{N}_{\mathrm{int}}\mathrm{M} \right\}$. The columns of $\mathbb{W}$ are called left singular vectors of ${\widetilde{\mathbb{S}}_T},$ and the columns of $\mathbb{B}$ are called right singular vectors of ${\widetilde{\mathbb{S}}_T}$.
To carry out a dimensionality reduction, the POD basis of rank $\mathrm{N} \ll r$ consists of the first $\mathrm{N}$ left singular vectors of ${\widetilde{\mathbb{S}}_T}$, and it has the property of minimizing the projection error defined by

\begin{equation}
\left\{\mathbf{w}_{1}, \cdots, \mathbf{w}_{\mathrm{N}}\right\} = \arg\min \left\{
\varepsilon\left(\tilde{\mathbf{w}}_{1}, \cdots, \tilde{\mathbf{w}}_{\mathrm{N}}\right)={\sum_{i=1}^{\mathrm{M}}\sum_{k=0}^{\mathrm{N}_\mathrm{int}} \left\|{T}_{h}\left(\cdot; t^{k}, \bm{\mu}^{(i)}\right)-\sum_{n=1}^{\mathrm{N}}\left({T}_{h}\left(\cdot; t^{k}, \bm{\mu}^{(i)}\right), \tilde{\mathbf{w}}_{n}\right)_{T} \tilde{\mathbf{w}}_{n}\right\|_{T}^{2}}
\right\}
\end{equation}

\noindent
among all the orthonormal bases $\left\{\tilde{\mathbf{w}}_{1}, \cdots, \tilde{\mathbf{w}}_{\mathrm{N}}\right\} {\subset \mathbb{R}^{N_{h}^T}}$. Here $(\cdot, \cdot)_T$ denotes an inner product for the temperature space, while $\left\|\cdot\right\|_T$ its induced norm.
The reduced basis space $\mathcal{T}_{\mathrm{N}}$ is then defined as the span of $\left\{\mathbf{w}_{1}, \cdots, \mathbf{w}_{\mathrm{N}}\right\}$. 
\end{enumerate}

\section{Information on the deep convolutional autoencoder} \label{sec:dc_used}

We provide a detailed architecture of the generator used in this study in Table \ref{tab:dcov}. In Table \ref{tab:dcov}, $\mathrm{B}$ is batch size, $\mathrm{C}$ is input channel, and we set $\mathrm{C}=1$ throughout this study. Furthermore, we note that we typically employ unstructured grids in the finite element solver or FOM presented in Appendix \ref{sec:fem}. In contrast, DC-AE is usually employed on a structured data set. Thus, we pre-process our snapshot data by interpolating the finite element result $T_h$ to structured grids, such as the finite element interpolation operator or cubic spline interpolation. We then replace the FOM dimension ${N}_h^T$, associated with the unstructured grid, with a pair $(\widetilde{N}_h^T, \widetilde{N}_h^T)$, associated to the structured grid. In practice, the value of $\widetilde{N}_h^T$ is often chosen independently on ${N}_h^T$. \par

The $1^{\mathrm{st}}$ convolutional layer is used to map $\mathrm{C}$ to hidden layer size ($\mathrm{H}$), and it does not subject to any activation function. We note that we set $\mathrm{H}=32$ throughout this work. The $2^{\mathrm{nd}}$ convolutional layer is used to map hidden layer size ($\mathrm{H}$) back to $\mathrm{C}$, and it subjects to the Sigmoid activation function as our data $T_h \left(\bm{\mu}^{ \left(1\right)}\right) , \cdots, T_h \left(\bm{\mu}^{\left(\mathrm{M}\right)}\right)$ is normalized to be in a range of $[0, 1]$ as follows

\begin{equation} \label{eq:norm_input_output}
\frac{{T}_{h}\left(\cdot; t^{k}, \bm{\mu}^{i}\right)-\min (T_h)}{\max (T_h)-\min (T_h)}.
\end{equation} 

\begin{table}[!ht]
\centering
\caption{Deep convolutional autoencoder used in this study (input and output sizes are represented by {[}$\mathrm{B}$, $\mathrm{C}$, $\widetilde{N}_h^T$, $\widetilde{N}_h^T${]}. We use hidden layers $\mathrm{H} = 32$.)}
\begin{tabular}{|l|c|c|c|c|}
\hline
\textbf{block}                  & \multicolumn{1}{l|}{\textbf{input size}} & \multicolumn{1}{l|}{\textbf{output size}} & \multicolumn{1}{l|}{\textbf{batch normalization}} & \multicolumn{1}{l|}{\textbf{dropout}} \\ \hline
$1^{\mathrm{st}}$ convolutional layer & {[}$\mathrm{B}$, 1, 128, 128{]}            & {[}$\mathrm{B}$, 32, 128, 128{]}            &                                          &                              \\ \hline
$1^{\mathrm{st}}$ contracting block   & {[}$\mathrm{B}$, 32, 128, 128{]}           & {[}$\mathrm{B}$, 64, 64, 64{]}              & \checkmark                                        & \checkmark                            \\ \hline
$2^{\mathrm{nd}}$ contracting block   & {[}$\mathrm{B}$, 64, 64, 64{]}             & {[}$\mathrm{B}$, 128, 32, 32{]}             & \checkmark                                        & \checkmark                            \\ \hline
$3^{\mathrm{rd}}$ contracting block   & {[}$\mathrm{B}$, 128, 32, 32{]}            & {[}$\mathrm{B}$, 256, 16, 16{]}             & \checkmark                                        & \checkmark                            \\ \hline
$4^{\mathrm{th}}$ contracting block   & {[}$\mathrm{B}$, 256, 16, 16{]}            & {[}$\mathrm{B}$, 512, 8, 8{]}               & \checkmark                                        &                              \\ \hline
$5^{\mathrm{th}}$ contracting block   & {[}$\mathrm{B}$, 512, 8, 8{]}              & {[}$\mathrm{B}$, 1024, 4, 4{]}              & \checkmark                                        &                              \\ \hline
$6^{\mathrm{th}}$ contracting block   & {[}$\mathrm{B}$, 1024, 4, 4{]}             & {[}$\mathrm{B}$, 2048, 2, 2{]}              & \checkmark                                        &                              \\ \hline
$7^{\mathrm{th}}$ contracting block   & {[}$\mathrm{B}$, 2048, 2, 2{]}             & {[}$\mathrm{B}$, 4196, 1, 1{]}              & \checkmark                                        &                              \\ \hline
$1^{\mathrm{st}}$ \textbf{bottleneck}     & {reshape([}$\mathrm{B}$, 4196{])}         & {[}$\mathrm{B}$, $\bm{z}$ {]}              &                                         &                              \\ \hline
$2^{\mathrm{nd}}$ \textbf{bottleneck}     & {[}$\mathrm{B}$, $\bm{z}$ {]}         & {reshape([}$\mathrm{B}$, 4196{])}              &                                         &                              \\ \hline
$1^{\mathrm{st}}$ expanding block     & {[}$\mathrm{B}$, 4196, 1, 1{]}       & {[}$\mathrm{B}$, 2048, 2, 2{]}            & \checkmark                                        &                              \\ \hline
$2^{\mathrm{nd}}$ expanding block     & {[}$\mathrm{B}$, 2048, 2, 2{]}         & {[}$\mathrm{B}$, 1024, 4, 4{]}              & \checkmark                                        &                              \\ \hline
$3^{\mathrm{rd}}$ expanding block    & {[}$\mathrm{B}$, 1024, 4, 4{]}         & {[}$\mathrm{B}$, 512, 8, 8{]}               & \checkmark                                        &                              \\ \hline
$4^{\mathrm{th}}$ expanding block     & {[}$\mathrm{B}$, 512, 8, 8{]}          & {[}$\mathrm{B}$, 256, 16, 16{]}             & \checkmark                                        &                              \\ \hline
$5^{\mathrm{th}}$ expanding block     & {[}$\mathrm{B}$, 256, 16, 16{]}      & {[}$\mathrm{B}$, 128, 32, 32{]}             & \checkmark                                        &                              \\ \hline
$6^{\mathrm{th}}$ expanding block     & {[}$\mathrm{B}$, 128, 32, 32{]}      & {[}$\mathrm{B}$, 64, 64, 64{]}              & \checkmark                                        &                              \\ \hline
$7^{\mathrm{th}}$ expanding block     & {[}$\mathrm{B}$, 64, 64, 64{]}       & {[}$\mathrm{B}$, 32, 128, 128{]}            & \checkmark                                        &                              \\ \hline
$2^{\mathrm{nd}}$ convolutional layer & {[}$\mathrm{B}$, 32, 128, 128{]}           & {[}$\mathrm{B}$, 1, 128, 128{]}             &                                          &                              \\ \hline
\end{tabular}
\label{tab:dcov}
\end{table}

\section{Information on training losses of approximators}\label{sec:sup_loss}

\begin{figure}[!ht]
   \centering
         \includegraphics[keepaspectratio, height=6.0cm]{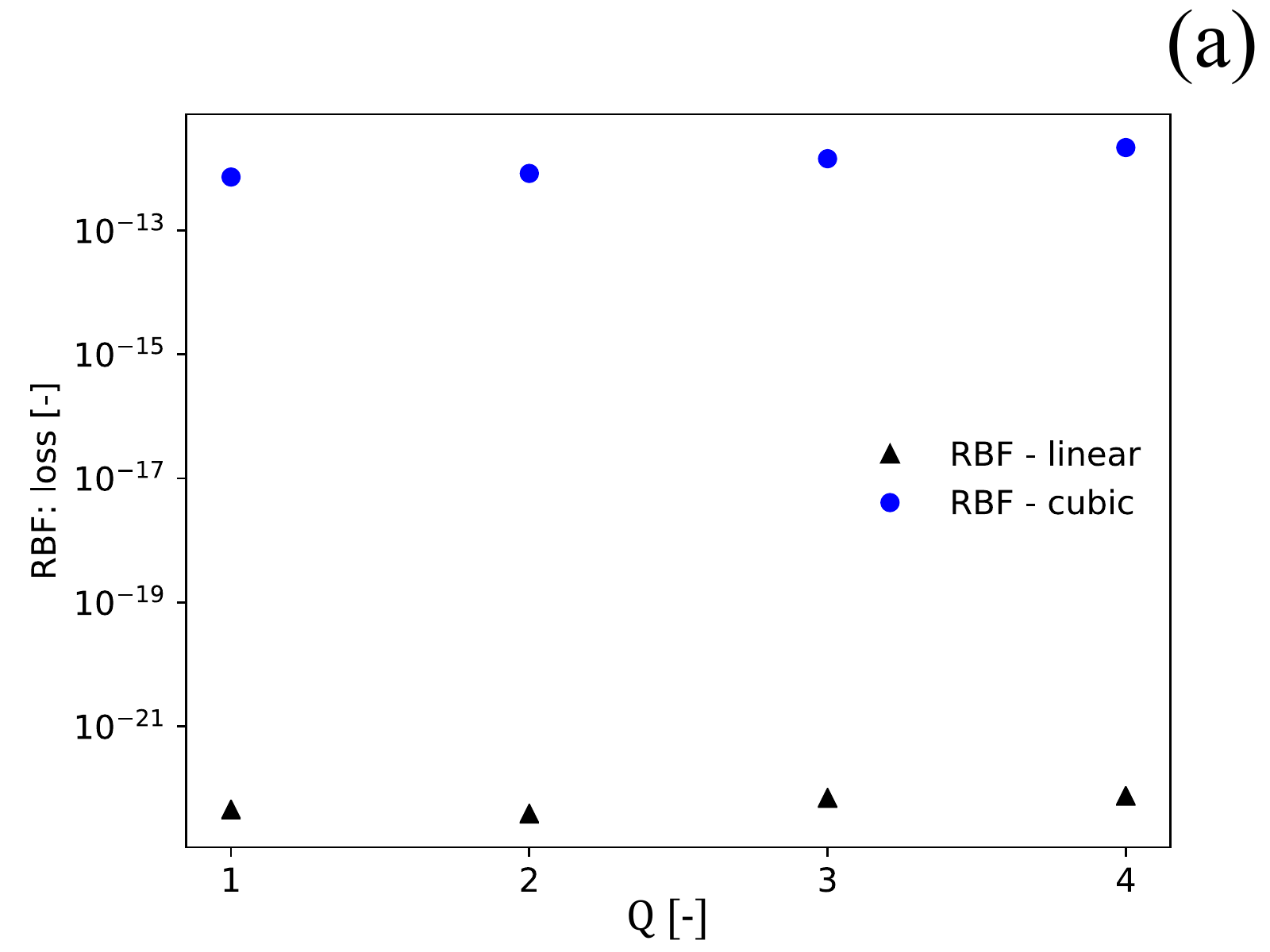}
         \includegraphics[keepaspectratio, height=6.0cm]{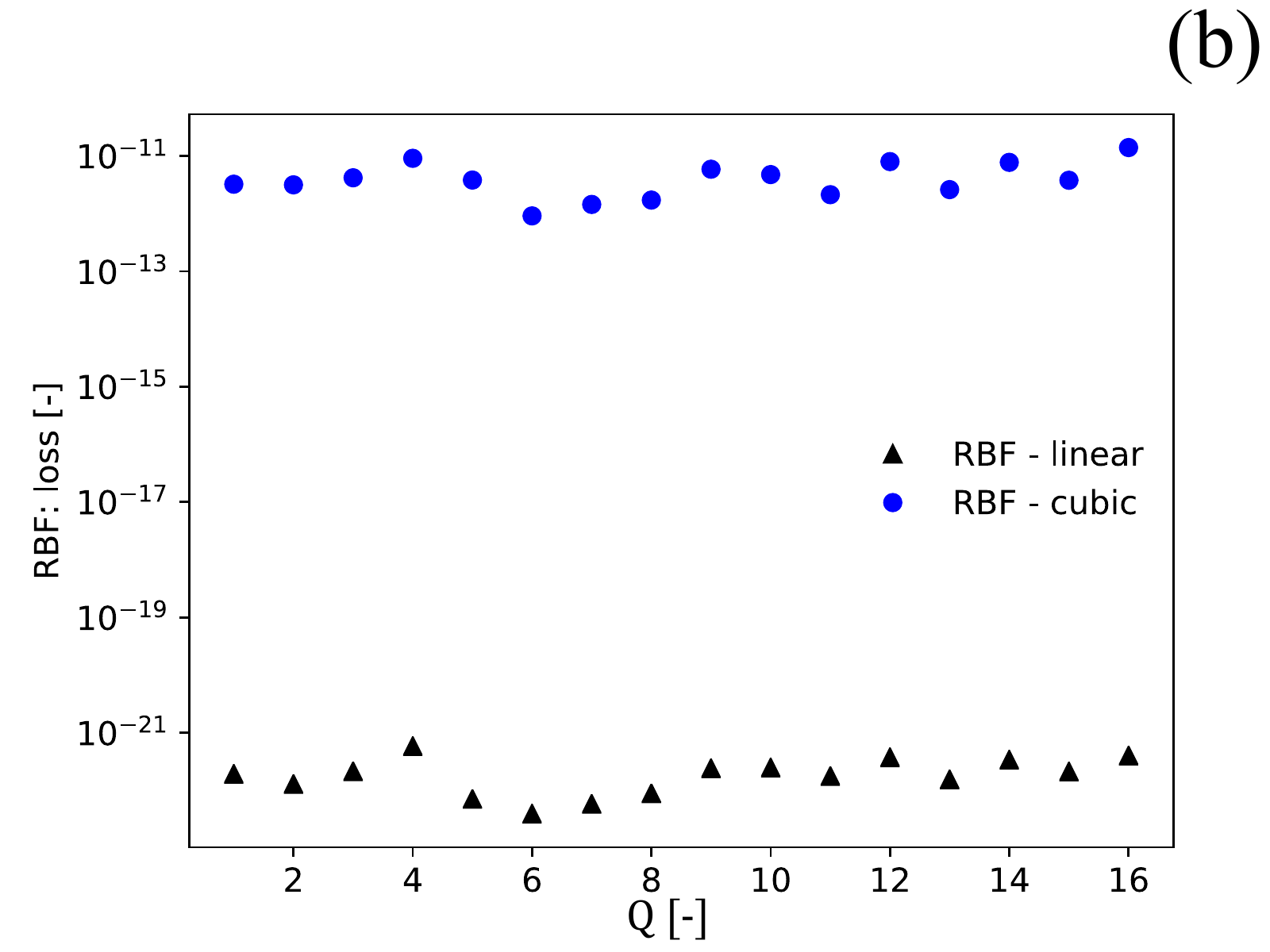}
          \includegraphics[keepaspectratio, height=6.0cm]{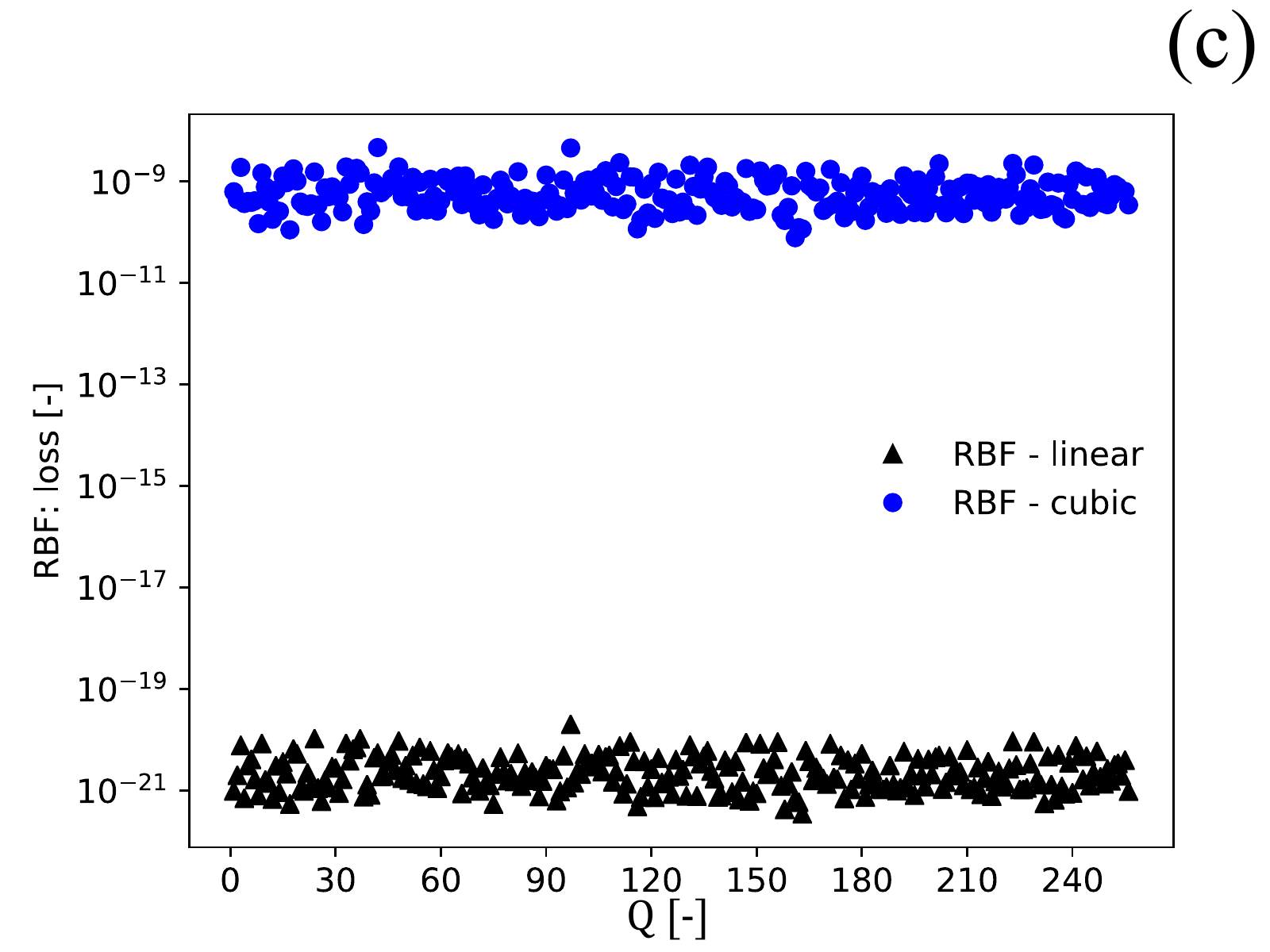}
   \caption{RBF: loss for nonlinear compression approach of Example 1: (a) $\mathrm{Q}=4$, (b) $\mathrm{Q}=16$, and (c) $\mathrm{Q}=256$.}
   \label{fig:si_training_loss_rbf_hfs}
\end{figure}

\begin{figure}[!ht]
   \centering
         \includegraphics[keepaspectratio, height=6.0cm]{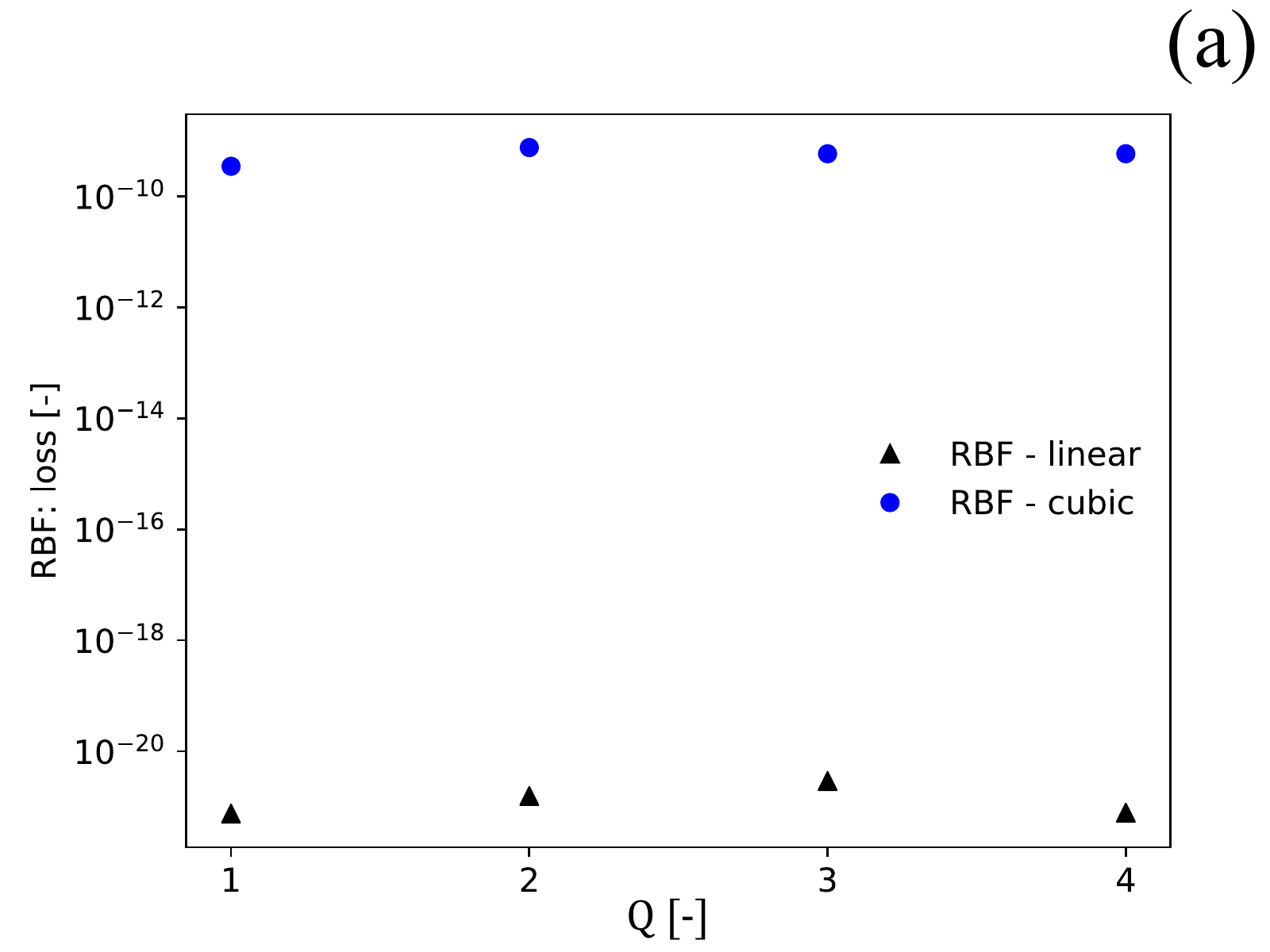}
         \includegraphics[keepaspectratio, height=6.0cm]{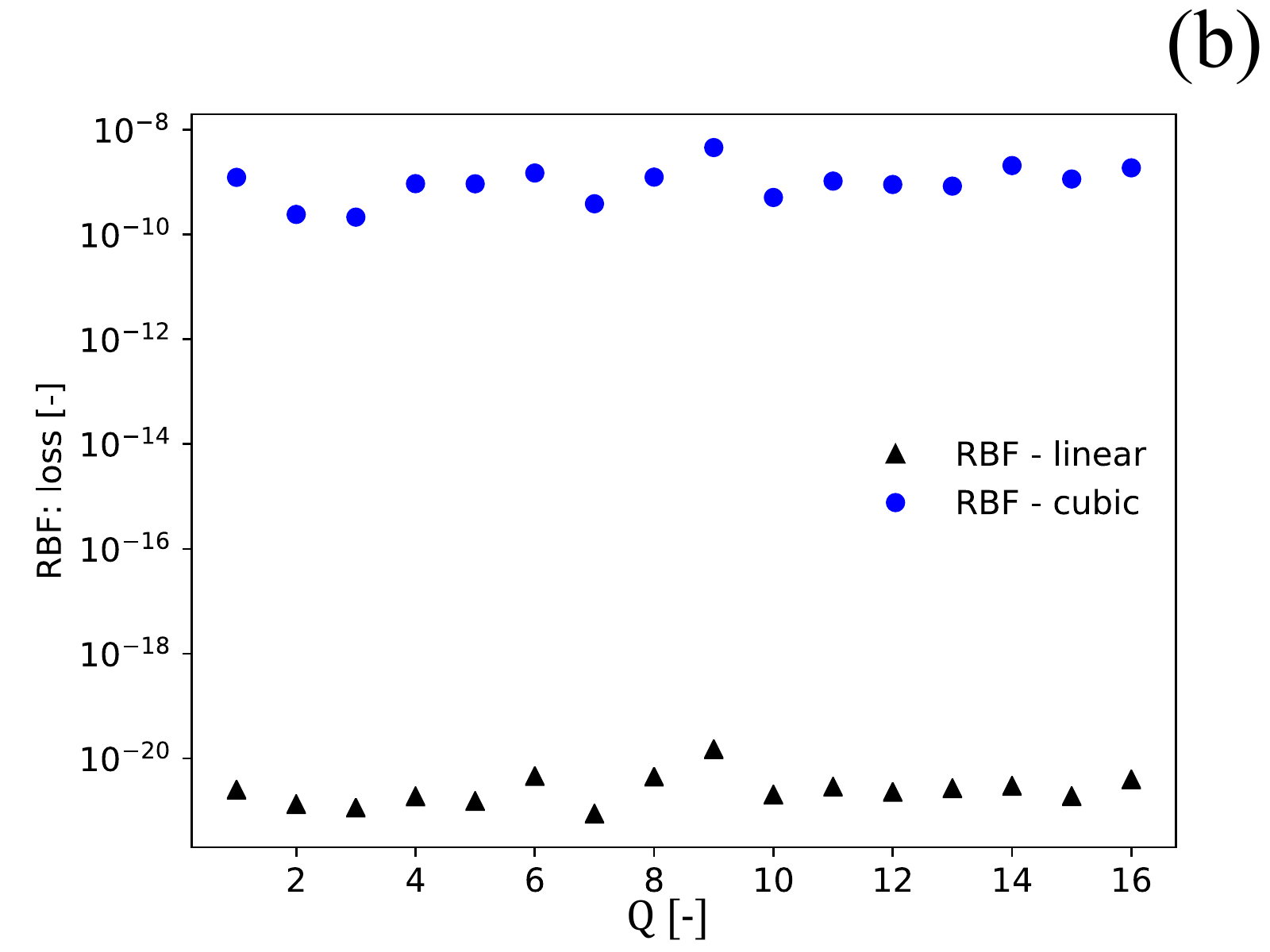}
          \includegraphics[keepaspectratio, height=6.0cm]{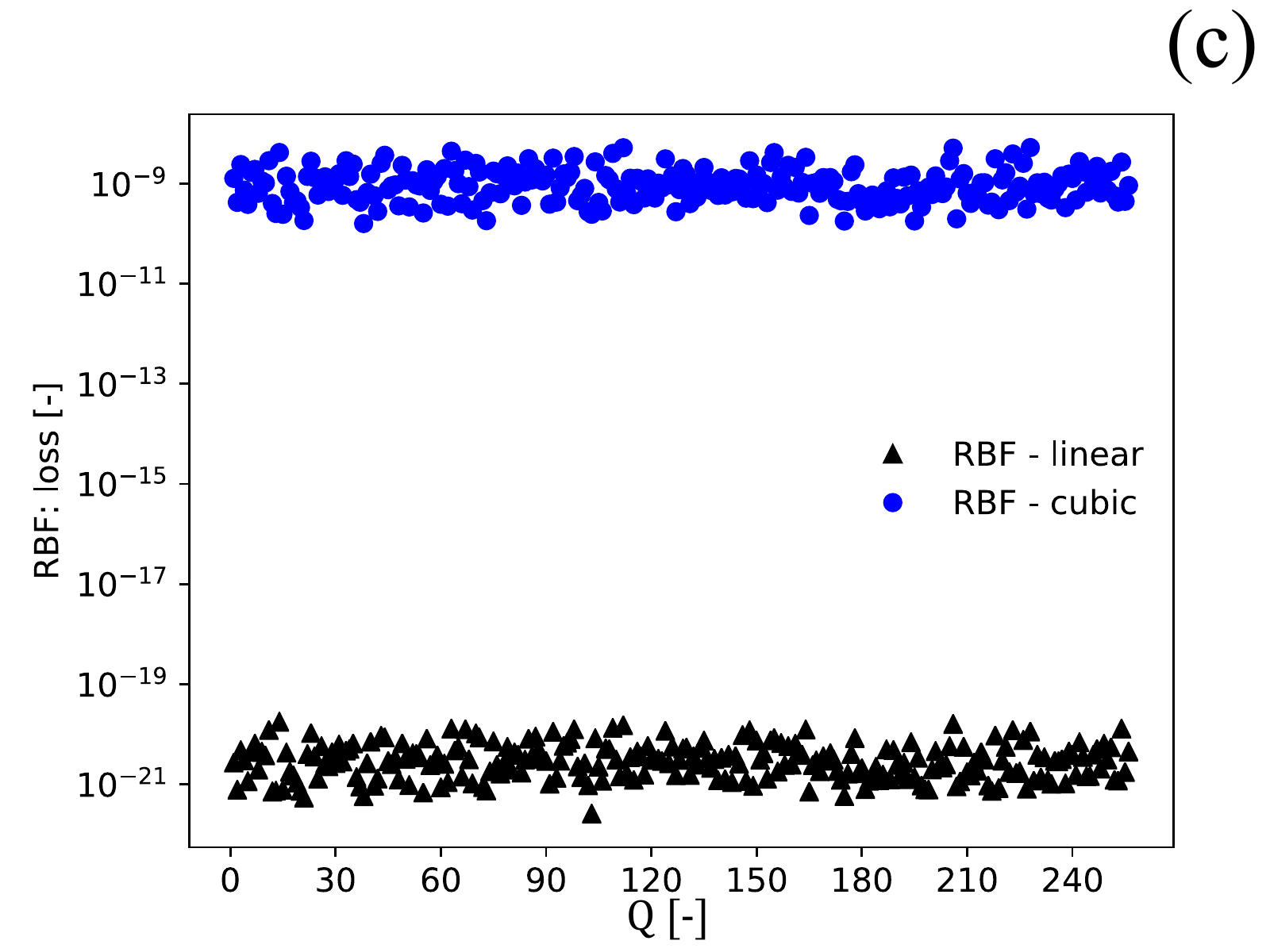}
   \caption{RBF: loss for nonlinear compression approach of Example 2: (a) $\mathrm{Q}=4$, (b) $\mathrm{Q}=16$, and (c) $\mathrm{Q}=256$.}
   \label{fig:si_training_loss_rbf_elder}
\end{figure}

\begin{figure}[!ht]
   \centering
         \includegraphics[keepaspectratio, height=6.0cm]{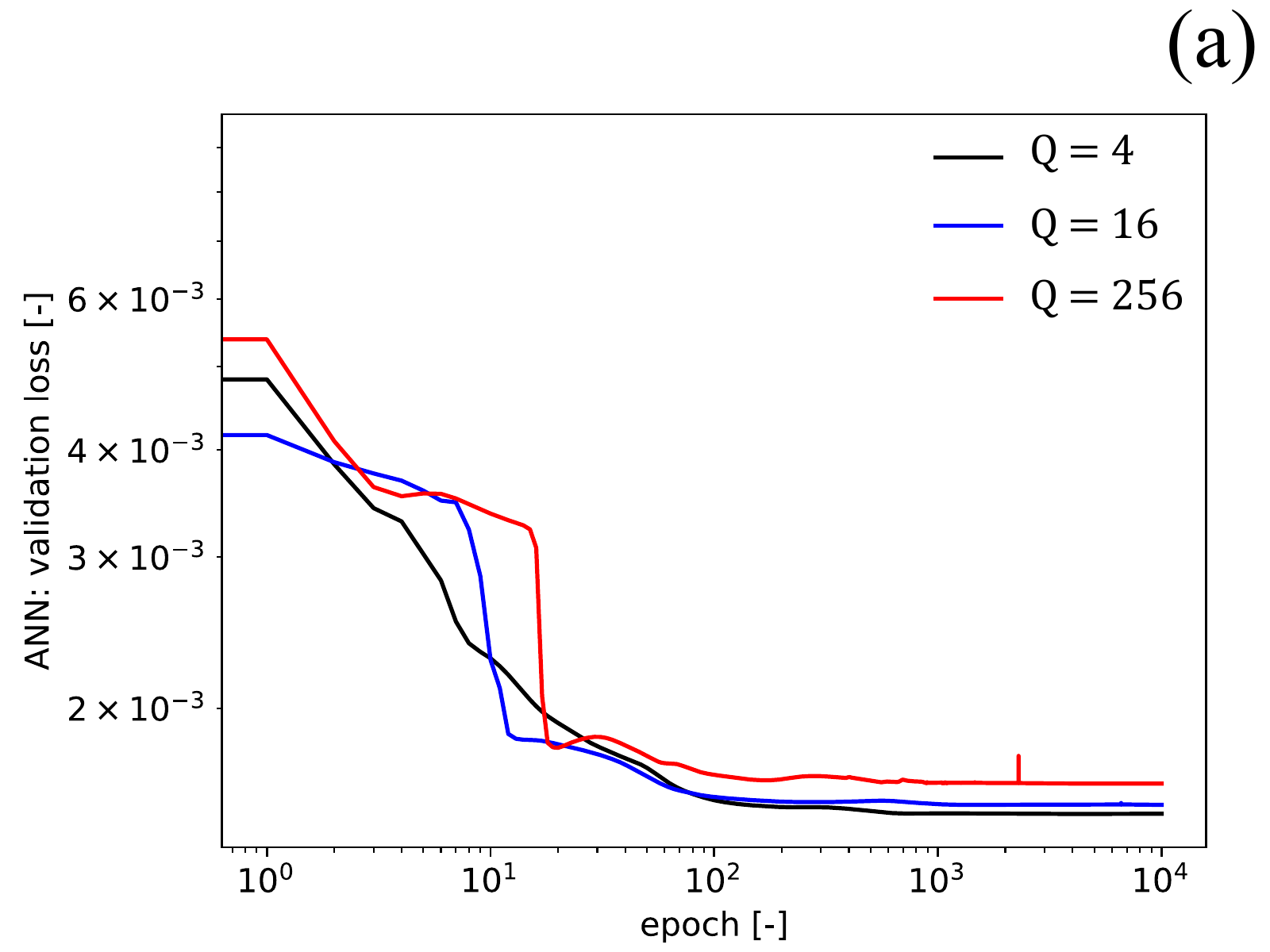}
         \includegraphics[keepaspectratio, height=6.0cm]{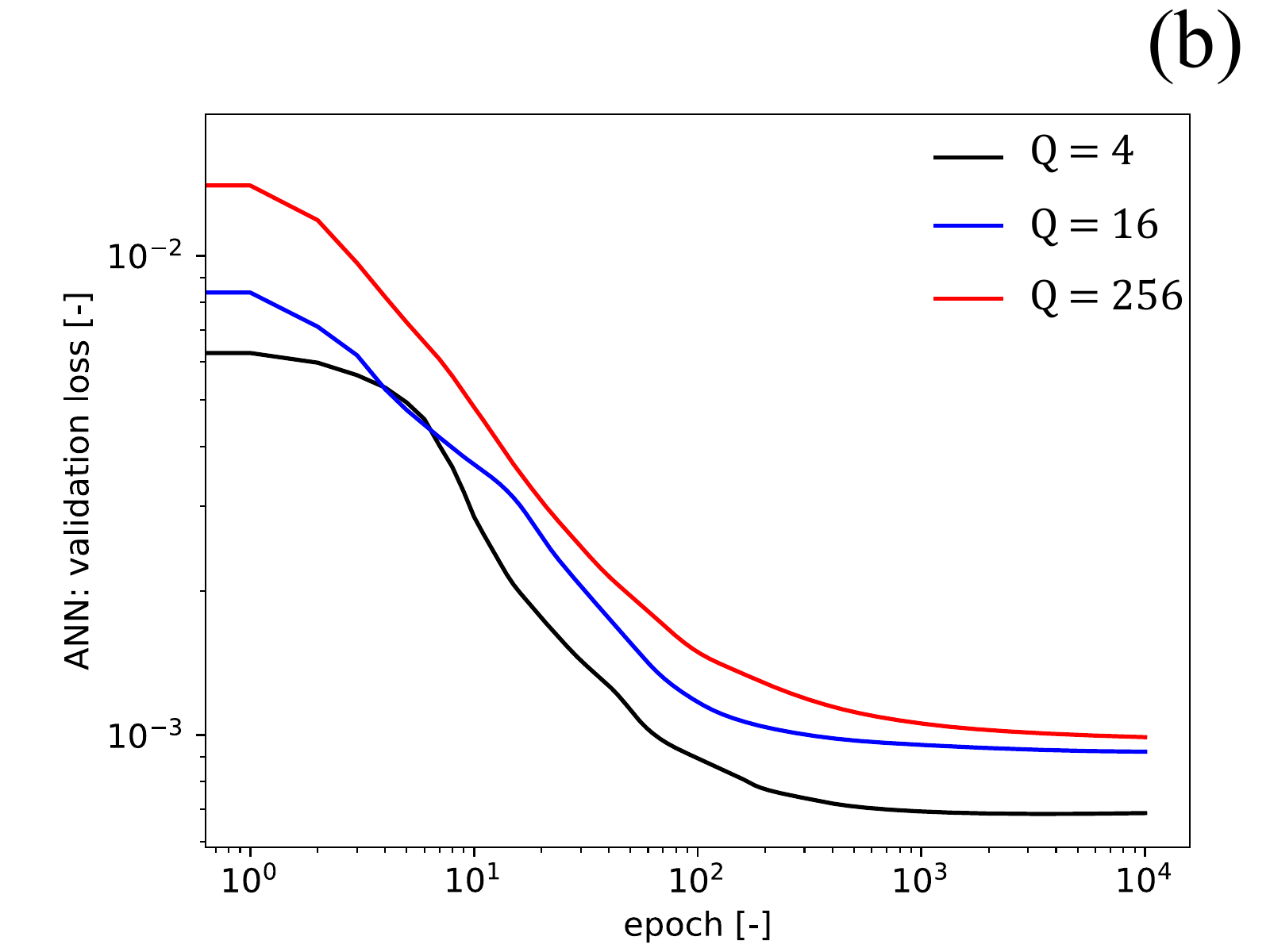}
          \includegraphics[keepaspectratio, height=6.0cm]{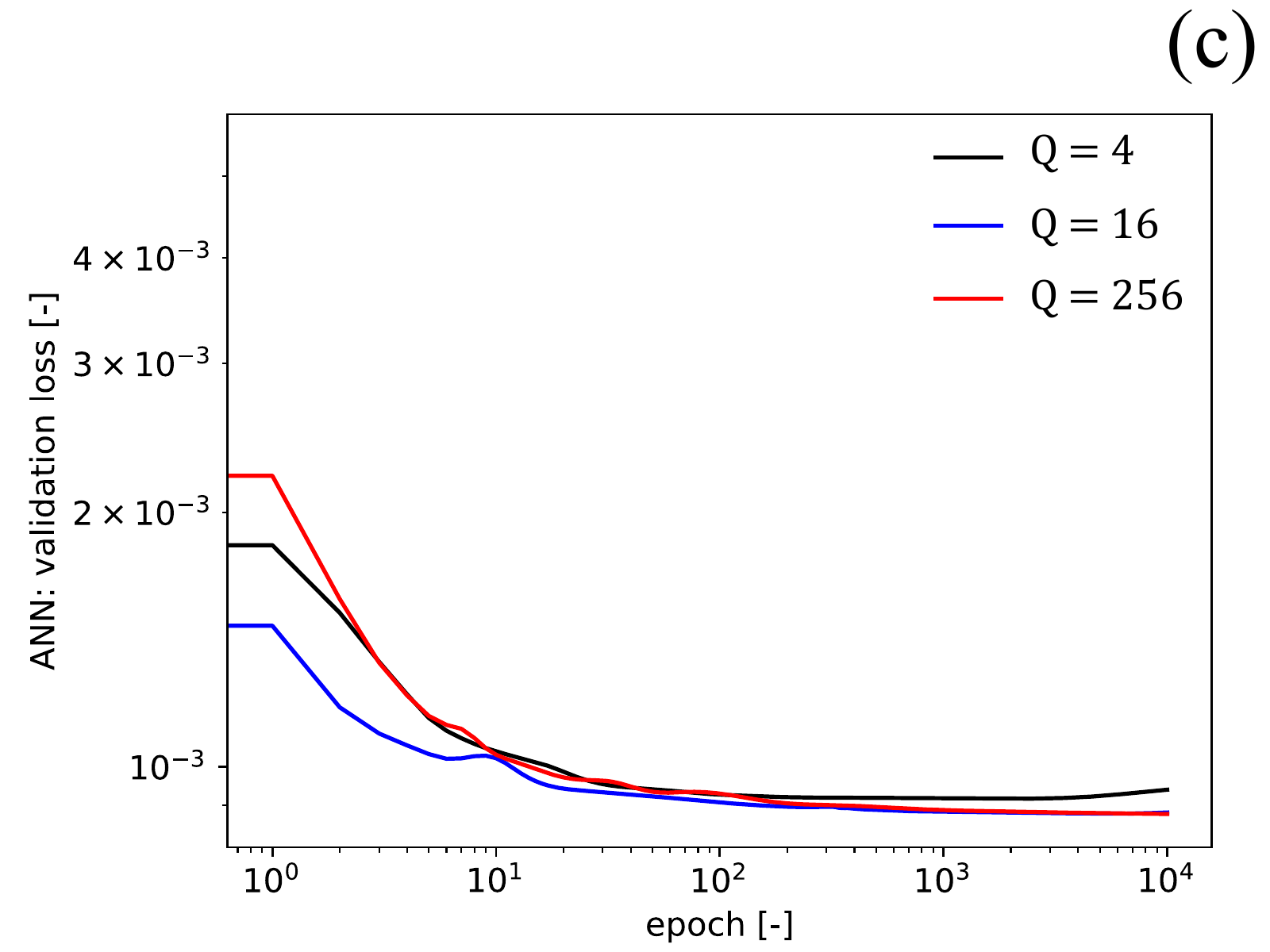}
   \caption{ ANN: validation loss for nonlinear compression approach of (a) Example 1, (b) Example 2, and (c) Example 3.}
   \label{fig:si_training_loss_ann}
\end{figure}

We present supplementary information on training losses of the approximators used for the nonlinear compression approach (see Section \ref{sec:nonlinear_com}). Figure \ref{fig:si_training_loss_rbf_hfs} contains RBF loss information for Example 1, Figure \ref{fig:si_training_loss_rbf_elder} contains RBF loss information for Example 2, and Figure \ref{fig:si_training_loss_ann} provides ANN validation loss during the training phase for Examples 1, 2, and 3. \par 

From Figure \ref{fig:si_training_loss_rbf_hfs}, we observe that the RBF with linear function has an extremely low loss (in an order of $10^{-20}$). The RBF models with cubic function have a higher loss, but still, they are pretty low (in order of $10^{-10}$). The similar results hold for Example 2 - see Figure \ref{fig:si_training_loss_rbf_elder}. The validation loss shown in Figure \ref{fig:si_training_loss_ann} is decreased as a function of epoch. We observe that the values of validation loss are not much different between each $\mathrm{Q}$ for each example. \par

\section{Example 2: Elder problem using autoencoder without convolutional layers}\label{sec:non_conv}

\begin{figure}[!ht]
   \centering
         \includegraphics[keepaspectratio, height=8.0cm]{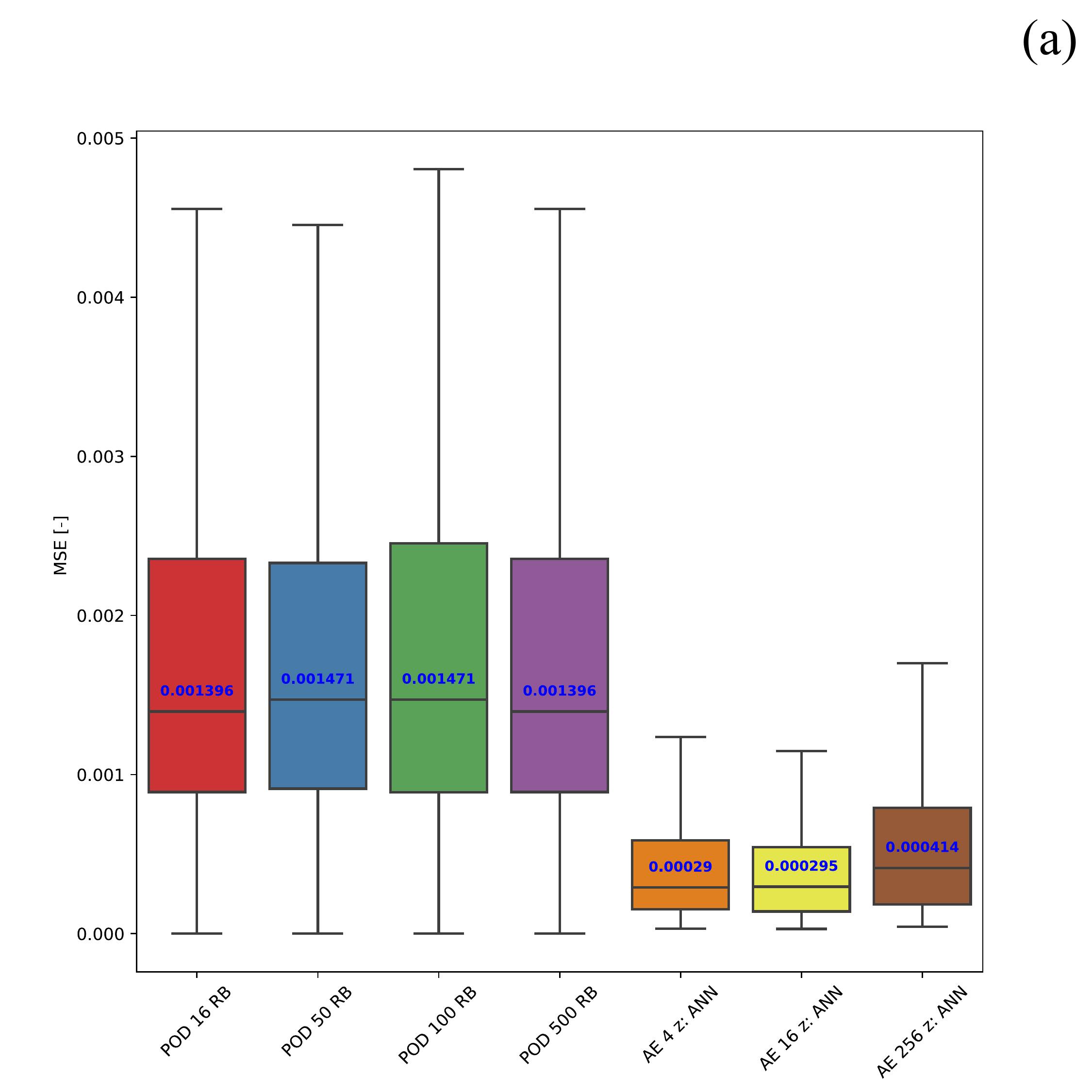}
         \includegraphics[keepaspectratio, height=8.0cm]{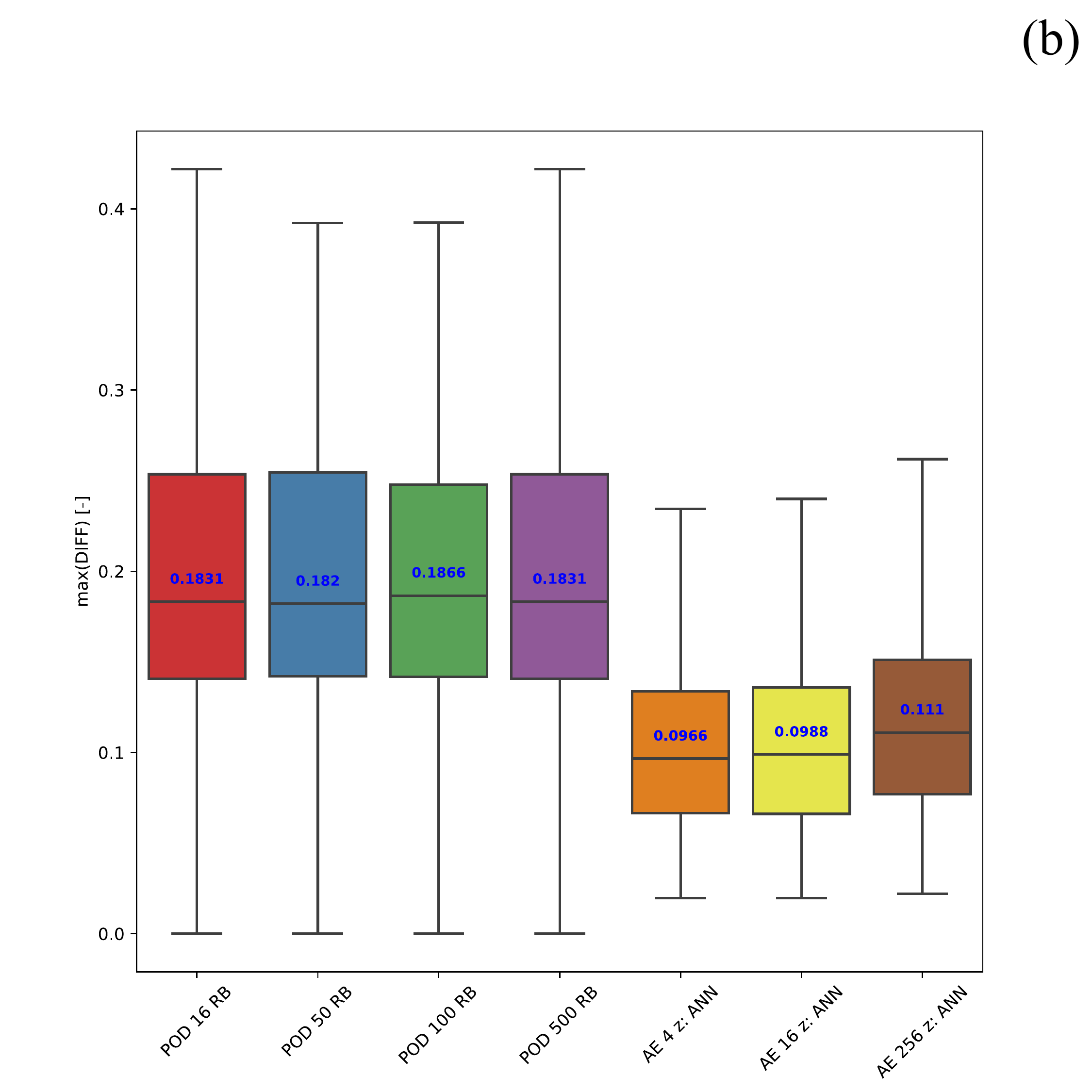}
   \caption{Example 2 - without convolutional layers: Each model's performance on a test set (a) mean square error (MSE) and (b) maximum DIFF (max(DIFF)). Blue text represents a mean value. 16 RB, 50 RB, 100 RB, and 500 RB represent linear compression models with $\left[\mathrm{N_{int}} = 16, \mathrm{N} = 16\right]$, $\left[\mathrm{N_{int}} = 50, \mathrm{N} = 50\right]$, $\left[\mathrm{N_{int}} = 100, \mathrm{N} = 100\right]$, and $\left[\mathrm{N_{int}} = 500, \mathrm{N} = 500\right]$, respectively.}
   \label{fig:ex2_test_no_conv}
\end{figure}

Here, we provide MSE and max(DIFF) results of Example 2 using an autoencoder without convolutional layers. The results are presented in Figure \ref{fig:ex2_test_no_conv}. We note that our autoencoder here utilizes seven linear hidden layers for its encoder as well as seven linear hidden layers for its decoder. Each hidden layer is subjected to \emph{tanh} activation function. Comparing to Figure \ref{fig:ex2_test}, it is clear that the models without convolutional layers have less accuracy than the ones that have. However, the autoencoder without convolutional layers still outperforms the linear compression approach for this Example 2. \par

\end{appendices}


\bibliographystyle{unsrtnat}
\bibliography{references} 

\end{document}